\documentclass[11pt]{article}
\usepackage{fullpage}
\usepackage{graphicx}
\usepackage{latexsym,amsmath,amssymb,amsthm,epic,eepic,multirow}
\usepackage{natbib}

\usepackage{multirow}
\usepackage{subfigure}
\usepackage{natbib}
\usepackage{algorithm}
\usepackage{algorithmic}

\usepackage{dsfont}

\newcommand{\PP}{\mathbb{P}}
\newcommand{\EE}{\mathbb{E}}
\newcommand{\R}{\mathbb{R}}

\newcommand{\eps}{\varepsilon}

\newcommand{\Var}{\mbox{Var}}

\newcommand{\bx}{\mathbf{X}}
\newcommand{\bz}{\mathbf{Z}}
\newcommand{\argmin}{\mathrm{argmin}}

\newcommand{\bel}{\begin{eqnarray}\label}
\newcommand{\eel}{\end{eqnarray}}
\newcommand{\bes}{\begin{eqnarray*}}
\newcommand{\ees}{\end{eqnarray*}}
\newcommand{\bei}{\begin{itemize}}
\newcommand{\eei}{\end{itemize}}

\def\E{{\mathbb{E}}}
\def\P{{\mathbb{P}}}
\def\veps{\varepsilon}
\def\hb{{\hat b}} 
\def\hbeta{{\hat \beta}}

\def\hsigma{{\hat \sigma}}

\def\sgn{\hbox{\rm sgn}}

\def\heps{{\hat\eps}}
\def\hepscent{{\hat\eps}_{\rm cent}}
\def\hX{{\hat X}}
\def\hY{{\hat Y}}
\def\hZ{{\hat Z}}

\newtheorem{theo}{Theorem}
\newtheorem{prop}{Proposition}

\usepackage{color}

\begin{document}

\title{High-dimensional simultaneous inference with the bootstrap}
  
\author{Ruben Dezeure\thanks{Partially supported by the Swiss National
    Science Foundation SNF 2-77991-14}, \
 Peter B\"uhlmann\thanks{Corresponding author: buhlmann@stat.math.ethz.ch}
 \ and Cun-Hui Zhang\thanks{Partially supported by NSF Grants DMS-12-09014
   and DMS-15-13378 and NSA Grant H98230-15-1-0040.}\\ 
Seminar for Statistics, ETH Z\"urich and Department of Statistics, Rutgers
University}


\maketitle

\begin{abstract}
We propose a residual and wild bootstrap methodology for individual and
simultaneous inference in high-dimensional linear models with possibly
non-Gaussian and heteroscedastic errors. We establish asymptotic
consistency for simultaneous inference for parameters in groups $G$,
where $p \gg n$, $s_0 = o(n^{1/2}/\{\log(p) \log(|G|)^{1/2}\})$  and
$\log(|G|) = o(n^{1/7})$, with  
$p$ the number of variables, $n$ the sample size and $s_0$ denoting the
sparsity. The theory is complemented by many
empirical results. Our proposed procedures are implemented in the
\texttt{R}-package \texttt{hdi} \citep{hdipackage}. 
\end{abstract}



\section{Introduction}
Recently, there has been growing interest for statistical inference,
hypothesis tests and confidence regions in high-dimensional models. In
fact, many applications nowadays involve high-dimensional models and thus,
accurate statistical inference methods and tools are very important. 
For general models and high-dimensional settings, sample splitting procedures
\citep{WR08,memepb09} and stability 
selection \citep{mebu10,shah13} provide some statistical 
error control and significance. For the case of a linear model
with homoscedastic and Gaussian errors, more recent and powerful techniques
have been proposed 
\citep{pb13,zhangzhang11,vdgetal13,jamo13b,meins13,foygcand14} and some of
these extend to generalized linear models. For a recent overview, see also
\citet{dezeureetal14}.  

We focus in this paper on a linear model 
\begin{eqnarray*}
Y = \bx \beta^0 + \eps,
\end{eqnarray*}
where we use the notation $Y$ for the $n \times 1$ response variable, $\bx$
for the $n \times p$ design matrix, $\beta^0$ for the vector of unknown
true regression coefficients, and $\eps$ for the errors; for more
assumptions see \eqref{lin.mod}. One goal is to
construct confidence intervals for individual coefficients $\beta_j^0$, for $j
\in \{1,\ldots ,p\}$, or corresponding statistical hypothesis tests of the form 
\begin{eqnarray*} 
H_{0,j}: \beta^0_j = 0\ \mbox{versus the alternative}\ H_{A,j}: \beta^0_j \neq 0\
  (j=1,\ldots ,p).
\end{eqnarray*}
More generally, for groups $G \subseteq \{1,\ldots ,p\}$ of variables, we consider 
\begin{eqnarray*}
H_{0,G}: \beta^0_j =
0\ \mbox{versus the alternative}\ H_{A,G}: \beta^0_j \neq 0\ \mbox{for
  some}\ j \in G,
\end{eqnarray*}
and of particular interest is also multiple testing 
adjustment when testing many individual or group hypotheses.  

In this work we will argue that the bootstrap is very
useful for individual and especially for simultaneous inference in
high-dimensional linear models, that is for testing individual or group
hypotheses $H_{0,j}$ or $H_{0,G}$, and for corresponding  
individual or simultaneous confidence regions. We thereby also
demonstrate its usefulness to deal with potentially heteroscedastic or
non-Gaussian errors. Instead of bootstrapping the Lasso estimator
directly (see also the comment in Section \ref{subsec.relwork}), we
propose to bootstrap the de-biased \citep{zhangzhang11} or de-sparsified
Lasso which is a regular non-sparse estimator achieving asymptotic efficiency
under certain assumptions \citep{vdgetal13}. This idea has been recently
also analyzed in \citet{ZhangCheng2016}: we will discuss the differences to
our work at the end of Section \ref{subsec.relwork}. 
We discuss several advantages
of bootstrapping the de-sparsified Lasso, including the issue of simultaneous
inference for large groups of variables and statistically efficient
multiple testing adjustment. These make our bootstrap approach a ``state of
the art tool'' for reliable inference in high-dimensional linear models
with potentially heteroscedastic and very non-Gaussian errors. The
resampling nature in general should further contribute additional stability and
robustness to statistical results and conclusions, cf. \citet{brei96b}.

From a computational point of view, the bootstrap scheme is feasible and
not substantially more expensive than the de-sparsified Lasso itself; 
especially when the number of variables is large, the extra cost of
bootstrapping is not very severe. 
The bootstrap procedures which we propose and discuss are implemented and
added to the \texttt{R}-package \texttt{hdi} \citep{hdipackage}. This supports
their use for practical analysis of high-dimensional data.

\subsection{Related work and our contribution}\label{subsec.relwork}

Besides the growing literature in assessing uncertainty in high-dimensional
statistical inference mentioned at the beginning of the introductory
section, the use of the bootstrap has been advocated in other
works. In particular, the recent contribution of
  \citet{ZhangCheng2016} is closely related to ours: more details are given
  below.
From a theoretical perspective, the results from
\citet{chernozhukov2013} are important for deriving results for
simultaneous inference based on the bootstrap. We extend their
  theory to analyze non-Gaussian (instead of Gaussian) multipliers in a
  wild bootstrap method: this extension seems worthwhile due to the
  advantages of non-Gaussian multipliers for wild bootstrapping
  \citep{mammenwild1993}.  
   
Bootstrapping the adaptive Lasso in high-dimensional linear models
has been put forward and analyzed by \citet{chatter11,chatter13}. A main
difference to our proposal is that their approach is for a sparse
Lasso-type estimator and they require a ``beta-min'' condition (saying that all
non-zero regression coefficients are sufficiently large in absolute value)
to ensure that the bootstrap captures the correct limiting distribution for
the non-zero parameters. We avoid a  ``beta-min'' assumption because it is
a main purpose of the inference method itself to find out which of the
underlying regression coefficients are sufficiently large or
not. Furthermore, from a practical perspective, bootstrapping a Lasso-type
(or other sparse) estimator will be severely exposed to the
super-efficiency phenomenon: it 
can be easily seen in numerical simulation studies, saying that inference
for non-zero regression coefficients can be very poor
\citep{dezeureetal14}. The bootstrap has 
also been used and studied in settings which are vaguely related to ours:
\citet{zhou14} presents an MCMC sampler for the distribution of an
augmented Lasso estimator which allows for some inferential tasks,
\citet{mckeague15} analyze the bootstrap for  marginal correlation
screening  for high-dimensional linear models, and \citet{shahpb15}
consider the use a bootstrap scheme for obtaining the exact distribution of
scaled residuals in a high-dimensional linear model with Gaussian errors,
which in turn enables to infer the distribution for any estimator or
function based on the scaled residuals.  

\paragraph{Recent work by \citet{ZhangCheng2016}, denoted here as
    ``ZC''.} These authors 
  have recently considered the idea of bootstrapping the de-sparsified
  Lasso; our contribution has been developed independently. Their work
  contains interesting results but we provide here a more general treatment
  which leads to wider applicability, better performance and weaker
  theoretical conditions. 

We discuss three different bootstrap methods: a residual bootstrap, a
multiplier wild bootstrap and a special version of a paired bootstrap
method, whereas ZC consider a Gaussian multiplier wild
bootstrap only. Our different procedures are motivated and carefully
discussed from the view point to deal with heteroscedastic errors while ZC
only deal with homoscedastic errors. We also allow for non-Gaussian
multipliers in the wild bootstrap, driven by the fact that non-Gaussian
multipliers are advantageous \citep{mammenwild1993}: this is in contrast to
ZC who consider Gaussian multipliers only and thus directly using results
from \citet{chernozhukov2013} for the Gaussian multiplier bootstrap. 

We advocate to bootstrap the entire de-sparsified Lasso estimator, using
the plug-in rule, whereas ZC only bootstrap the linearized part of the
estimator. In the presented theories, there is no need to bootstrap the
non-linear asymptotically negligible part of the estimator: finite sample
results though speak much in favor to bootstrap the entire estimator (as we
do): see ???. Bootstrapping the entire procedure also makes the ``RLDPE''
version of the de-sparsified unnecessary which has been in introduced by
\citet{zhangzhang11} to improve coverage of nominal confidence while paying a
price for efficiency; see Sections \ref{subsubsec:homoscedastic-gaussian}
and \ref{subsubsect:homoscedasticnongaus}.  

Regarding theory, our condition on the sparsity of the design is much
weaker than in ZC. We 
require an $\ell_1$-norm condition for the rows of the inverse covariance
matrix while they require a much more stringent $\ell_0$-sparsity
condition. The details are as follows: we require an $\ell_1$-norm
condition in the second part of (B2) which is implied by the
$\ell_0$-sparsity condition $s_j =
o(n/\log(p))$ where $s_j = \sum_{k\neq j}I(\Sigma_X^{-1})_{jk} \neq 0)$;
due to $\|\gamma_j\|_1 \le O(1)\sqrt{s_j}$ when 
$\lambda_{\min}(\Sigma) > c >0$. In
contrast, ZC require $s_j = o(\sqrt{n/\log(p)})$. For
details of notation see Section \ref{sec.consistency}.

\medskip
Our contribution here can be seen as a very general development of
bootstrap methods for the 
de-biased or de-sparsified Lasso for confidence intervals and hypotheses
testing in high-dimensional linear models with
potentially heteroscedastic and non-Gaussian errors, with a 
particular emphasis on simultaneous inference and multiple testing
adjustment. 
Our aim is to establish, by theory and empirical results, the
practical usefulness and reliability of the bootstrap for 
high-dimensional inference: our procedures are implemented in the
\texttt{R}-package \texttt{hdi} \citep{hdipackage}. 
  
%
%
%
%

\section{High-dimensional linear model and the de-sparsified
  Lasso}

We consider in this work a high-dimensional linear model
\begin{eqnarray}\label{lin.mod}
Y = \bx \beta^0 + \eps,
\end{eqnarray}
with $n \times 1$ response vector $Y$, $n \times p$ fixed design matrix
$\bx$, $p \times 1$ vector $\beta^0$ of the true underlying unknown regression
parameters and $n \times 1$ vector of error terms. The $n \times 1$ columns
vectors of $\bx$ are denoted by $X_j$ ($j=1,\ldots ,p$).   
The errors are assumed to be independent with mean $\EE[\eps_i] = 0$ but
potentially 
heteroscedastic with variances $\EE[\eps_i^2] = \sigma_i^2$. We note that
the case of fixed design arises when conditioning on the
covariables. We focus on the high-dimensional regime where the dimension $p
\gg n$ is much larger than sample size $n$. Then, the linearity itself is
not a real restriction, as discussed in Section \ref{subsec.modelmisspec}. 
The goal in this paper is
inference for the unknown parameter vector $\beta^0$, in particular in
terms of statistical hypothesis tests and confidence intervals. 
 
We propose to do such inference based on non-sparse estimators. The
non-sparsity of an  
estimator typically induces ``regularity'' and avoids the phenomenon of
super-efficiency: we believe that this classical viewpoint
\citep[cf.]{bicketal98} is
actually important and leads to much better performance for constructing
confidence intervals for non-zero parameters. Regularity typically enables
asymptotic 
normality and efficiency, and it is also advantageous for consistency of the
bootstrap due to fundamental results by \citet{gine1989} and \citet{gine1990}.

\subsection{The de-sparsified Lasso}

The de-biased Lasso \citep{zhangzhang11}, also called the de-sparsified
Lasso \citep{vdgetal13}, can be considered as a generalization of the
ordinary least squares approach to the high-dimensional setting. 

In the low-dimensional $p<n$ setting with $\bx$ having full rank $p$,
denote by $V_j$ the residual vector when doing an ordinary least
squares regression of $X_j$ versus $\textbf{X}_{-j}$. Then, the ordinary least
squares estimator for $\beta^0$ can be written as 
\begin{equation*}
  \hat{\beta}_j^{OLS}= \frac{V_j^T Y}{V_j^TX_j}.
\end{equation*}

When $p>n$, the $V_j$'s are zero vectors and we cannot use such a
construction. Instead, we consider the residuals $Z_j$ from a Lasso
regression of $X_j$ versus all others variables in $\textbf{X}_{-j}$:
\begin{eqnarray*}
& &\hat{\gamma}_j = \hat{\gamma}_j(\lambda_X) = \argmin_{\gamma_j} (\|X_j -
  \bx_{-j} \gamma_j\|_2^2/n + \lambda_X \|\gamma_j\|_1),\\
& &Z_j = X_j - \bx_{-j} \hat{\gamma}_j.
\end{eqnarray*}
We then project on these regularized residuals while introducing a bias:
\begin{eqnarray*}
\hat{\beta}_j^{'}=\frac{Z_j^T Y}{Z_j^TX_j}
               =\beta_j^0 + \sum_{k \neq j}\frac{Z_j^T
                 X_k}{Z_j^TX_j}\beta_k^0 + \frac{Z_j^T
                 \varepsilon}{Z_j^TX_j}. 
\end{eqnarray*}
The introduced bias $\sum_{k \neq j}\frac{Z_j^T X_k}{Z_j^TX_j}\beta_k^0$ can
be estimated and corrected for by plugging in the Lasso from a regression
of $Y$ versus $\bx$:
\begin{eqnarray*}
  \hat{\beta} = \hat{\beta}(\lambda) = \argmin_{\beta} (\|Y - \bx
  \beta\|_2^2/n + \lambda \|\beta\|_1).
\end{eqnarray*}
This gives us the de-biased or de-sparsified Lasso:
\begin{eqnarray}\label{desparsLasso-constr}
\hat{b}_j = \hat{\beta}_j^{'} - \sum_{k \neq j}\frac{Z_j^T
                X_k}{Z_j^TX_j}\hat{\beta}_k = \beta_j^0 + \sum_{k \neq
  j}\frac{Z_j^T 
                X_k}{Z_j^TX_j}(\beta_k^0 - \hat{\beta}_k) +
                \frac{Z_j^T \varepsilon}{Z_j^TX_j}.
\end{eqnarray}
The estimator $\hat{b}_j$ is not sparse anymore, and hence the name
de-sparsified Lasso \citep{vdgetal13}; we can also write it as 
\begin{eqnarray*}
\hat{b}_j = \hat{\beta}_j + \frac{Z_j^T (Y - \bx \hat{\beta})}{Z_j^T X_j},
\end{eqnarray*}
which means that it equals the Lasso plus a one step bias correction, and
hence the alternative name de-biased Lasso \citep{zhangzhang11}. In
the sequel, we use the terminology de-sparsified Lasso. 

When interested in all $j=1,\ldots ,p$, the procedure requires one to run the 
Lasso with tuning parameter $\lambda$ for the regression of $Y$ versus
$\bx$; and the nodewise Lasso \citep{mebu06} which means the Lasso for
every regression of $X_j$ versus $\bx_{-j}$ ($j=1,\ldots ,p$) with tuning
parameter $\lambda_X$ (the same for all $j$). The total
computational requirement is thus to run $p+1$ Lasso regressions which can be
substantial if $p$ is large. Luckily, parallel computation can be done very
easily, as implemented in \texttt{hdi} \citep{hdipackage,dezeureetal14}. 

It has been shown first by \citet{zhangzhang11}, for homoscedastic
errors, that under some conditions,
\begin{eqnarray}\label{asympivot}
(\hat{b}_j - \beta^0_j)/s.e._j \Rightarrow {\cal N}(0,1)\ \
  (j=1,\ldots,p),
\end{eqnarray}
with the approximate standard error given in Theorem \ref{th1b} or
\ref{th2} for the case of homoscedastic or heteroscedastic errors,
respectively. The convergence is understood as both $p \ge n \to
\infty$. For the homoscedastic case, the asymptotic variance reaches 
the semiparametric information bound \citep{vdgetal13}. 

%
%

Estimation of the standard error is discussed below in
Section \ref{subsec.se}. With an approximate pivot at hand, we can
construct confidence intervals and hypothesis tests: for homoscedastic
errors, this has been
pursued by various authors and \citet{dezeureetal14} present a review
and description how inference based on such pivots can be done with the
\texttt{R}-package \texttt{hdi} \citep{hdipackage}. 

In this work we will argue that bootstrapping the de-sparsified Lasso
$\hat{b}$ will bring additional benefits over the asymptotic inference
based on a Gaussian limiting distribution arising in \eqref{asympivot}. 






%
%
%
%
%
%
%
%
 
\subsection{Estimation of the standard error and robustness for
  heteroscedastic errors}\label{subsec.se}

Based on the developed theoretical results in Section
\ref{sec.consistency}, one can show that the asymptotic standard error of
the de-sparsified estimator behaves like 
\begin{eqnarray*}
s.e._j =  
n^{-1/2} \frac{\sqrt{\Var(n^{-1/2}
  \sum_{i=1}^n Z_{j;i} \eps_i)}}{|Z_j^T X_j/n|}
\end{eqnarray*}

For the case of homoscedastic i.i.d. errors with $\Var(\eps_i) =
\sigma_{\eps}^2$, the inverse of the standard error is then asymptotically
behaving like 
\begin{eqnarray*}
s.e._j = n^{-1/2} \frac{\sigma_{\eps} \|Z_j\|_2/\sqrt{n}}{|Z_j^T X_j/n|}.
\end{eqnarray*}
This suggests to use as an estimate 
\begin{eqnarray}\label{est.vareps}
& &\widehat{s.e.}_j = n^{-1/2} \frac{\hat{\sigma}_{\eps} \|Z_j\|_2/\sqrt{n}}{|Z_j^T X_j/n|},\nonumber\\
& &\hat{\sigma}_{\eps}^2 = \frac{1}{n-\hat{s}} \|Y - \bx \hat{\beta}\|_2^2,
\end{eqnarray}
with $\hat{s}$ the number of nonzero coefficients in the estimate
$\hat{\beta}$. This choice of $\hat{\sigma}_{\eps}^2$ is based on
the recommendation of 
\citet{reidtibsh13} and supported by our own empirical experience with
different variance estimators. This standard error estimate is
implemented in the \texttt{R}-package \texttt{hdi} \citep{hdipackage}.

For heteroscedastic but independent errors with $\Var(\eps_i) =
\sigma_i^2$, the asymptotic standard error behaves as 
\begin{eqnarray*}
& &s.e._{\mathrm{robust},j} = n^{-1/2} \frac{\omega_j}{|Z_j^T X_j/n|},\\
& &\omega_j^2 = n^{-1} \sum_{i=1}^n Z_{j;i}^2 \sigma_i^2.
\end{eqnarray*}
We then propose the robust estimator 
\begin{eqnarray}\label{se-robust}
& &\widehat{s.e.}_{\mathrm{robust},j} =  n^{-1/2}
    \frac{\hat{\omega}_j}{|Z_j^T X_j/n|},\nonumber\\ 
& &\hat{\omega}_j^2 = \frac{1}{n - \hat{s}}\sum_{i=1}^n (\hat{\eps}_i
    Z_{j;i} - n^{-1} 
    \sum_{r=1}^n \hat{\eps}_r Z_{j;r})^2,\ \ \hat{\eps} = Y - X
    \hat{\beta},
\end{eqnarray}
which has been used in \citet{misspecified-pbvdg2015} for the different
context of misspecified linear models with random design. We prove that
under some conditions, 
$\widehat{s.e.}_j/\Var(\hat{b}_j)^{1/2} = 1 + o_P(1)$ (Theorem \ref{th1b}
for the homoscedastic case) and
$\widehat{s.e.}_{\mathrm{robust},j}/\Var(\hat{b}_j)^{1/2} = 1 + o_P(1)$ 
(Theorem \ref{th2} for the heteroscedastic case). In fact, the robust
standard error estimator is consistent for both the homo- and
heteroscedastic case for the error terms: 
therefore, it is robust against heteroscedasticity which explains its
name. The phenomenon is closely related to the robust sandwich
estimator for the standard error of the MLE in low dimensional models
\citep{eicker1967limit,huber1967behavior,white1980heteroskedasticity,freedman1981}.  

We point out that the result 
\begin{eqnarray*}
(\hat{b}_j - \beta^0_j)/\widehat{s.e.}_{\mathrm{robust},j} \Longrightarrow {\cal
  N}(0,1),
\end{eqnarray*}
presented later in Theorem \ref{th2} is a new extension which covers the
case with heteroscedastic errors. All what is conceptually needed is the
robust standard error estimate $\widehat{s.e.}_{\mathrm{robust},j}$. 

\section{Bootstrapping the de-sparsified Lasso} 

We consider first a residual bootstrap procedure. Two alternative bootstrap
methods are discussed in Sections \ref{subsec.multiplbtrp} and
\ref{subsec.hetresbootstrap}. We use the Lasso 
for computing residuals $\hat{\eps} = Y - \bx
\hat{\beta}$ and centered residuals $\hat{\eps}_{\mathrm{cent},i} = \hat{\eps}_i -
\overline{\hat{\eps}}\ (i=1,\ldots ,n)$, where $\overline{\hat{\eps}} =
n^{-1} \sum \hat{\eps}_i$. The bootstrapped errors are then
constructed from the 
\paragraph{Residual bootstrap:} 
\begin{eqnarray*}
\eps_1^*,\ldots ,\eps_n^*\ \mbox{i.i.d. (re-)sampled from the centered
  residuals}\ \hat{\eps}_{\mathrm{cent},i}\ (i=1,\ldots ,n).
\end{eqnarray*}

\medskip
We then construct the bootstrapped response
variables as 
\begin{equation}
\label{eq:res.boot}
  Y^* = \bx \hat{\beta} + \eps^*.
\end{equation}
and the bootstrap sample is $\{(\bx_i,Y_i^*)\}_{i=1}^n$, reflecting the fact
of fixed (non-random) design. Here and in the sequel $\bx_i$ denotes the $p
\times 1$ row vectors of $\bx$ ($i=1,\ldots,n$). 

\subsection{Individual inference}

We aim to estimate the distribution of the asymptotic pivot (see Theorem
\ref{th1b} and \ref{th2}) 
\begin{equation}\label{pivot-robust}
  T_j = \frac{\hat{b}_j - \beta_j^0}{\widehat{s.e.}_{\mathrm{robust},j}},
\end{equation}
where $\hat{b}_j$ is the de-sparsified estimator and
$\widehat{s.e.}_{\mathrm{robust},j}$ is the robust standard error in
\eqref{se-robust}. We propose to always use this robust standard error in
practice because it automatically provides protection (robustness) against
heteroscedastic errors. At some places, we also discuss the use of the more
usual standard error formula $\widehat{s.e.}_j$ from \eqref{est.vareps} for
the case with homoscedastic errors: but this serves mainly for explaining
some conceptual differences. For estimating the distribution in
\eqref{pivot-robust}, we use the 
bootstrap distribution of 
\begin{equation}\label{pivot-bootstrap-robust}
  T_j^* = \frac{\hat{b}^*_j - \hat{\beta}_j}{\widehat{s.e.}^*_{\mathrm{robust},j}},
\end{equation}
where $\hat{b}^*_j$ and $\widehat{s.e.}^*_{\mathrm{robust},j}$ are computed
by plugging in the bootstrap sample instead of the original data points
(alternatively, when using the non--robust standard error, we would also
use the bootstrap for the non-robust version). Denote by $q^*_{j;\nu}$ the
$\nu$-quantile of the bootstrap 
distribution of $T_j^*$. We then construct two-sided $100(1-\alpha) \%$
confidence intervals for the $j$th coefficient $\beta^0_j$ as 
\begin{eqnarray}\label{bootCI}
\mathrm{CI}_j = [\hat{b}_j - q^*_{j;1-\alpha/2} \: \widehat{s.e.}_{\mathrm{robust},j}
  ,\hat{b}_j - q^*_{j;\alpha/2} \:  \widehat{s.e.}_{\mathrm{robust},j}].
\end{eqnarray}
Corresponding p-values for the null-hypothesis $H_{0,j}$ versus the two-sided alternative
$H_{A,j}$ can then be computed by duality. Bootstrapping pivots in
classical low-dimensional settings is known to improve the level of
accuracy of confidence intervals and hypothesis tests
\citep{hallwilson1991}. 


%

\subsection{Simultaneous confidence regions, intervals
  and p-values for groups}

We can construct simultaneous confidence regions over a
group of variables $G$. Rather than using the sup-norm, we build the region
\begin{eqnarray*}
C(1-\alpha) = \{b \in \R^p;\max_{j \in G} T_j \le q^*_{\mathrm{max;G}}(1-\alpha/2)\ \mbox{and}\ \min_{j \in G} T_j \ge q^*_{\mathrm{min;G}}(\alpha/2)\},
\end{eqnarray*}
where $q_{\mathrm{max};G}^*(\nu)$ is the $\nu$-quantile of the bootstrap
distribution of $\max_{j \in G} T_j^*$ and $q_{\mathrm{min};G}^*(\nu)$ the
$\nu$-quantile of the bootstrap distribution of $\min_{j \in G} T_j^*$,
respectively. 
If the group $G$ is large, a more informative view is to take the
componentwise version of $C(1-\alpha)$: for each component $j \in G$ we
consider the confidence interval for $\beta^0_j$ of the form
\begin{eqnarray}\label{bootsimCI}
\mathrm{CI}_{\mathrm{simult},j} = [\hat{b}_j -
  \widehat{s.e.}_{\mathrm{robust},j} q_{\mathrm{max};G}^*(1-
  \alpha/2),\hat{b}_j - \widehat{s.e.}_{\mathrm{robust},j}
  q_{\mathrm{min};G}^*(\alpha/2)].   
\end{eqnarray}
We may also replace $q_{\mathrm{max};G}^*(1-\alpha/2)$ and $q_{\mathrm{min};G}^*(\alpha/2)$ 
by $\pm q_{\mathrm{abs};G}^*(1-\alpha)$, where $q_{\mathrm{abs};G}^*(\nu)$ 
is the $\nu$-quantile of the bootstrap distribution of $\max_{j \in G}|T_j^*|$, resulting in slightly 
narrower simultaneous confidence intervals. In contrast to the confidence
intervals in \eqref{bootCI}, the intervals in 
\eqref{bootsimCI} are simultaneous and hence wider, providing approximate 
coverage in the form of 
\begin{eqnarray*}
\PP[\beta^0_j \in \mathrm{CI}_{\mathrm{simult},j}\ \mbox{for all}\ j \in G] \approx
  1- \alpha.
\end{eqnarray*}
Of  particular interest is the case with $G = \{1,\ldots ,p\}$. This
construction often provides shorter intervals than using a Bonferroni
correction, especially in presence of positive dependence. See also the
empirical results in Section \ref{subsec:closerlookmulttest} for the
related problem of adjustment for multiple testing.  

We might also be interested in p-values for testing the null-hypothesis
\begin{eqnarray*}
H_{0,G}: \beta^0_j = 0\ \mbox{for all}\ j \in G,
\end{eqnarray*}
against the alternative $H_{A,G}:\ \beta^0_j \neq 0\ \mbox{for some}\ j \in
G$. We consider the max-type statistics $\max_{j \in G} |T_j|$ which should
be powerful for detecting sparse alternatives. We can use the bootstrap
under $H_{0,G}$, or  
alternatively under the complete null hypothesis, $H_{0,\mathrm{complete}}:
\beta^0_j = 0\ \forall j= 1,\ldots ,p$, by exploiting (asymptotic)
restricted subset pivotality. The details are given in Section
\ref{sec:westfallyoung}. Resampling under $H_{0,\mathrm{complete}}$ is 
computationally much more attractive when considering many groups since we
can use the same bootstrap distribution to compute the p-values for many 
groups. The p-value is then given by 
\begin{eqnarray*}
P_G = \PP^{*0}[\max_{j \in G} |T_j^{*0}| > \max_{j \in G} |t_j|],
\end{eqnarray*}
where the asterisk ``$^{*0}$'' emphasizes that the bootstrap is constructed
under the complete null hypothesis $H_{0,\mathrm{complete}}$ and $t_j$ is
the observed realized value of the studentized statistics $T_j$. 

In the presence of heteroscedasticity, the residual bootstrap is
inconsistent for simultaneous inference, and the wild bootstrap or a
  paired bootstrap scheme described in Sections \ref{subsec.multiplbtrp}
  and \ref{subsec.hetresbootstrap} should be used instead.  

\subsection{Consistency of the residual bootstrap}\label{sec.consistency}

For deriving the asymptotic consistency of the bootstrap, we make the
following assumptions. 
\begin{description}
\item[(A1)] $\|\hat{\beta} - \beta^0\|_1 =
  o_P(1/\sqrt{\log(p) \log(1+|G|)})$. 
\item[(A2)] $\lambda_X \asymp \sqrt{\log(p)/n}$, 
$\|Z_j\|_2^2/n \ge L_Z$, 
$\|Z_j\|_{2+ \delta}^{2 + \delta} = o(\|Z_j\|_2^{2+ \delta})$, $j\in G$. 
\item[(A3)] $\eps_1,\ldots ,\eps_n$ independent, 
  $\EE[\eps] = 0$, $\EE\|\eps\|_2^2/n = \sigma_{\eps}^2$, 
  $L \le \EE|\eps_i|^2 = \sigma_{i}^2$, $\EE|\eps_i|^{2+ \delta} \le C$,
  for all $i$.  
\item[(A4)] $\|\hat{\beta}^* - \hat{\beta}\|_1 =
  o_{P^*}(1/\sqrt{\log(p) \log(1+|G|)})$ in probability.
\item[(A5)] $\max_{ij} |X_{ij}| \le C_X$.
\item[(A6)] 
$\max_{j\in G}\|Z_j\|_\infty \le K, \delta=2$, i.e. bounded 4th moment of $\eps$, $\log(|G|)=o(n^{1/7})$.
\end{description}
Here $\sigma_\eps$, $\delta$, $L$, $C$, $C_X$, $L_Z$ and $K$ are 
positive constants uniformly bounded away from 0 and $\infty$, 
and $G\subseteq \{1,\ldots,p\}$ indicates a set of variables of interest, 
e.g. $G=\{j\}$ for inference of a single $\beta_j$. 
As our theoretical results require no more than the fourth moment of $\eps$, 
we set $\delta \in (0,2]$ for simplicity without loss of generality. 
The constant $\delta$ is the same in (A2), (A3) and (A6), e.g. $\delta=2$ in (A3) when (A6) 
is imposed. Unless otherwise stated, (A2) is imposed with an arbitrarily small $\delta>0$ when $|G|=O(1)$, 
and strengthened with (A6) when $|G|\to\infty$.

 
\paragraph{Justification of (A1), (A2), (A4) and (A6).}
Sufficient assumptions for (A1), (A2), (A4) and (A6) (and choosing $\lambda_X \asymp
\sqrt{\log(p)/n}$) are as follows. 
\begin{description} 
\item[(B1)] the rows of the design matrix are i.i.d. realizations from a
  distribution with covariance matrix $\Sigma_X$, and the smallest
  eigenvalue of $\Sigma_X$ is larger than some $M > 0$. Furthermore: for
  some constants $C_1, C_2$,
$0 < C_1 \le 
\tau_j^2 
= 1/(\Sigma_X^{-1})_{jj} \le C_2 < \infty$.
\item[(B2)] $s_0 = o(\sqrt{n}/\{\log(p)\sqrt{\log(|G|)}\})$, 
$\sum_{k\neq j}|(\Sigma_X^{-1})_{jk}| \le o(\sqrt{n/\log p})$.  
\item[(B3)] The smallest sparse eigenvalue of $\bx^T
\bx/n$, with sparsity of the order $s_0$, is bounded from below by a positive
constant. 
\end{description}
Assumptions (B1, only the first requirement), (B2, only the first requirement) and (A5) imply that 
with high probability (w.r.t. i.i.d. sampling the rows of the design
matrix), (B3) and the compatibility condition for the set $S_0$ hold
\citep[Cor.6.8]{pbvdg11}.  
Alternatively, by Maurey's empirical method \citep{RudelsonZ13}, (B3) and (A5) 
directly imply the compatibility condition for deterministic design. 
It is known \citep[Th.6.1 and Ex14.3]{pbvdg11} that with the
compatibility condition for $S_0$ and $\lambda \ge 2\|\bx^T\eps/n\|_\infty$
we have that $\|\hat{\beta} - \beta^0\|_1 = O_P(s_0 \sqrt{\log(p)/n})$ and
thus, (B2) implies (A1). 

Let $\gamma^0_j$ be the population regression coefficients of $X_j$ versus
$\bx_{-j}$  and $Z_j^0 = X_j - \sum_{k\neq j}X_k(\gamma_j^0)_k$. 
By Nemirovski's inequality, (B1) and (A5) imply
$2\max_{k\neq j}|X_k^TZ_j^0/n|  \le \lambda_X$ with large probability 
for a certain $\lambda_X= O_P(\sqrt{\log(p)/n})$. 
For such $\lambda_X$, the second part of (B2) implies 
$$
\|Z_j - Z_j^0\|_2^2/n + 2^{-1}\lambda_X\|\hat{\gamma}_j\|_1 \le (3/2)\lambda_X\|\gamma^0_j\|_1=o(1). 
$$ 
As $\|Z_j^0\|_\infty\le C_X(1+\|\gamma^0_j\|_1) =o(\sqrt{n/\log p})$ by (A5), the Bernstein inequality gives
$$
\max_{j\le p}\left|\tau_j^2 -\|Z_j^0\|_2^2/n\right|=o_P(1).
$$ 
Thus, due to the second part of (B1) we have proved the requirement on
$\|Z_j\|_2^2/n$ in (A2) and (A6). 
Moreover, as $\|Z_j - Z_j^0\|_{2+\delta} \le \|Z_j - Z_j^0\|_{2} = o(n^{1/2})$,
$$
\|Z_j\|_{2+ \delta}^{2 + \delta} 
\le 2^{1+\delta}\Big(\|Z_j^0\|_2^2\|Z_j^0\|_\infty^\delta + \|Z_j - Z_j^0\|_{2}^{2+\delta}\Big)
\ll n^{1+\delta/2} \asymp \|Z_j\|_2^{2+\delta},
$$
which proves the last statement in (A2). If the second requirement of (B2)
is strengthened to  
$\max_{j\le p}\|\gamma_j^0\|_1 =C_{\Sigma}$, the $\ell_\infty$ bound in
(A6) follows from  
$\|Z_j\|_\infty \le (1+\|\hat{\gamma}_j\|_1)C_X\le (1+3C_{\Sigma})C_X$. 

Assumption (A4) holds when assuming (B1, only the first requirement), (B2,
only the first requirement) and (A5) (and these
assumptions imply the compatibility
condition as mentioned earlier), ensuring
that $\hat{s}_0 = \|\hat{\beta}\|_0 = O_P(s_0) =
o_P(\sqrt{n}/\log(p))$. The latter  
holds under a sparse eigenvalue condition on the design \citep{ZhangH08} or
when using 
e.g. the adaptive or thresholded Lasso in the construction of the bootstrap
samples \citep{geer11} and \citep[Ch.7.8-7.9]{pbvdg11}. 


\subsubsection{Homoscedastic errors}

The bootstrap is used to estimate the distribution of the studentized statistic 
\begin{eqnarray*} 
& & (\hat{b}_j - \beta^0_j)/\widehat{s.e.}_j,\\
& &1/\widehat{s.e.}_j = \sqrt{n} \frac{|Z_j^T
  X_j/n|}{\hat{\sigma}_{\eps} \|Z_j\|_2/\sqrt{n}}, 
\end{eqnarray*}
where $\widehat{s.e.}_j$ is the approximate standard error for $\hat{b}_j$ 
when the Lasso is nearly fully de-biased, with the estimated standard
deviation of the error. 

\begin{theo}\label{th1b}
Assume (A1)-(A5) with common $\E\,\eps_i^2=\sigma^2_\eps$ throughout the
theorem. Let $\P^*$ represent the residual bootstrap. Then,
\begin{eqnarray*}
& & T_j = (\hat{b}_j - \beta^0_j)/\widehat{s.e.}_j \Longrightarrow {\cal
  N}(0,1),\\
& & T^*_j = (\hat{b}^*_j - \hat{\beta}_j)/\widehat{s.e.}^*_j \buildrel {\cal
    D}^* \over \Longrightarrow {\cal
  N}(0,1)\ \mbox{in probability},
\end{eqnarray*}
for each $j\in G$. If $|G|=O(1)$, then, 
\bes
\sup_{(t_j, j\in G)} \left|\P^*\left[ T^*_j 
\le t_j, j\in G\right] 
- \P\left[ T_j 
\le t_j, j\in G\right] \right|=o_P(1). 
\ees
If (A6) holds, then 
\bes
\sup_{c \in \R}\left|\PP^*\left[\max_{j \in G}h\left(T^*_j 
\right) \le c\right] 
   - \PP\left[\max_{j \in G}h\left(T_j 
   \right) \le c\right]\right| = o_P(1)
\ees
for $h(t)=t$, $h(t)=-t$ and $h(t)=|t|$.
\end{theo}

A proof is given in Section \ref{subsec.proofth1b}. We note that Theorem
\ref{th1b} only requires a  
weak form of homoscedasticity in the sense of equal variance, 
instead of the stronger assumption of equal distribution, and that under
this weak homoscedasticity, the original and the bootstrap distributions
have asymptotically the same (estimated) standard 
errors
\begin{eqnarray*}
\widehat{s.e.}_j \sim \sqrt{\hbox{\rm Asymp.Var}(\hat{b}_j)}
\sim \sqrt{\hbox{\rm Asymp.Var}^*(\hat{b}^*_j)} \sim
  \widehat{s.e.}_j^*,
\end{eqnarray*}
where we omit that the statements are with high probability (in $P^*$
and/or in $P$). See also after the proof of Theorem \ref{th1b} in Section
\ref{subsec.proofth1b}. 

\subsubsection{Heteroscedastic errors}\label{subsec.heterogerr}

Consider the inverse of the robust standard error formula:
\begin{eqnarray*}
& &1/\widehat{s.e.}_{\mathrm{robust},j} = \sqrt{n} \frac{|Z_j^T
    X_j/n|}{\hat{\omega}_j},\\ 
& &\hat{\omega}_j^2 = n^{-1} \sum_{i=1}^n (\hat{\eps}_i Z_{j;i} - n^{-1}
    \sum_{r=1}^n \hat{\eps}_r Z_{j;r})^2.
\end{eqnarray*}
For deriving the consistency of the bootstrap in presence of
heteroscedastic errors, we remove the homoscedasticity assumption on the
variance, $\E\eps_i^2=\sigma^2_\eps$, imposed in Theorem \ref{th1b}.


\begin{theo}\label{th2}
Assume (A1)-(A5).  Let $\P^*$ represent the residual bootstrap. Then, for
each $j\in G$, 
\begin{eqnarray*}
& &(\hat{b}_j - \beta^0_j)/\widehat{s.e.}_{\mathrm{robust},j} \Longrightarrow
  {\cal N}(0,1),\\
& &(\hat{b}^*_j - \hat{\beta}_j)/\widehat{s.e.}^*_{\mathrm{robust},j}
    \buildrel {\cal
    D}^* \over \Longrightarrow
  {\cal N}(0,1)\ \mbox{in probability}. 
\end{eqnarray*}
\end{theo}
A proof is given in Section \ref{subsec.proofth2}. Different than for the
  homoscedastic case, the original and the
bootstrap distribution have asymptotically different (estimated) standard
errors
\begin{eqnarray*}
\widehat{s.e.}_{\mathrm{robust},j} \sim \sqrt{\hbox{\rm
  Asymp.Var}(\hat{b}_j)} \not\sim \sqrt{\hbox{\rm
  Asymp.Var}^*(\hat{b}^*_j)} \sim \widehat{s.e.}_{\mathrm{robust},j}^*,
\end{eqnarray*}
where we omit that the statements are with high probability (in $P^*$ and/or in $P$). Similarly, the residual bootstrap does not provide consistent estimation 
of the correlation between different $\hb_j$ to justify simultaneous inference 
as considered in Theorem \ref{th1b}.  
The reason is that the bootstrap constructs i.i.d. errors and does
not mimic the heteroscedastic structure in the original sample. See also
the sentences after the proof of Theorem \ref{th2} in Section
\ref{subsec.proofth2}. Simultaneous inference with heteroscedastic
  errors is treated in the following section.  

\section{Simultaneous inference with the bootstrap}\label{sec.multsiminf}

We discuss here the advantages of the bootstrap for simultaneous inference
and multiple testing adjustment in the presence of heteroscedasticity.  
Of particular interest here is the problem of simultaneous inference over a group $G \subseteq \{1,\ldots
,p\}$ of components of the regression parameter $\beta$, including the case
where $G = \{1,\ldots ,p\}$ is very large and includes all components. More
precisely, we want to estimate the distribution of 
\begin{eqnarray}\label{max-pivot}
\max_{j \in G} h(T_j),\ T_j = (\hat{b}_j -
  \beta^0_j)/\widehat{s.e.}_{\mathrm{robust},j}, 
\end{eqnarray}
by using the bootstrap for $h(t) = t$, $h(t)=-t$ and $h(t)=|t|$. 

We propose below bootstrap schemes which are consistent and work well for
either homoscedastic or heteroscedastic errors. 

\subsection{The multiplier wild bootstrap}\label{subsec.multiplbtrp}

We introduce a multiplier wild bootstrap
\citep{wu1986jackknife,liu1992efficiency,mammenwild1993}. Consider the 
centered residuals $\hat{\eps}_{\mathrm{cent}} =\hat{\eps} -
\overline{\hat{\eps}}$, where $\hat{\eps} = Y - \bx 
\hat{\beta}$, and construct the multiplier bootstrapped residuals
as 
\begin{eqnarray}\label{multipl-boot}
& &\eps_{i}^{*W} = W_i\, \hat{\eps}_{\mathrm{cent},i}\ (i=1,\ldots ,n),\nonumber\\
& &W_1,\ldots ,W_n\ \mbox{i.i.d. independent of the data with $\E W_i=0, \E W_i^2=1$ and $\E W_i^4<\infty$.}
\end{eqnarray}
We then proceed as with the standard residual bootstrap for constructing
$Y^* = \bx \hat{\beta} + \eps^{*W}$, and the bootstrap sample is then
$\{(X_i,Y^*_i)\}_{i=1}^n$ as input to compute the bootstrapped estimator 
$T_j^* = (\hat{b}_j^* - \hat{\beta}_j)/\widehat{s.e.}^*_{\mathrm{robust},j}$, i.e. using the
plug-in rule of the bootstrap sample to the estimator. 

This wild bootstrap scheme is asymptotically consistent for simultaneous
inference with heteroscedastic (as well as homoscedastic) errors, see
Section \ref{subsec.wildconsist}.

\subsection{The xyz-paired bootstrap}\label{subsec.hetresbootstrap} 

We modify here the paired bootstrap for regression \citep{efron1979,liu1992efficiency} to deal with the case of
heteroscedastic errors \citep{freedman1981}. 
As re-computation of $Z_j$ with bootstrap data would be expensive, we propose 
to append z-variables to the xy-matrix as additional regressors 
and bootstrap the entire rows of the xyz-matrix. 
However, to create an unbiased regression model for the bootstrap, the variables have to be correctly 
centered to assure $\E^*[(X_j^*)^T\eps^*]=\E^*[(Z_j^*)^T\eps^*]=0$.  
We note that this is not a problem in the low-dimensional case because the 
residual vector in the least squares estimation is automatically orthogonal to all design vectors. 
The wild bootstrap does not have a 
centering problem either because the newly generated multiplier variables $W_i$ all have zero mean. 
For the paired bootstrap, we propose to i.i.d. sample rows of the
$n\times(2p+1)$ matrix $({\hat\bx},\hY,{\hat\bz})$, and hence the
  name xyz-paired bootstrap,
\bes
\hX_j = X_j - \frac{X_j^T{\hat\eps}_{\rm cent}}{\|{\hat\eps}_{\rm cent}\|_2^2}{\hat\eps}_{\rm cent},\ 
\hY = {\hat\bx}\hbeta + \hepscent,\ 
\hZ_j = Z_j - \frac{Z_j^T{\hat\eps}_{\rm cent}}{\|{\hat\eps}_{\rm cent}\|_2^2}{\hat\eps}_{\rm cent}, 
\ees
where $\hepscent$ is as in the residual bootstrap. 
Indeed, for the resulting $(\bx^*,Y^*,\bz^*)$,  
\bes
\E^*[\eps^*] = \E^*[(X_j^*)^T\eps^*]=\E^*[(Z_j^*)^T\eps^*]=0\ \hbox{ with } \eps^*=Y^* - \bx^*\hbeta = (\hepscent)^*. 
\ees
The bootstrapped estimators $\hb_j^*$, ${\hat\omega}_j^*$ and $\widehat{s.e.}_{\mathrm{robust},j}^*$ are 
then defined by the plug-in rule as in wild bootstrap, with 
$T_j^* = (\hat{b}_j^* -\hbeta_j)/\widehat{s.e.}^*_{\mathrm{robust},j}$.


The xyz-paired bootstrap is shown to be consistent for simultaneous
inference with heteroscedastic errors, see Section
  \ref{subsec.wildconsist}. However, limited empirical  
results (not shown in the paper) suggested that it may not be competitive in
comparison to the Gaussian multiplier wild bootstrap from Section
\ref{subsec.multiplbtrp}. 


\subsection{The Westfall-Young procedure for multiple testing adjustment}\label{sec:westfallyoung}

The Westfall-Young procedure \citep{westyoung93} is a very attractive
powerful approach for multiple testing adjustment based on resampling. It uses
the bootstrap to approximate joint distributions of p-values and
test statistics, therefore taking their dependencies into account. This in
turn leads to efficiency gains: the procedure has been proven for certain
settings to be (nearly) optimal for controlling the familywise error rate
\citep{memabu11}.   

A standard assumption for the Westfall-Young procedure is the so-called
subset pivotality for the statistics $T_j =
\hat{b}_j/\widehat{s.e.}_{\mathrm{robust},j}$ (or using the version for the
homoscedastic case with $\widehat{s.e.}_{j}$). Note that in this subsection,
$T_j$ is without the centering at $\hat{\beta}_j$.  
\begin{description}
\item[(subs-piv)] Subset pivotality holds if, for every possible subset G,
  the marginal distribution for $\{T_j\ j \in G\}$   
  remains the same under the restriction $H_{0,G}:\ \beta_j = 0$ for all $j
  \in G$ and $H_{0,\mathrm{complete}}:\ \beta_j = 0$ for all $j=1,\ldots ,p$.
\end{description}
When focusing specifically on a max-type statistics, we can weaken subset
pivotality to a restricted form. 
\begin{description}
\item[(restricted subs-piv)] Restricted subset pivotality holds if, for
  every possible subset G, the 
  distribution of $\max_{j \in G}|T_j|$ remains the same under the
  restriction $H_{0,G}:\ 
  \beta_j = 0$ for all $j \in G$ and $H_{0,\mathrm{complete}}:\ \beta_j =
  0$ for all $j=1,\ldots ,p$. 
\end{description}
Subset pivotality can be justified in an asymptotic sense. For groups $G$
with finite cardinality, Theorem \ref{th1b} and Theorem \ref{th.max} presented below 
imply that asymptotic
subset pivotality holds. For large groups $G$ (with $|G|$ as large as $p \gg
n$) and assuming Gaussian errors, the restricted form of subset pivotality
holds, see e.g. \citet{zhangzhang11}. For large groups and non-Gaussian
errors, restricted subset pivotality can be established under the
conditions in Theorems \ref{th1b} and \ref{th.max} 
(the proof of these theorems implies the restricted subset pivotality, by using
arguments from \citet{chernozhukov2013}). 

Assuming restricted subset pivotality (in an asymptotic sense) we
immediately obtain that for any group $G \subseteq \{1,\ldots ,p\}$:  
\begin{eqnarray}\label{restr-subset-pivot}
\PP[\max_{j \in G} |T_j| \le c| H_{0,G}] = \PP[\max_{j \in G} |T_j| \le c|
  H_{0,\mathrm{complete}}]\ \ (c \in \R),
\end{eqnarray}
or its asymptotic version with approximate equality. 
This suggests to approximate the distribution of $\max_{j \in G} |T_j|$
under the complete null-hypothesis $H_{0,\mathrm{complete}}$ by using a bootstrap
scheme under the complete null-hypothesis $H_{0,\mathrm{complete}}$. We use 
\begin{eqnarray}\label{bootcomplete}
Y^{*0} = \eps^{*W}\ \mbox{for the multiplier wild bootstrap in
  \eqref{multipl-boot}}, 
\end{eqnarray}
that is, the construction as before but replacing $\hat{\beta}$ by the zero
vector. For the heteroscedastic residual bootstrap, this means that we
perform i.i.d. resampling of the rows of
$(\hat{\eps}_{\mathrm{cent}},X,Z_j)$. We notationally emphasize the
bootstrap under $H_{0,\mathrm{complete}}$ by the 
asterisk ``$^{*0}$''. The bootstrap approximation is then as follows:
\begin{eqnarray*}
\PP^{*0}[\max_{j \in G} |T_j^{*0}| \le c] \approx \PP[\max_{j \in G} |T_j| \le c|
  H_{0,\mathrm{complete}}]\ \ (c \in \R),
\end{eqnarray*}
and when invoking \eqref{restr-subset-pivot} we obtain that $\PP^{*0}[\max_{j
  \in G} |T_j^{*0}| \le c] \approx \PP[\max_{j \in G} |T_j| \le c|
H_{0,G}]$. A rigorous justification for this approximation and the parallel approximation by the xyz-paired bootstrap is given in Theorem \ref{th.max} below. 

We then easily obtain multiplicity adjusted p-values which approximately
control the familywise error rate for testing all the hypotheses $H_{0,j}:
  \beta^0_j = 0$ for all $j=1,\ldots ,p$: 
\begin{equation*}
P_{j,corr} = \PP^{*0}[\max_{k \in \{1,\ldots ,p\}} |T_k^{*0}| > |t_j|], 
\end{equation*}
where $T_k = \hat{b}_k/\widehat{s.e.}_{\mathrm{robust},k}$ (or using the
non-robust version $\widehat{s.e.}_k$), $T_k^{*0}$ its bootstrapped version
under $H_{0,\mathrm{complete}}$ using \eqref{bootcomplete} and $t_j$ is the
observed, realized value of the test statistic $T_j$. 
Because the bootstrap is
constructed under the complete $H_{0,\mathrm{complete}}$ we can compute
the bootstrap distribution of $\max_{k \in \{1,\ldots ,p\}} |T_k^{*0}|$ once
  and then use it to calibrate the p-values for all components $j =
  1,\ldots,p$: obviously, this is computationally very efficient. 


As described in \citet{westyoung93}, this method improves upon
Bonferroni-style and Sidak adjustments, mainly because the bootstrap is
taking dependence among the test statistics into account and hence is not
overly conservative like the Bonferroni-type or Sidak
correction. Furthermore, the Westfall-Young method doesn't rely on the
assumption that the p-values are uniformly distributed under $H_{0;j}$, for
all $j$. Finally, a Bonferroni-type correction goes far into the tails of
the distributions of the individual test statistics, in particular if $p$
is large: one typically would
need some importance sampling for a computationally efficient bootstrap
approximation of a single test statistics in the tails. We found that
the Westfall-Young method is much less exposed to this issue (because
the maximum statistics is directly bootstrapped without doing additional
corrections in the tail).  

\subsection{Consistency of the multiplier wild and xyz-paired bootstrap}\label{subsec.wildconsist}

We discuss under which assumptions the multiplier wild and
xyz-paired bootstrap 
schemes achieve 
consistency for estimating the distribution of $T_j = (\hat{b}_j - \beta^0_j)/\widehat{s.e.}_{\mathrm{robust},j},\ 
\max_{j \in G}(\pm T_j)$, and $\max_{j\in G}|T_j|$,
where $G \subseteq \{1,\ldots ,p\}$. 
The centered and standardized bootstrapped estimator is $T_j^* 
= (\hat{b}_j^* - 
\hat{\beta}_j)/\widehat{s.e.}^*_{\mathrm{robust},j}$. 

\begin{theo}\label{th.max}
Assume (A1)-(A5) (and thus allowing for heteroscedastic errors). Let $\P^*$ represent the multiplier wild
bootstrap. Then,
\begin{eqnarray*}
& & T_j = (\hat{b}_j - \beta^0_j)/\widehat{s.e.}_{\mathrm{robust},j} \Longrightarrow {\cal
  N}(0,1),\\
& & T^*_j = (\hat{b}^*_j - \hat{\beta}_j)/\widehat{s.e.}^*_{\mathrm{robust},j} \buildrel {\cal
    D}^* \over \Longrightarrow {\cal
  N}(0,1)\ \mbox{in probability},
\end{eqnarray*}
for each $j\in G$. If $|G|=O(1)$, then, 
\bes
\sup_{(t_j, j\in G)} \left|\P^*\left[ T^*_j 
\le t_j, j\in G\right] 
- \P\left[ T_j 
\le t_j, j\in G\right] \right|=o_P(1). 
\ees
If (A6) holds, then 
\bes
\sup_{c \in \R}\left|\PP^*\left[\max_{j \in G}h\left(T^*_j 
\right) \le c\right] 
   - \PP\left[\max_{j \in G}h\left(T_j 
   \right) \le c\right]\right| = o_P(1)
\ees
for $h(t)=t$, $h(t)=-t$ and $h(t)=|t|$.  

Moreover, all the above statements also hold when $\P^*$ represents
  the xyz-paired bootstrap, provided that 
$\delta=2$ in (A2) and (A3), $\log p = o(n^{1/2})$ and 
$\max_{j\in G}(\|Z_j\|_2/|Z_j^TX_j|) = o_P(1/\sqrt{\log(2|G|)})$.
\end{theo}

A proof is given in Section \ref{subsec.proofthmax}. We note that the
assumption (A4) is meant to be with respect to the multiplier wild or the
paired xyz-bootstrap, respectively: it is ensured by the same conditions as
outlined in Section \ref{sec.consistency} and \ref{subsec.heterogerr}. 

For the xyz-paired bootstrap, the additional condition $\log p = o(n^{1/2})$ 
is a consequence of (A1) and the $\ell_1$ minimax rate of the Lasso 
\citep{YeZ10}, and upper bounds of the form 
$\max_j \|Z_j\|_2/|Z_j^TX_j| = O_P(n^{-1/2})$, implying the requirement in Theorem \ref{th.max} 
and the uniform $n^{-1/2}$ rate for the standard error of $\hb_j$
can be found in 
\cite{zhangzhang11} and \cite{vdgetal13}.
 

\subsubsection{Conceptual differences between the multiplier wild and
  residual bootstrap}\label{subsec.multiplbtrpdisc}

We briefly discuss some conceptual differences between the multiplier and
residual bootstrap while (mostly) not distinguishing whether the inference is
simultaneous or for individual parameters (the residual bootstrap also
works for simultaneous inference as discussed in Theorem \ref{th1b}.

The multiplier wild bootstrap leads to the correct standard error of the
estimator for both cases of either homo- or heteroscedastic errors, i.e., 
\begin{eqnarray*}
\widehat{s.e.}^*_{\mathrm{robust},j} \sim \sqrt{\hbox{Asym.Var}^*(\hat{b}_j^*)} \sim
  \sqrt{\hbox{Asym.Var}(\hat{b}_j)} \sim \widehat{s.e.}_{\mathrm{robust},j}.
\end{eqnarray*}  
The asymptotic equivalence $\hbox{Asym.Var}^*(\hat{b}_j^*) \sim
  \hbox{Asym.Var}(\hat{b}_j)$ 
  does not hold for the residual bootstrap in the case of
  heteroscedastic errors. However, this property is not needed when
  constructing the inference based on the pivots as in \eqref{bootCI}, and
  the absence of the asymptotic equivalence between 
  studentized $\hb_j^*$ and $\hb_j$ 
  is theoretically supported by Theorem 
  \ref{th2}. Nevertheless, the fact that the residual bootstrap does not
  capture the correct asymptotic variance in the non-standardized case,
  which has been a major reason to introduce the wild bootstrap
  \citep{mammenwild1993}, might remain a disadvantage for the residual
  bootstrap. 

When the multiplier variables $W_i$ are i.i.d. ${\cal N}(0,1)$, the wild
bootstrap as in \eqref{multipl-boot} induces an
\emph{exact} Gaussian distribution (given the data) for the linear part $Z_j^T
\eps^*/Z_j^T X_j$, the leading term of $\hat{b}^*_j$. 
This is considered in \citet{ZhangCheng2016}.
For the finite sample
case with non-Gaussian errors, the distribution of the original quantity $Z_j^T
\eps/Z_j^T X_j$ is non-Gaussian: by construction, the Gaussian multiplier bootstrap
cannot capture such a non-Gaussianity. The residual bootstrap is better
tailored to potentially pick-up such non-Gaussianity and hence might have an
advantage over the Gaussian multiplier wild bootstrap. 
Still, if heteroscedasticity is a concern, one should use non-Gaussian
multipliers  as advocated in \citet{mammenwild1993} and justified in
Theorem \ref{th.max}. 

Our limited empirical results suggest that the residual and Gaussian
multiplier wild bootstrap lead to very similar empirical results in terms
of type I (actual level of significance for tests, and actual confidence
coverage) and type II errors (power of tests, and size of confidence
regions) for (i) the case of homoscedastic errors and for individual and
simultaneous inference, (ii) the case of heteroscedastic errors and
individual inference when using the robust standard error formula for the
residual bootstrap. For the case of heteroscedastic errors and simultaneous
inference, the wild bootstrap seems to be the preferred method. 
Some supporting empirical results are given in Section
\ref{app:multiplier-quantileest0p95} and \ref{app:additionalmultiplier}. 

\section{Empirical results}\label{sec:empresults}

We compare the bootstrapped estimator to the original de-sparsified Lasso
in terms of single testing confidence intervals and multiple testing
corrected p-values. We also consider the restricted low-dimensional
projection estimator (RLDPE) which has been introduced by
\citet{zhangzhang11} as a version of the de-biased (or de-sparsified) Lasso
to enhance reliability of coverage while paying a price for
efficiency; and we also compare with the ZC approach from
\citet{ZhangCheng2016} which applies the bootstrap only to the linear part
of the de-sparsified estimator without bootstrapping the estimated bias
correction term. We always consider the residual bootstrap, unless
  explicitly specified that the wild bootstrap (with Gaussian multipliers)
  is used. Moreover, when considering scenarios with homoscedastic errors,
  we always studentize with the non-robust estimator $\widehat{s.e.}_j$ and
  for heteroscedastic errors, we always studentize with the robust
  estimator $\widehat{s.e.}_{\mathrm{robust},j}$ (unless specified differently).
\begin{figure}[!htb]
\centering
\includegraphics[scale=0.7]{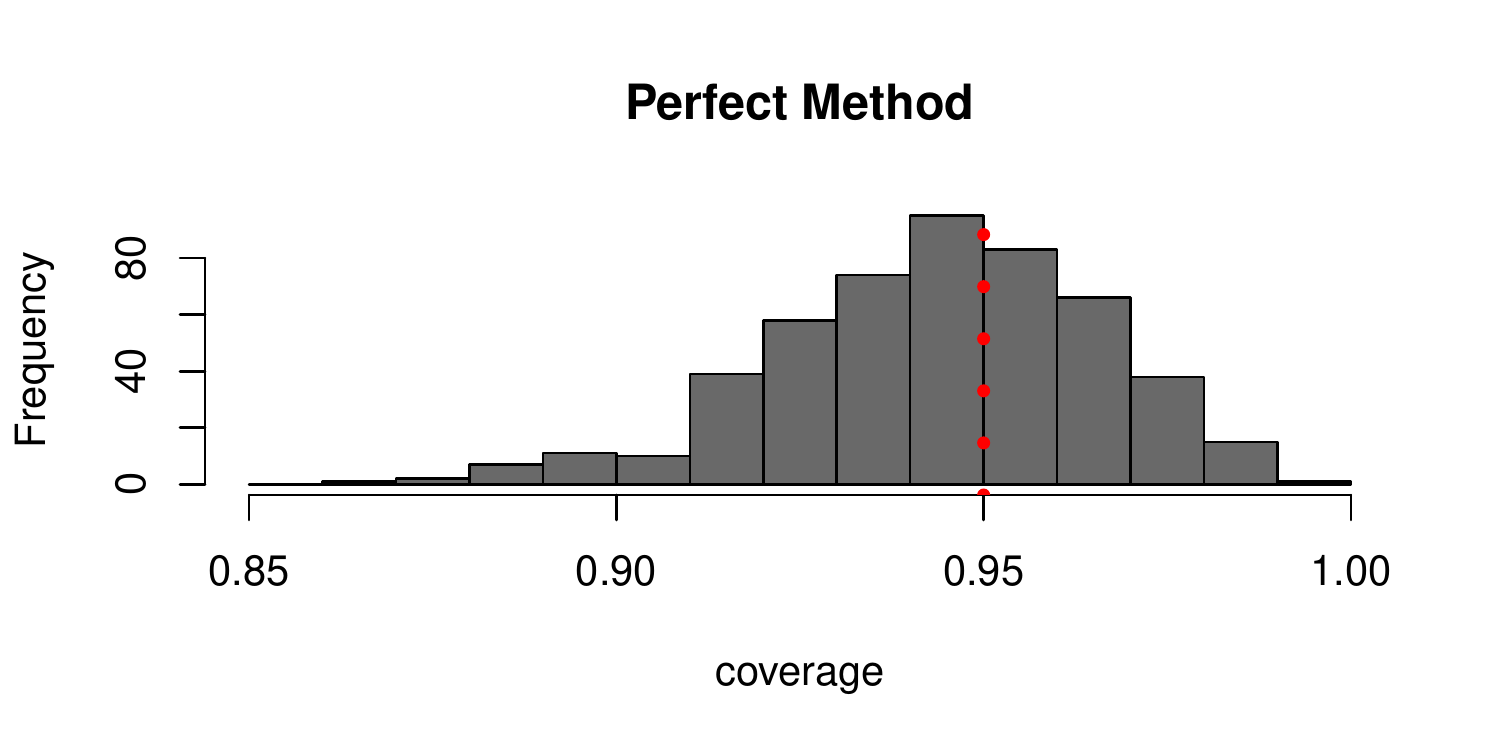}
\caption{Histogram of the coverage probabilities of two sided 95\%
  confidence intervals for  500 parameters. It illustrates how the results
 look like for a perfectly correct method for creating confidence 
 intervals and one uses only 
 100 realizations to compute the probabilities.}
\label{fig:perfectci-hist}
\end{figure}
\begin{figure}[!htb]
\centering
\includegraphics[scale=0.7]{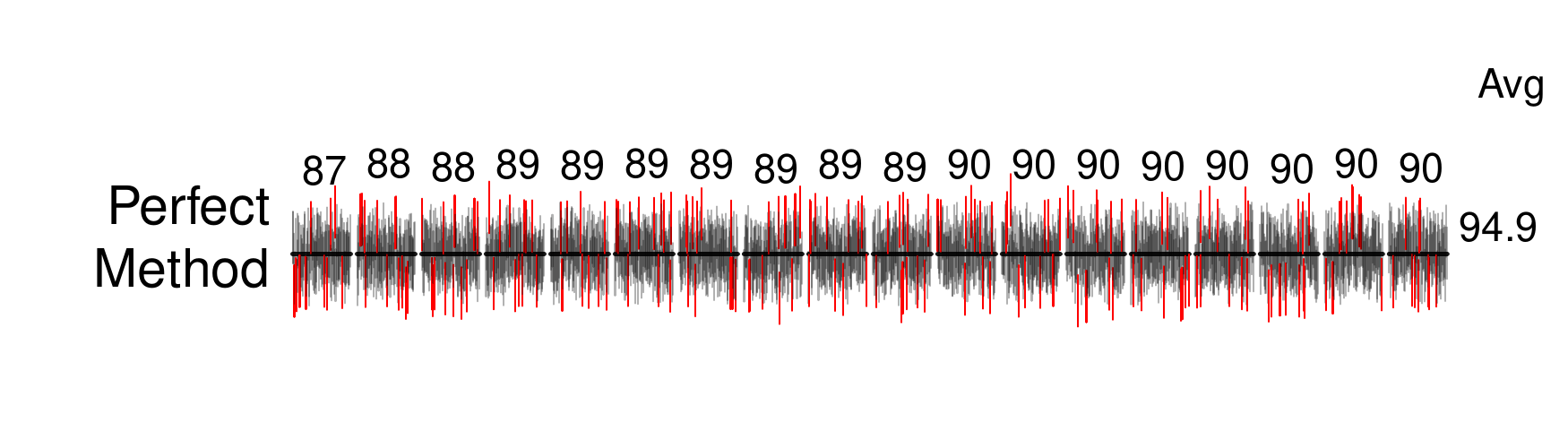}
\caption{Plot of two sided 95\% confidence
 intervals. It illustrates how the results would look like for a correct
 method for creating confidence intervals when one only computes 100
 confidence intervals. 18 coefficients are chosen and are drawn in 18
 columns from left to right with a black horizontal bar indicating the
 coefficient size. If any 
 coefficients differ from zero then they are plotted first from the left
 (in order of decreasing magnitude). This particular example doesn't 
 exhibit any of those non-zero coefficients. The other coefficients are
 chosen to be those with 
 the lowest coverage such that we can investigate potential causes for this poor
 coverage.
 The 100 computed confidence intervals are drawn from left to right in the
 column for the corresponding coefficient. The line segment is colored
 black in case it contains the truth, red otherwise. The number of
 confidence intervals that cover the truth for a particular coefficient
 is written above the confidence intervals in the respective column. The
 overall average coverage probability over all coefficients is displayed
 in the right-most column. }
\label{fig:perfectci}
\end{figure}
\begin{figure}[!htb]
\centering
\includegraphics[scale=0.5]{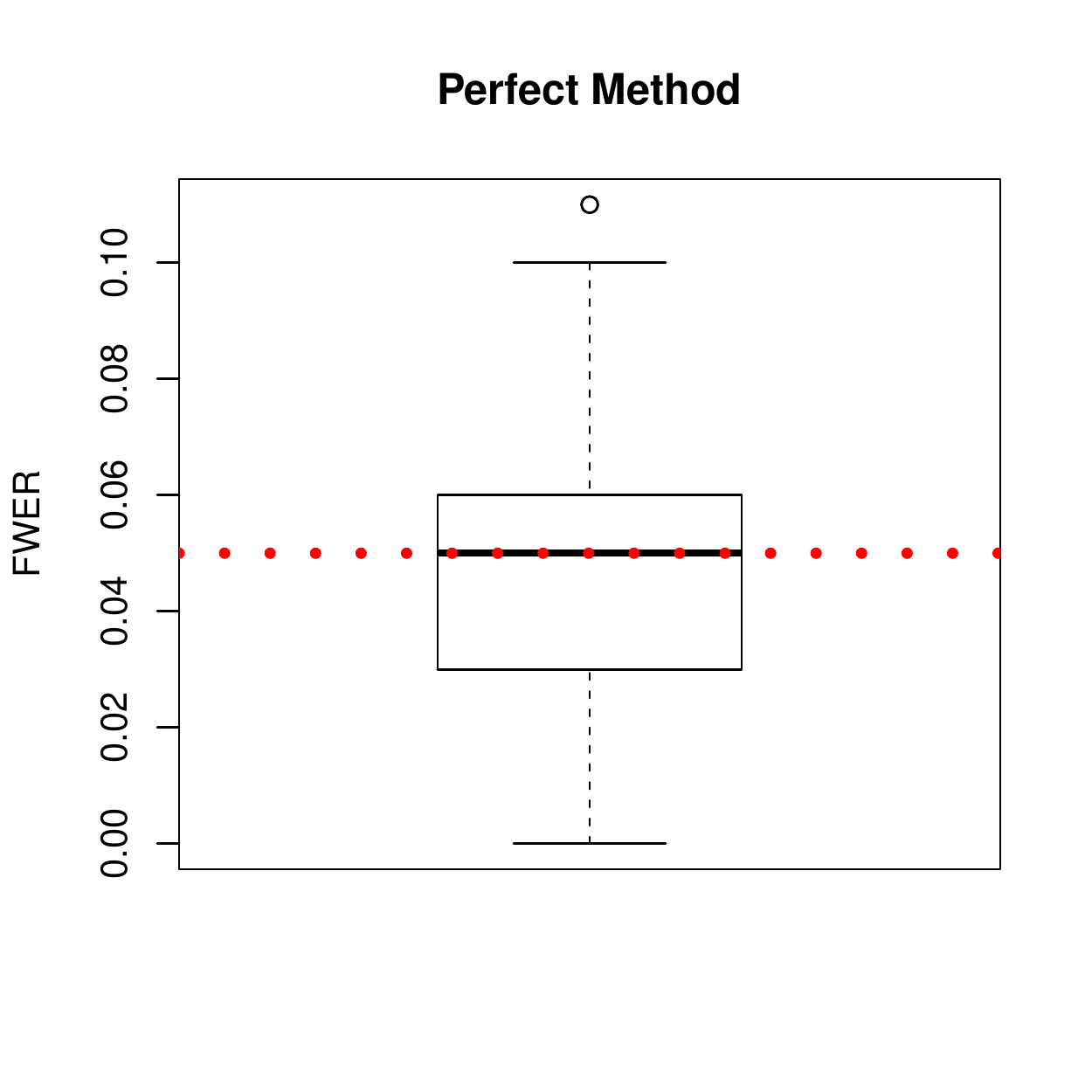}
\caption{Plot of multiple testing performance in
 terms of familywise error rate (FWER) control. It illustrates how the results
 look like for a correct method for multiple testing
 correction, if one computes the error rates over 100 realizations of
 the model. The target is controlling the FWER at level $0.05$. This
 target is highlighted by a red-dotted horizontal line. We sample  
 independent and identically distributed p-values $p_j \sim U(0,1)$, for
 $j=1,\dots,500$, and compute the familywise error rate over 100
 realizations when using the rejection threshold
 $\alpha=0.05/500=0.0001$. The boxplot based on 300 data points is the
 result of repeating this experiment 300 times.} 
\label{fig:perfectfwer}
\end{figure}

Of
particular interest is the accuracy of the bootstrap 
when dealing with non-Gaussian and even heteroscedastic
errors. For multiple testing, one would like to find out how much there is
to gain when using the Westfall-Young procedure over a method that
doesn't exploit dependencies between the outcomes of the tests, such as
Bonferroni-Holm. To this end, it is 
interesting to look at a variety of dependency structures for the design
matrix and to look at real data as well.

For confidence intervals, we visualize the overall average 
coverage probability as well as the occurrence of too high or too low coverage
probabilities. We work with histograms of the coverage
probabilities for all coefficients in the model, as in example Figure
\ref{fig:perfectci-hist}. These probabilities are always computed based on 100
realizations of the corresponding linear model. For those cases where
coverage is too 
low, we visualize the confidence intervals themselves to illustrate the
poor coverage. An example of the plot we'll work with can be found in Figure
\ref{fig:perfectci}. 

For multiple testing, we look at the power and the familywise
error rate, 
\begin{eqnarray*}
\mbox{Power} = \sum_{j \in S_0} \PP[H_{0,j}\mbox{ is rejected}]/s_0,\\
\mbox{FWER} = \PP[\exists j \in S_0^c : H_{0,j}\mbox{ is rejected}],
\end{eqnarray*}
where the probabilities are computed based on 100 realizations of the linear
model.

We use boxplots to visualize the power and error rates, similar to
Figure \ref{fig:perfectfwer}, where each data point is the result
of the probability calculation described above. In order to generate interesting and
representative data points, we look at different choices for the signal and
different seeds for the data generation. As a rule, results for different design types are
put in separate plots.


\subsection{Varying the distribution of the errors}
\label{subsec:varyingepsdist}
We first consider the performance of the bootstrap when varying the distribution
of the errors for simulated data. 

The design matrix will be generated $\sim
\mathcal{N}_p(0,\Sigma)$ with a covariance matrix $\Sigma$ of two possible
types (although mainly of Toeplitz type):   
\begin{eqnarray*}
\mbox{Toeplitz:}& &\ \Sigma_{j,k} = 0.9^{|j-k|}.\\
\mbox{Independence:}& &\ \Sigma = I_p.
\end{eqnarray*}

In case the model contains signal, the coefficient vector will have $s_0=3$
coefficients that differ from zero. The coefficients are picked in 6
different ways: 
\begin{eqnarray*}
\mbox{Randomly generated}:& \mbox{U(0,2), U(0,4), U(-2,2),}\\
\mbox{A fixed value}:& \mbox{1, 2 or 10.}\\
\end{eqnarray*}


\subsubsection{Homoscedastic Gaussian errors}
\label{subsubsec:homoscedastic-gaussian}

Data is generated from a linear model with Toeplitz design matrix and
homoscedastic Gaussian errors of variance $\sigma^2 = 1$, $\varepsilon \sim
\mathcal{N}_n(0,I_n)$. The sample size is chosen to be $n=100$, the number
of parameters $p=500$. 

\begin{figure}[!htb]
 \centering
\includegraphics[scale=0.7]{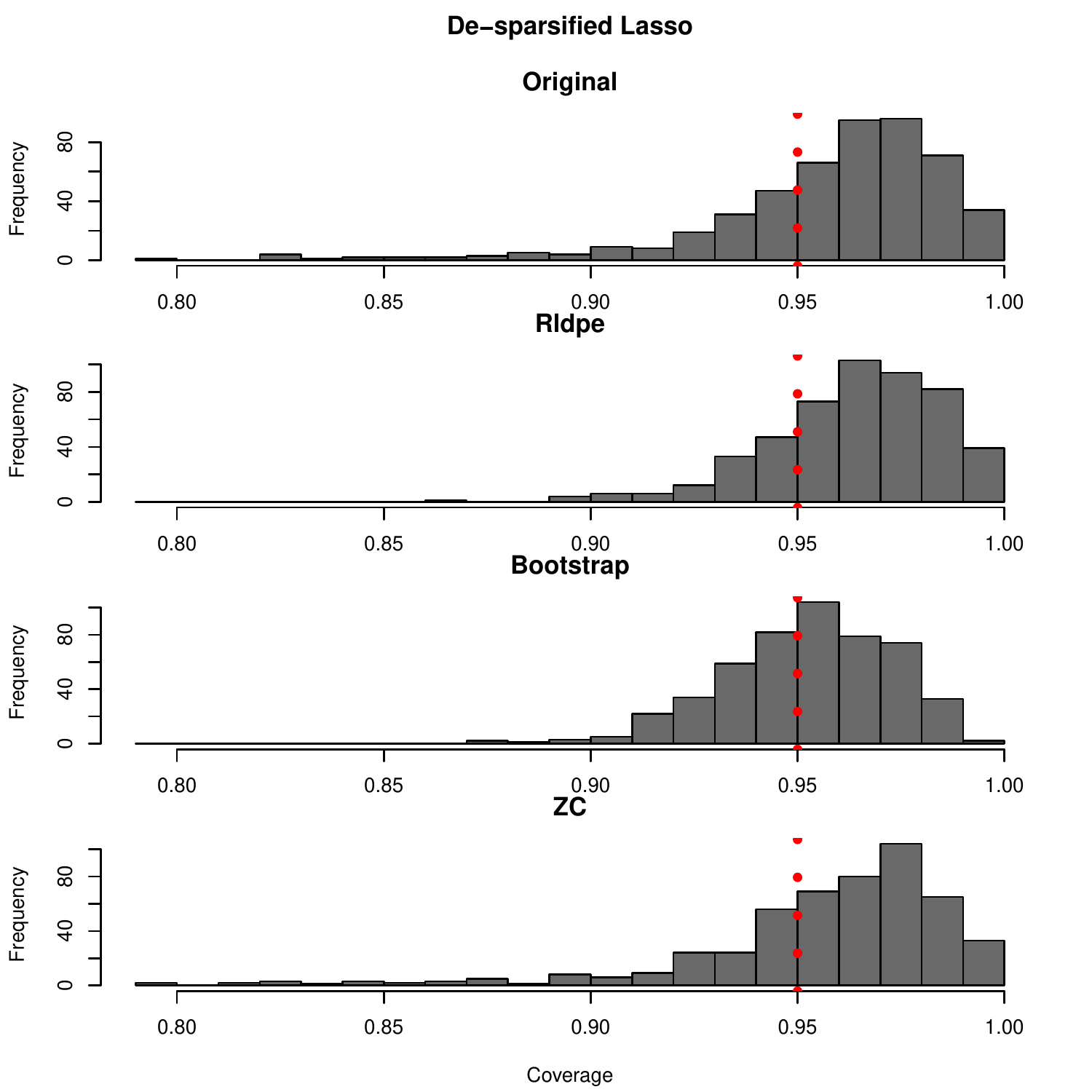}
 \caption{Histograms of the coverage probabilities of two-sided 95\%
   confidence intervals for all $500$ parameters in a linear model ($n=100,
   p=500$), computed from 100 independent replications.  Perfect
   performance would look like Figure \ref{fig:perfectci-hist}. The fixed
   design matrix is of Toeplitz type, the single coefficient 
   vector of type $U(-2,2)$ and 
   \textbf{homoscedastic Gaussian errors}. The original estimator has more
   over-coverage and under-coverage than the bootstrapped estimator. The
   RLDPE estimator has little under-coverage, like the 
   bootstrapped estimator, but it has too high coverage probabilities
   overall. The ZC approach to bootstrapping,
     which only bootstraps the linearized part of the estimator, doesn't
     show any improvements over the original de-sparsified Lasso.}
 \label{fig:cihist-toeplitzgauss}
\end{figure}

\begin{figure}[!htb]
 \centering
 \includegraphics[scale=0.7]{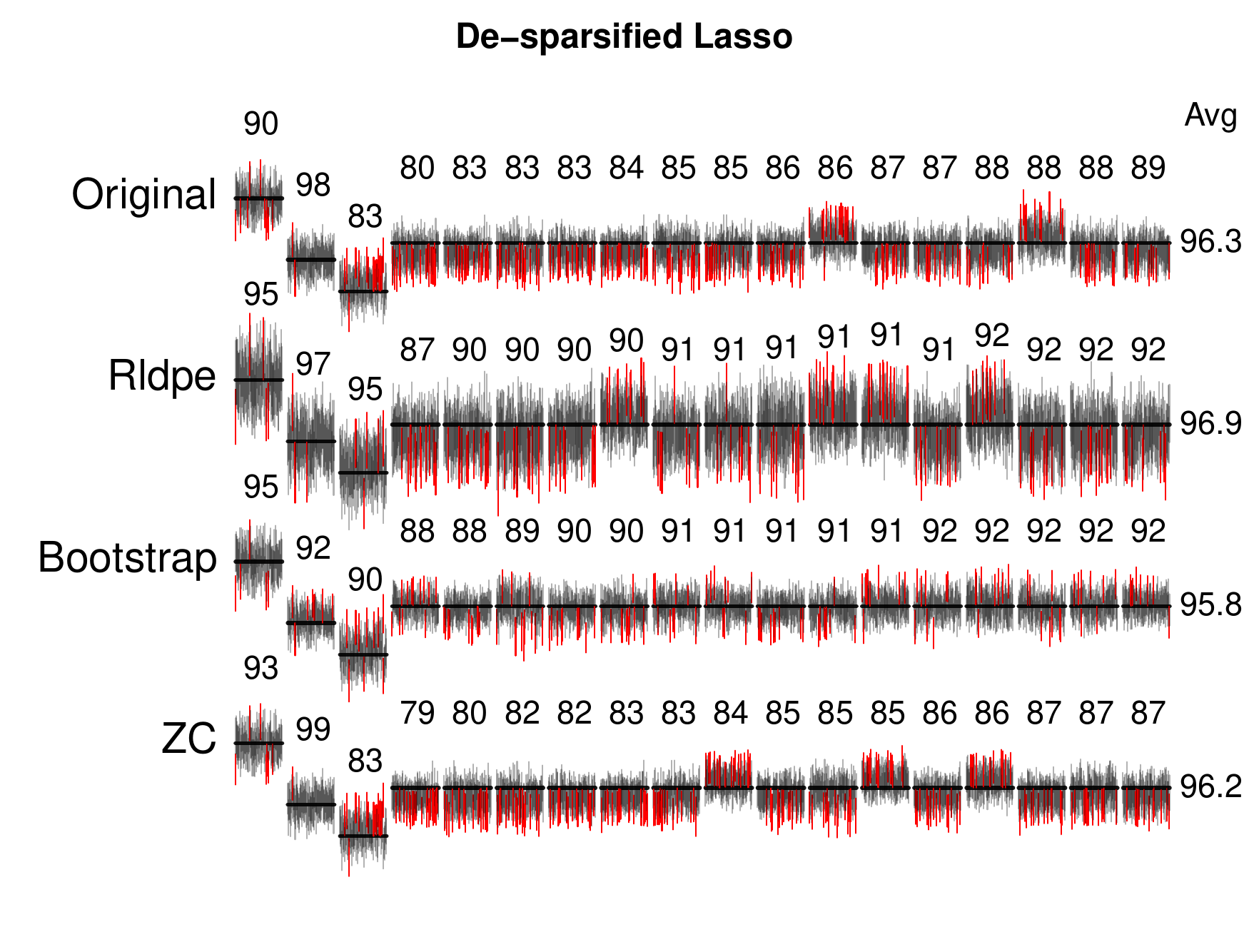}
 \caption{Two-sided 95\% confidence intervals for the de-sparsified Lasso
   estimator. From left to right 18 coefficients are shown with a black
   horizontal bar of a certain height 
   illustrating the value of the coefficient. Only the first three
   coefficients differ from zero. The other 15 coefficients presented are
   those with the lowest confidence interval coverage for that particular
   method (in decreasing order from left to right).
   100 response vectors were
     generated for a linear model with 
     \textbf{homoscedastic Gaussian errors}, fixed design of type Toeplitz,
     a single coefficient vector of type $U(-2,2)$, sample size
     $n=100$ and dimension $p=500$.
   Each of these realizations was fitted to produce a confidence
   interval for each coefficient in the model. The 100 confidence
   intervals are drawn as vertical lines and ordered from left to right in the column
   corresponding to that particular coefficient. The line segments are colored
   black if they cover the true coefficient and colored red otherwise. The
   number above each coefficient corresponds to the number of confidence
   intervals, out of 100, which end up covering the truth. The average
   coverage probability over all coefficients is
   provided in a column to the right of all coefficients. The original
   estimator has some bias for a few coefficients, which results in a
   lower than desired coverage for those coefficients. The RLDPE estimator
   has wider confidence intervals exhibiting over-coverage. The ZC
     approach to bootstrapping, which only bootstraps the linearized part
     of the estimator, doesn't show any improvements over the original
     de-sparsified Lasso.}
 \label{fig:ci-toeplitzgauss}
\end{figure}

For confidence intervals, we focus on one generated design
matrix $\bx$ and one generated coefficient vector of type $U(-2,2)$. The
histograms for the coverage probabilities can be found in Figure
\ref{fig:cihist-toeplitzgauss}. The coverage probabilities are more correct
for the bootstrapped 
estimator. The original estimator has a bias for quite a few coefficients resulting in low coverage,
as can be seen in Figure \ref{fig:ci-toeplitzgauss}. In addition, it tends
to have too high coverage for many coefficients. The conservative RLDPE
estimator has much wider confidence intervals  
which addresses the problem of low coverage but results in too high overall
coverage. 

\begin{figure}[!htb]
 \centering
\includegraphics[scale=0.65]{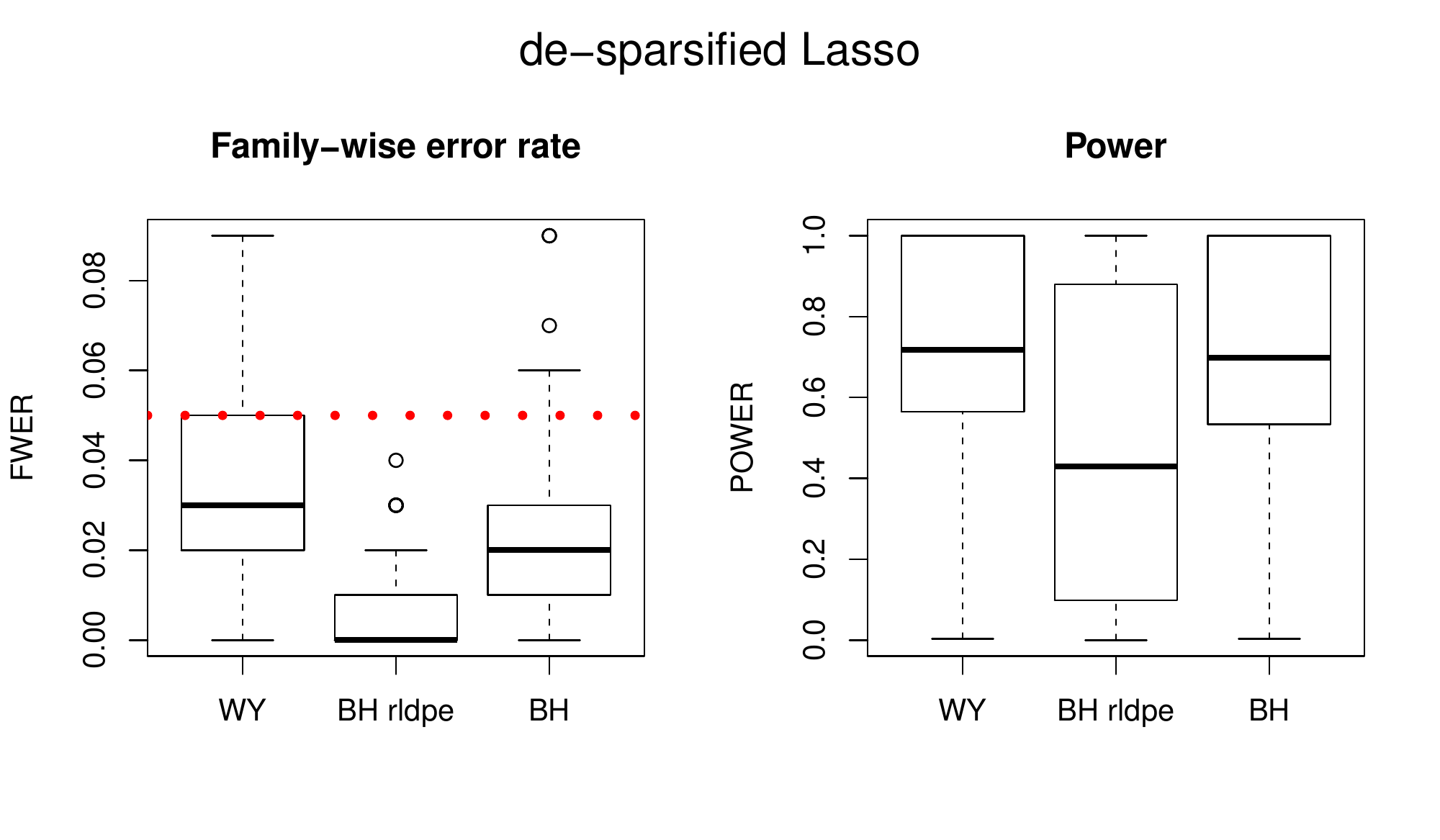}
 \caption{Boxplot of the familywise error rate and the power for
   multiple testing for the de-sparsified Lasso. The target is
   controlling the FWER at level $0.05$, highlighted by
   a red-dotted horizontal line. Two different approaches
   for multiple testing correction are compared, Westfall-Young (WY) and
   Bonferroni-Holm (BH). For Bonferroni-Holm, we make the distinction
   between the original method and the RLDPE approach. 300 linear models
   are investigated in total, 
   where 50 Toeplitz design matrices are combined with 50 coefficient
   vectors for each of the 6 types $U(0,2), U(0,4), U(-2,2), \mathrm{fixed
     \:} 1, \mathrm{fixed \:} 2, \mathrm{fixed \:} 10$. The variables
   belonging to the active set are chosen randomly. The errors in the
   linear model were  
   chosen to be \textbf{homoscedastic Gaussian}. Each of the
   models has a data point for the error rate and the power in the
   boxplot. The error rate and power probabilities 
   were calculated by averaging over 100 realizations.}

 \label{fig:powerfwer-toeplitzgauss}
\end{figure}

For multiple testing, we generate 50 Toeplitz design matrices $\bx$ which
are combined with 50 coefficient vectors for each coefficient type $U(0,2),
U(0,4), U(-2,2), \mathrm{fixed \:} 1, \mathrm{fixed \:} 2$ and $\mathrm{fixed
  \:} 10$. For each of these $300$ linear models, the coefficient vector
undergoes a different random permutation. A value for the familywise
error rate and power is then computed by generating 100 realizations of
the linear models, as described in the introduction of Section
\ref{sec:empresults}. The boxplots of the power and familywise error rate
can be found in Figure \ref{fig:powerfwer-toeplitzgauss}.
The bootstrap is the least conservative option. In addition, one can
conclude that it still has proper error control by comparing the results to
perfect error control in Figure \ref{fig:perfectfwer}. One would expect
to see a difference in power, but there doesn't seem to be a visible
difference between the bootstrap approach and the original estimator for
our dataset. The RLDPE estimator, on the other hand, does turn out to be
more conservative.

\subsubsection{Homoscedastic non-Gaussian errors}
\label{subsubsect:homoscedasticnongaus}
Data is generated from a linear model with Toeplitz design matrix and
homoscedastic centered chi-squared errors $\eps_1,\ldots ,\eps_n,$ of variance
$\sigma^2=1$, 
\begin{eqnarray*}
  \zeta_1,\ldots, \zeta_n \: \mbox{i.i.d.}  \: \sim \chi_1^2,\ \ \eps_i =
                            \frac{\zeta_i-1}{\sqrt{2}}, \: \: \: i=1,\dots,n.
\end{eqnarray*}
The sample size is chosen to be $n=100$, the number of parameters $p=500$. 

\begin{figure}[!htb]
 \centering
\includegraphics[scale=0.6]{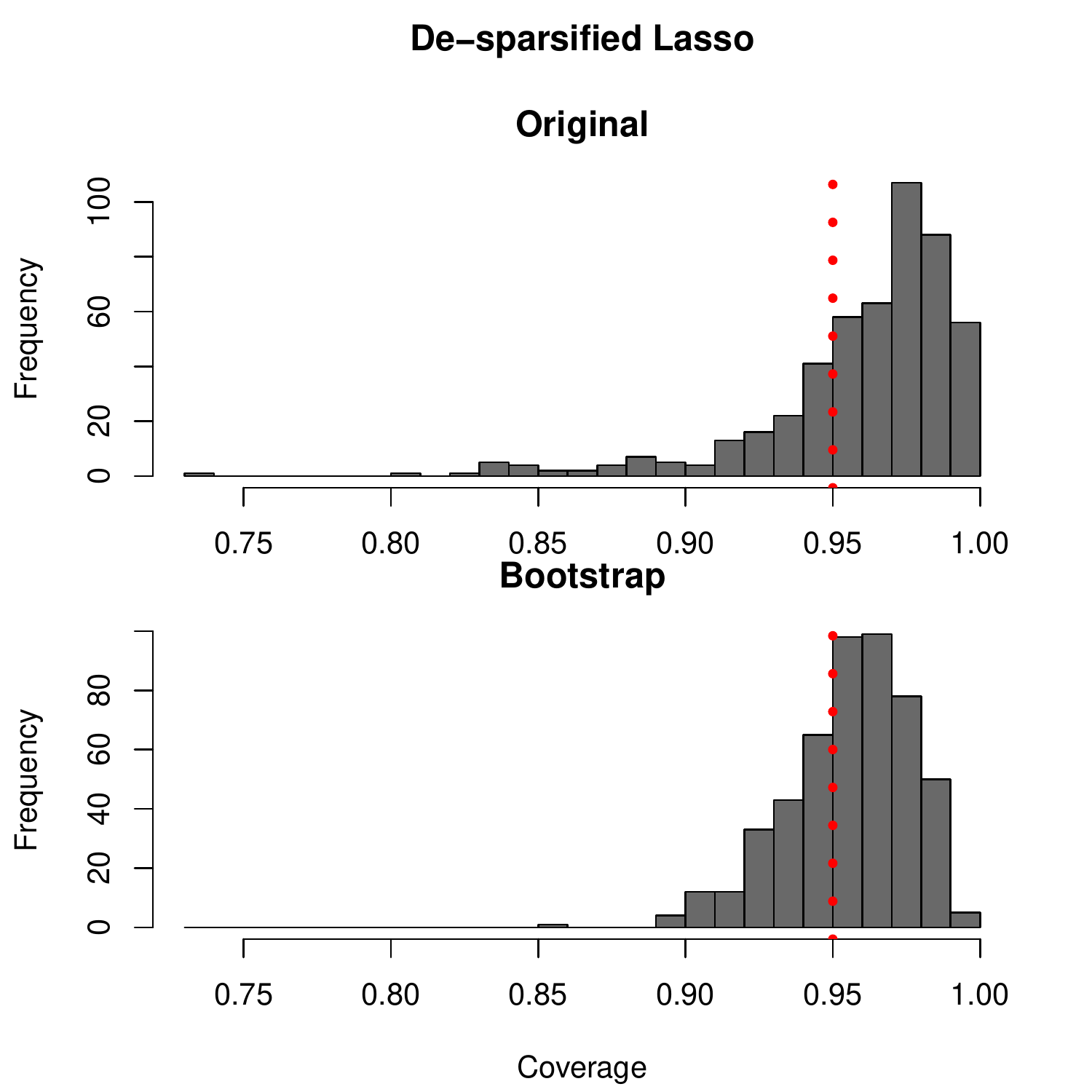}
 \caption{The same plot as Figure \ref{fig:cihist-toeplitzgauss} but for
   \textbf{homoscedastic chi-squared errors}. 
   The bootstrapped estimator has better coverage properties.} 
 \label{fig:cihist-toeplitzchisq}
\end{figure}

For confidence intervals, we focus on one generated design
matrix $\bx$ and one generated coefficient vector of type $U(-2,2)$. The
histograms for the coverage probabilities can be found in Figure
\ref{fig:cihist-toeplitzchisq}. 

\begin{figure}[!htb]
 \centering
\includegraphics[scale=0.7]{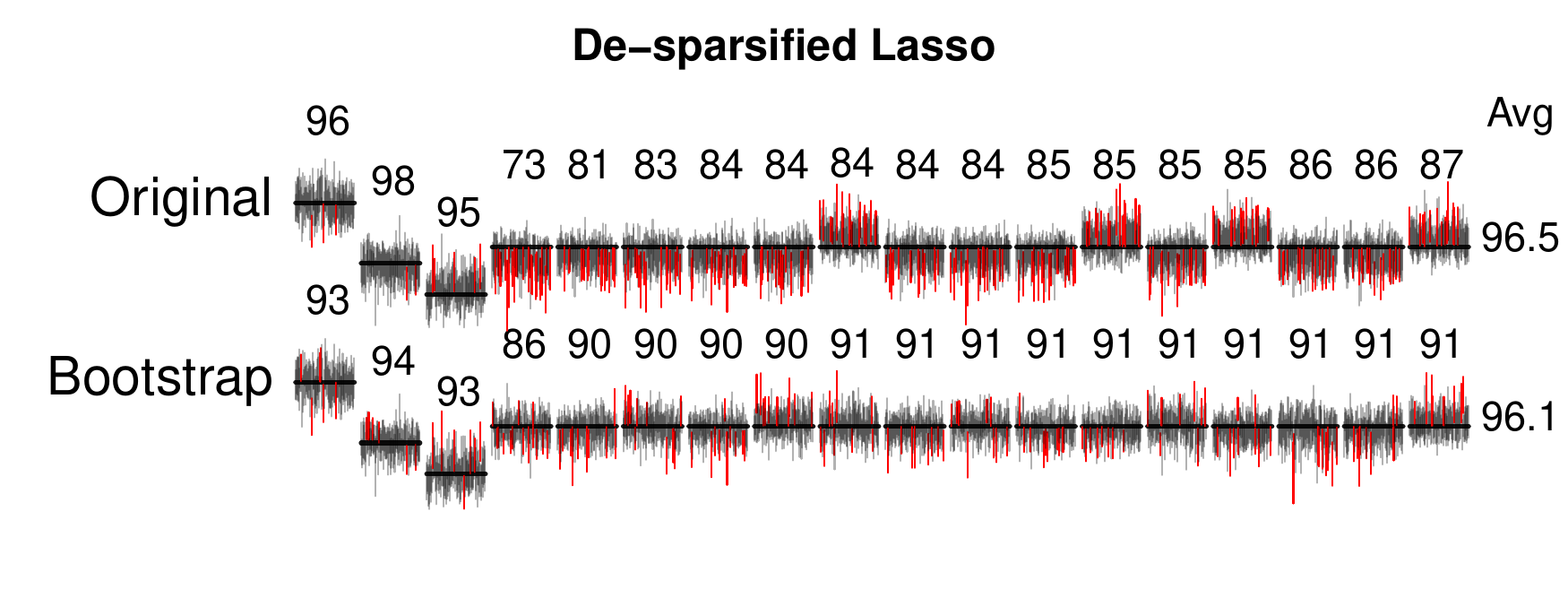}
 \caption{The same plot as Figure \ref{fig:ci-toeplitzgauss} but for
   \textbf{homoscedastic chi-squared errors}.  The original estimator has
   quite some bias for a few coefficients, which 
   results in a lower than desired coverage for those coefficients.} 
 \label{fig:ci-toeplitzchisq}
\end{figure}

The performance for the confidence intervals looks similar to that for
Gaussian errors, only the under coverage of the original estimator is even
more pronounced. The 
coverage for the bootstrapped estimator looks as good as in the
Gaussian case. As can be seen in Figure \ref{fig:ci-toeplitzchisq}, the cause for the
poor coverage of the non-bootstrapped estimator is again bias. Using
the robust standard error estimation doesn't impact the results, as can be
seen in Appendix \ref{app:additionalsimulations}.

\begin{figure}[!htb]
 \centering
\includegraphics[scale=0.65]{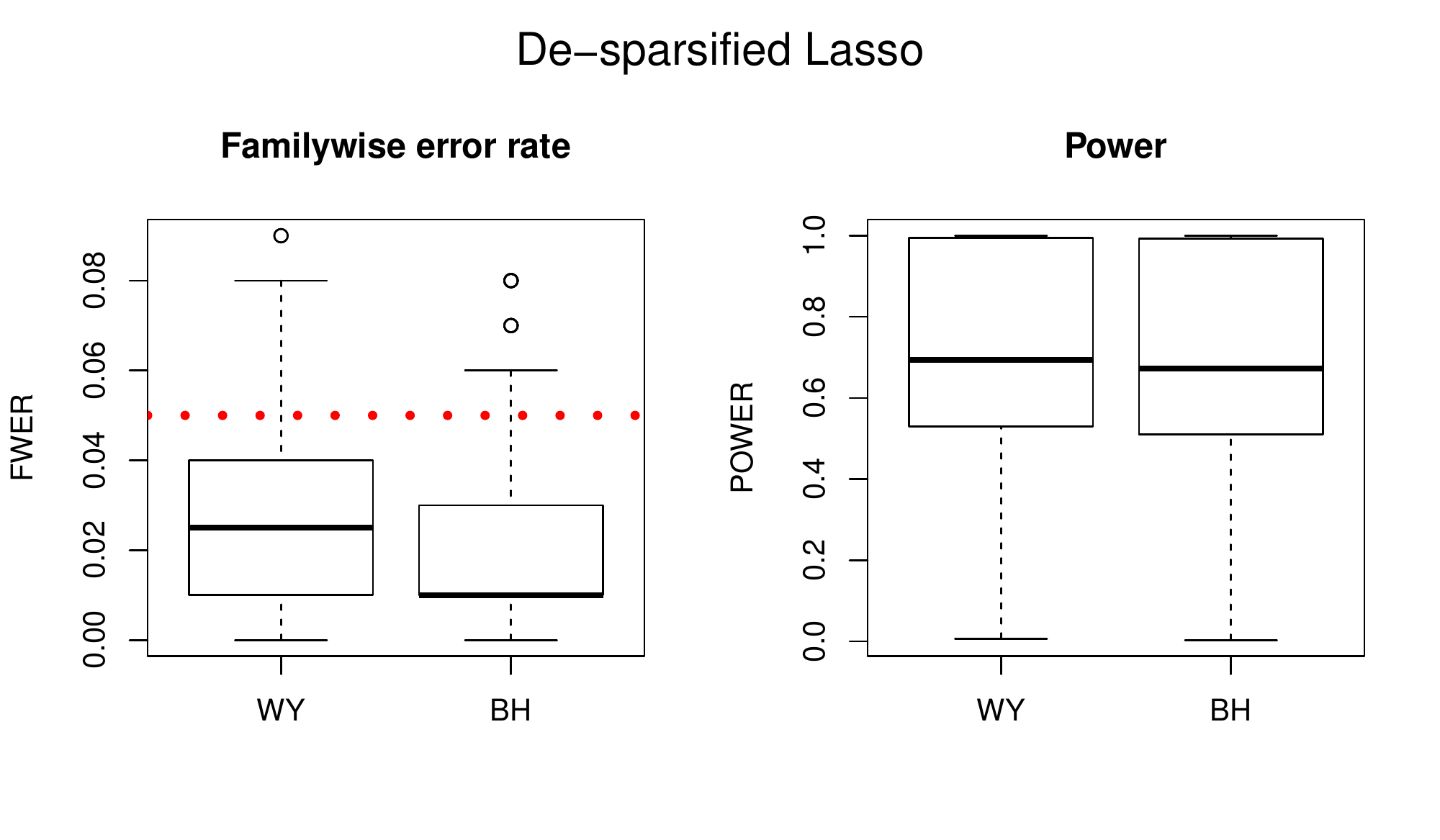}
 \caption{The same plot as Figure \ref{fig:powerfwer-toeplitzgauss} but for
   \textbf{homoscedastic chi-squared errors}.}
 \label{fig:powerfwer-toeplitzchisq}
\end{figure}

For multiple testing, the same setups were looked at as in Section
\ref{subsubsec:homoscedastic-gaussian} but now with the different 
errors. As can be seen in Figure \ref{fig:powerfwer-toeplitzchisq}, the
poor single testing confidence interval coverage does not translate into
poor multiple testing error control. The original method with
Bonferroni-Holm is on the conservative side, while the bootstrap is
slightly closer to the correct level.

\subsubsection{Heteroscedastic non-Gaussian errors}
\label{subsubsect:heteroscedasticnongaus}
Data is generated from a linear model with heteroscedastic non-Gaussian
errors. The example is taken from \citet{mammenwild1993} with sample
size $n=50$, but where we increased the number of parameters to $p=250$ from
the original $p=5$. The model has no
signal $\beta_1^0=\beta_2^0=\dots=\beta_p^0=0$ and introduces 
heteroscedasticity while still maintaining the correctness of the linear
model.

Each row of the design matrix $\bx$ is generated independently $\sim
\mathcal{N}_p(0,I_p)$ and then given a different variance. Each row is
multiplied with the value
$Z_i/2$, where the $\{Z_1,\dots,Z_n\}$ are chosen i.i.d. $\sim U(1,3)$.

The errors $\varepsilon_i$ are chosen to be a mixture of normal
distributions
\begin{eqnarray*}
& &\eps_1,\ldots ,\eps_n\ \mbox{i.i.d},\ \eps_i = l_i \zeta_i + (l_i-1)
    \eta_i,\\ 
& &l_i \sim \operatorname{Bernoulli}(0.5),\ \zeta_i \sim
    \mathcal{N}(1/2, (1.2)^2),\ \eta_i \sim \mathcal{N}(-1/2,(0.7)^2),
\end{eqnarray*}
with $l_i,\ \zeta_i,\ \eta_i$ independent of each other. 
The responses are generated by introducing heteroscedasticity in the errors
\begin{equation*}
  Y_i = Q_i \varepsilon_i + \varepsilon_i \: \: \: \: \: \: \forall i = 1,\ldots,n
\end{equation*}
with $Q_i = X_{i,1}^2 + X_{i,2}^2 + X_{i,3}^2 + X_{i,4}^2 + X_{i,5}^2 - \EE[Z_i^2]$.

\begin{figure}[!htb]
 \centering
\includegraphics[scale=0.6]{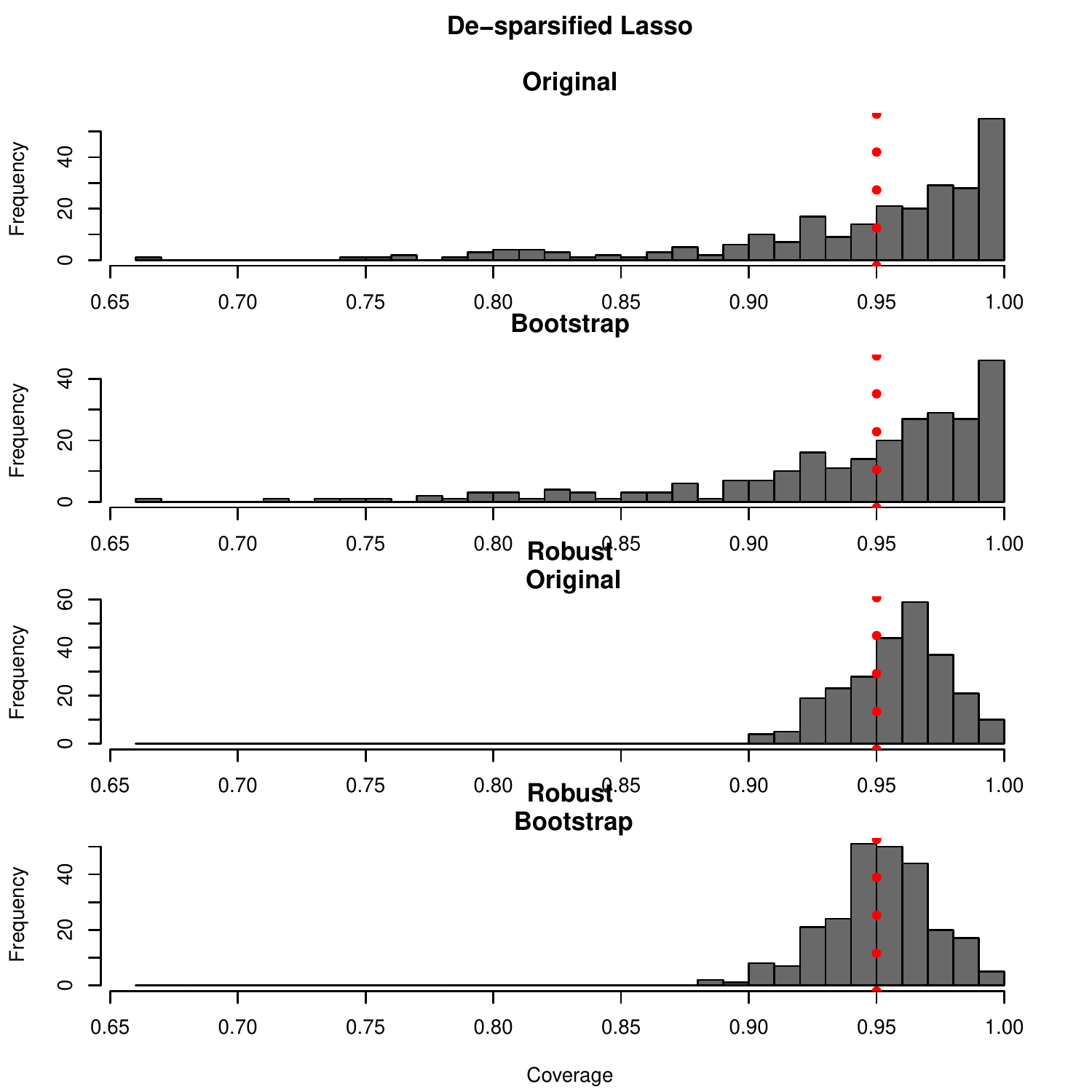}
 \caption{The same plot as Figure \ref{fig:cihist-toeplitzgauss} but for
   \textbf{heteroscedastic non-Gaussian errors} and without signal.
   The robust standard error estimation clearly outperforms the non-robust
   version. There seems to be hardly any difference between the bootstrap
   and the original estimator after choosing the standard error estimation.} 
 \label{fig:cihist-heterosced}
\end{figure}

\begin{figure}[!htb]
 \centering
\includegraphics[scale=0.7]{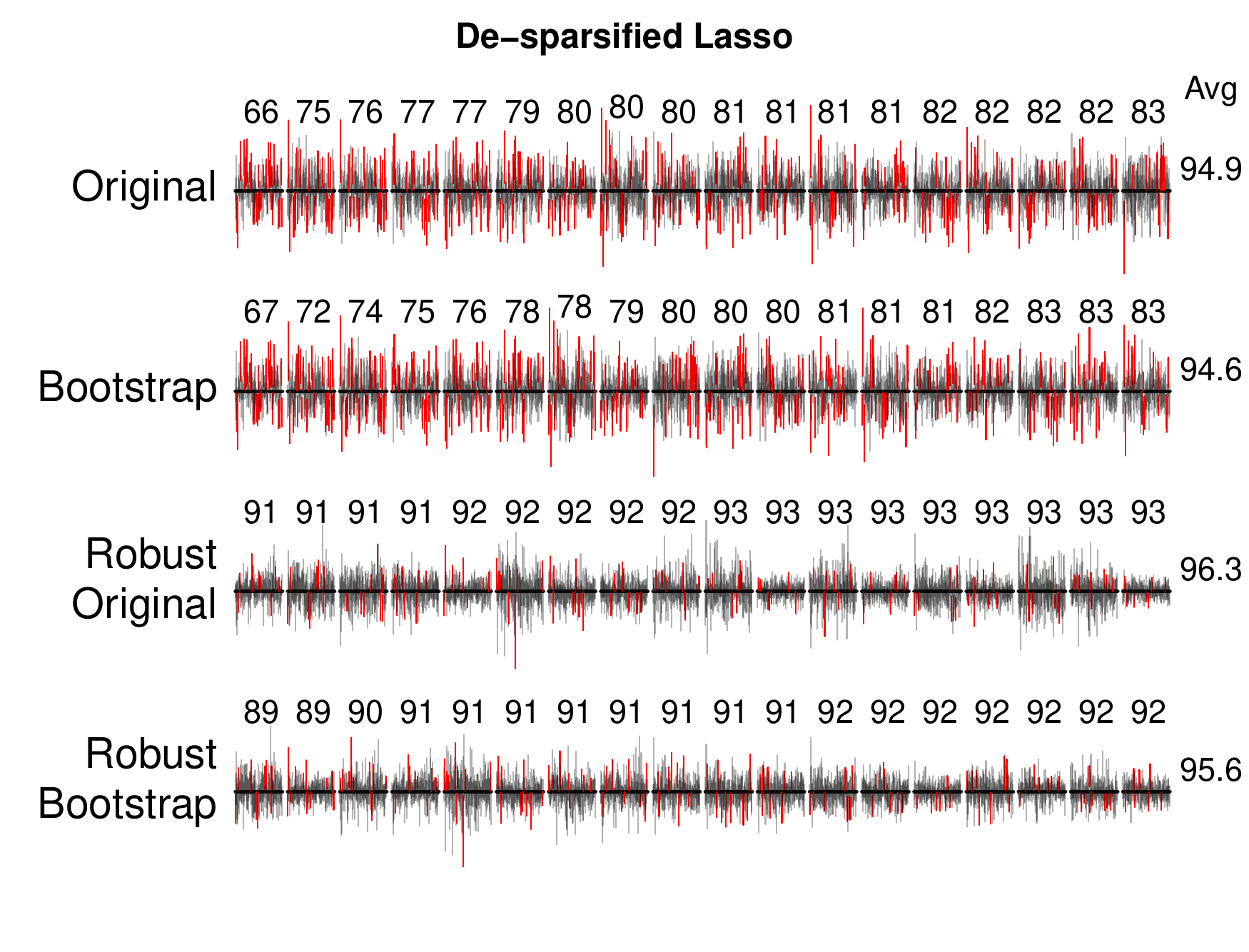}
 \caption{The same plot as Figure \ref{fig:ci-toeplitzgauss} but for
   \textbf{heteroscedastic non-Gaussian errors} and without signal.
   The non-robust 
   estimators have low coverage for many coefficients. Unlike the other
   setups, there doesn't seem to be a bias in the original estimator for
   this dataset.} 
 \label{fig:ci-heterosced}
\end{figure}

For confidence intervals, we focus on one generated design matrix
$\bx$. The histograms for the coverage probabilities can be found in Figure
\ref{fig:cihist-heterosced}, the plot of the actual confidence intervals
is shown in Figure \ref{fig:ci-heterosced}.
What is immediately clear from Figure \ref{fig:cihist-heterosced} is that 
it makes a big difference if one uses the robust version of the 
standard error estimation or not. The coverage is very poor for the
non-robust methods, while for the robust methods the performance looks like
perfect coverage (Figure \ref{fig:perfectci-hist}). 

There isn't any benefit for the bootstrap over the original 
estimator for this dataset. The robust original estimator doesn't show any 
bias in Figure \ref{fig:ci-heterosced} and has great coverage already. The
overall average coverage is slightly more correct for the bootstrap with a
value of 95.1 versus 95.9. 

In contrast to the single testing confidence intervals, all methods (robust
and non-robust) perform adequately for multiple testing as can be seen in
Figure \ref{fig:powerfwer-toeplitzchisq}. Due to the lack of signal in the
dataset, we can only investigate error rates. 50 different design matrices
were generated to produce the 50 data points in the boxplots.
The bootstrap is less conservative and has actual error rate closer to the
true level. 

\begin{figure}[!htb]
 \centerline{
   \includegraphics[scale=0.65]{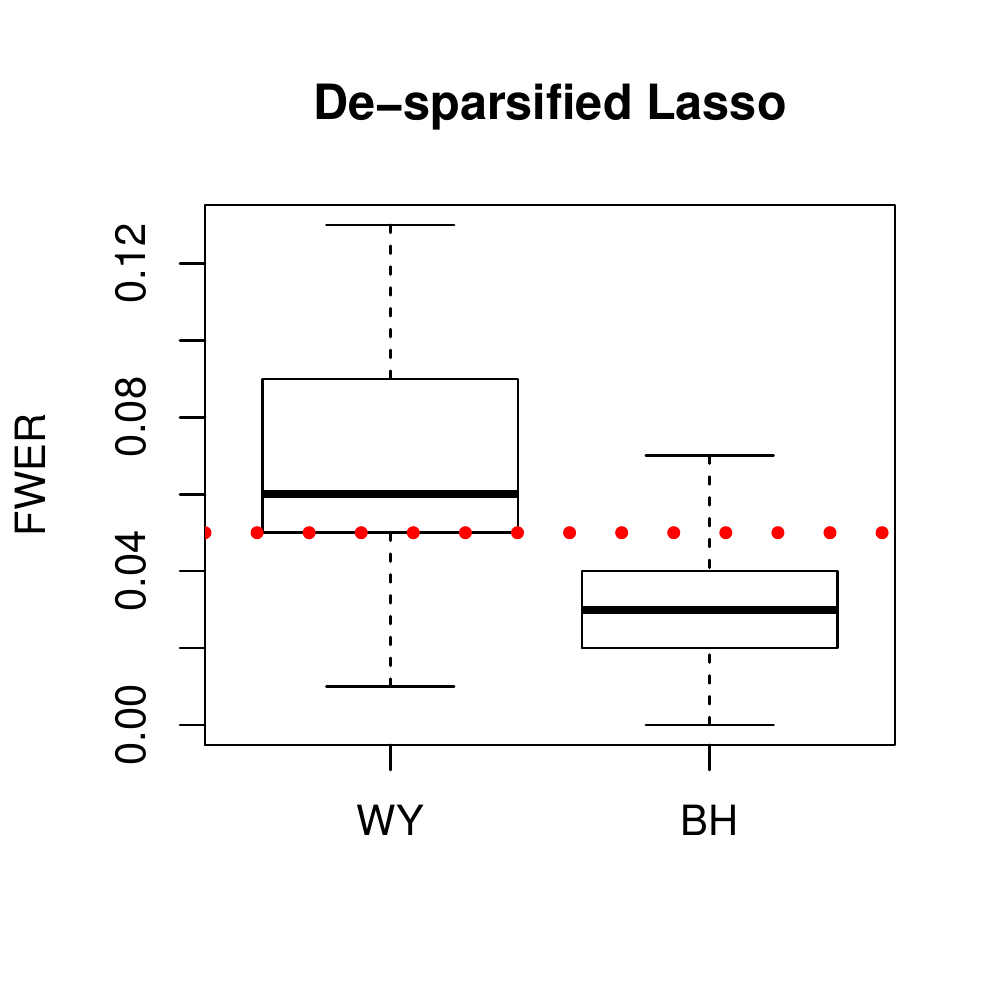}
   \includegraphics[scale=0.65]{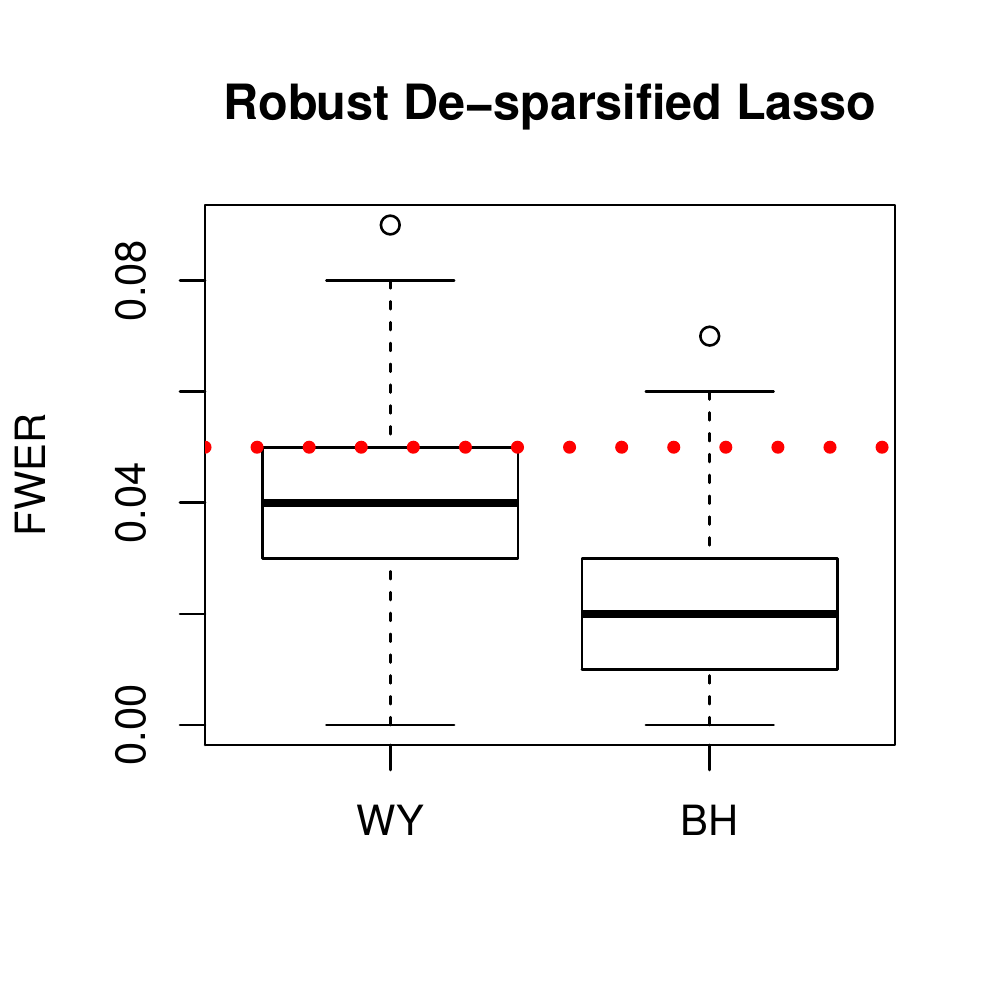}
 }
 \caption{The same plot as Figure \ref{fig:powerfwer-toeplitzgauss} but for
   \textbf{heteroscedastic non-Gaussian errors} and without signal.
   We only report the error
   rate because all null hypotheses are true for the generated
   dataset. The plot on the left is for the non-robust methods, the one on
   the right for the robust ones.}
 \label{fig:powerfwer-heterosced}
\end{figure}


\subsubsection{Discussion}
Bootstrapping the de-sparsified Lasso turns out to improve the coverage of 
confidence intervals without increasing the confidence interval
lengths (that is, without loosing efficiency). The use of the conservative
RLDPE \citep{zhangzhang11} is not necessary: the bootstrap achieves
reliable coverage, while for the original de-sparsified Lasso, the RLDPE
seems worthwhile to achieve reasonable coverage while paying a price in
terms of efficiency. Furthermore, bootstrapping only the linearized part of
the de-sparsified estimator as proposed by \citet{ZhangCheng2016} is
clearly sub-ideal in comparison to bootstrapping the entire estimator and
using the plug-in principle as advocated here. 

For multiple testing, the bootstrapped estimator had familywise error rates
that were closer to the target level while Bonferroni-Holm adjustment is
too conservative. This finding was not reflected in any
noticeable power improvements but some gains are found, see Section
\ref{subsec:closerlookmulttest} below. 

The robust standard error turned out to be critical when dealing with
heteroscedastic errors. Therefore, we recommend the
bootstrapped estimator with robust standard error estimation as the method
to be used: if the errors were homoscedastic, we pay a price of
  efficiency; see also the sentence at the end of Section
  \ref{subsubsec:realXpequiv}.

As can be seen in Appendix \ref{app:additionalmultiplier}, the Gaussian
multiplier bootstrap also performs well. The performance is very
similar to the residual bootstrap and, as one would expect, it handles
heteroscedastic errors as good as the robust standard error bootstrap approach. 


\subsection{A closer look at multiple testing}
\label{subsec:closerlookmulttest}

The examples from Section \ref{subsec:varyingepsdist} showed little to no
power difference between the bootstrap and the original estimator. One
straightforward explanation for this is that the signal in the simulated
datasets didn't fall into the (possibly small) differences in rejection
regions.  
\begin{figure}[!htb]
  \centering
\includegraphics[scale=0.5]{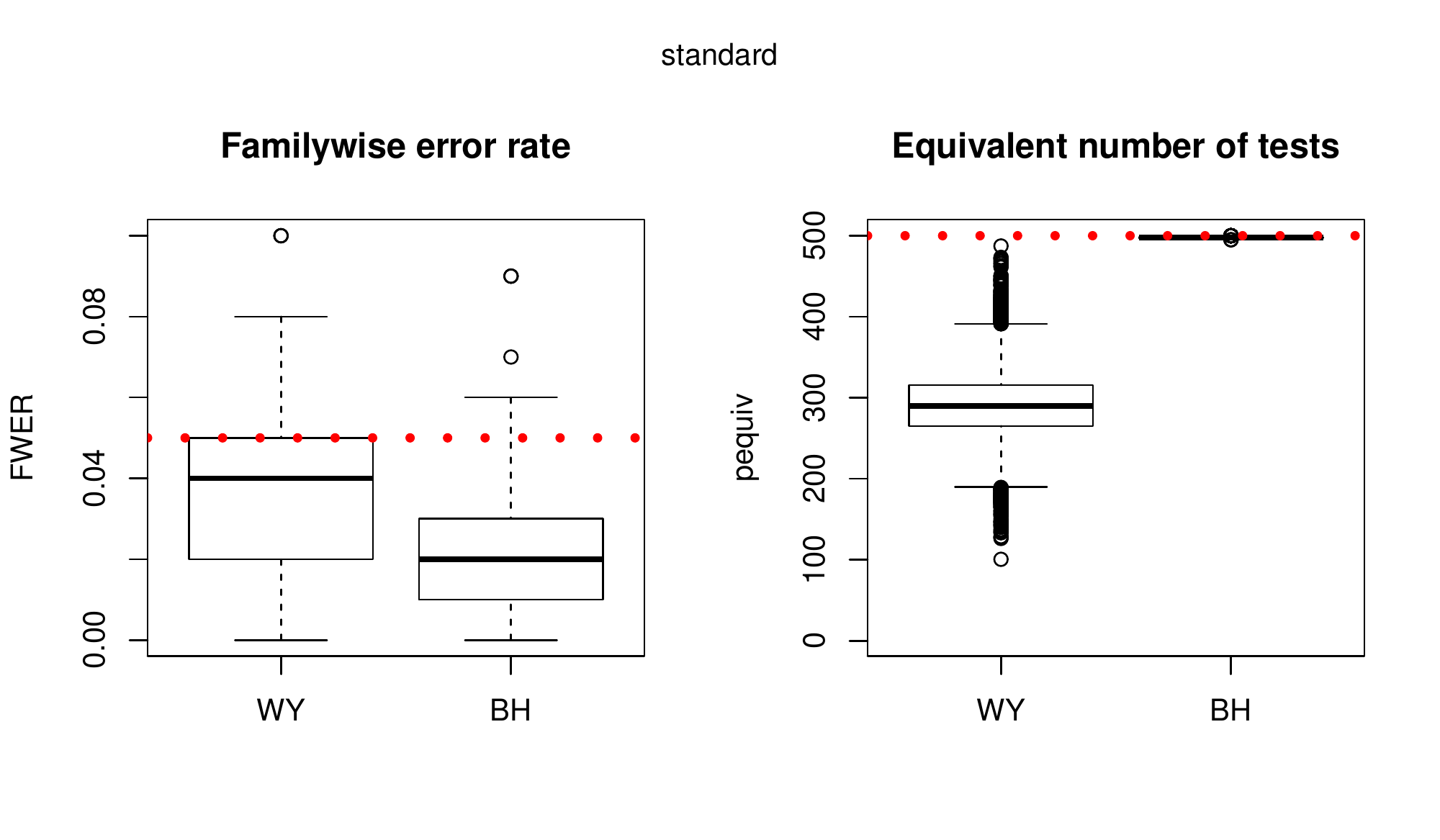}
\includegraphics[scale=0.5]{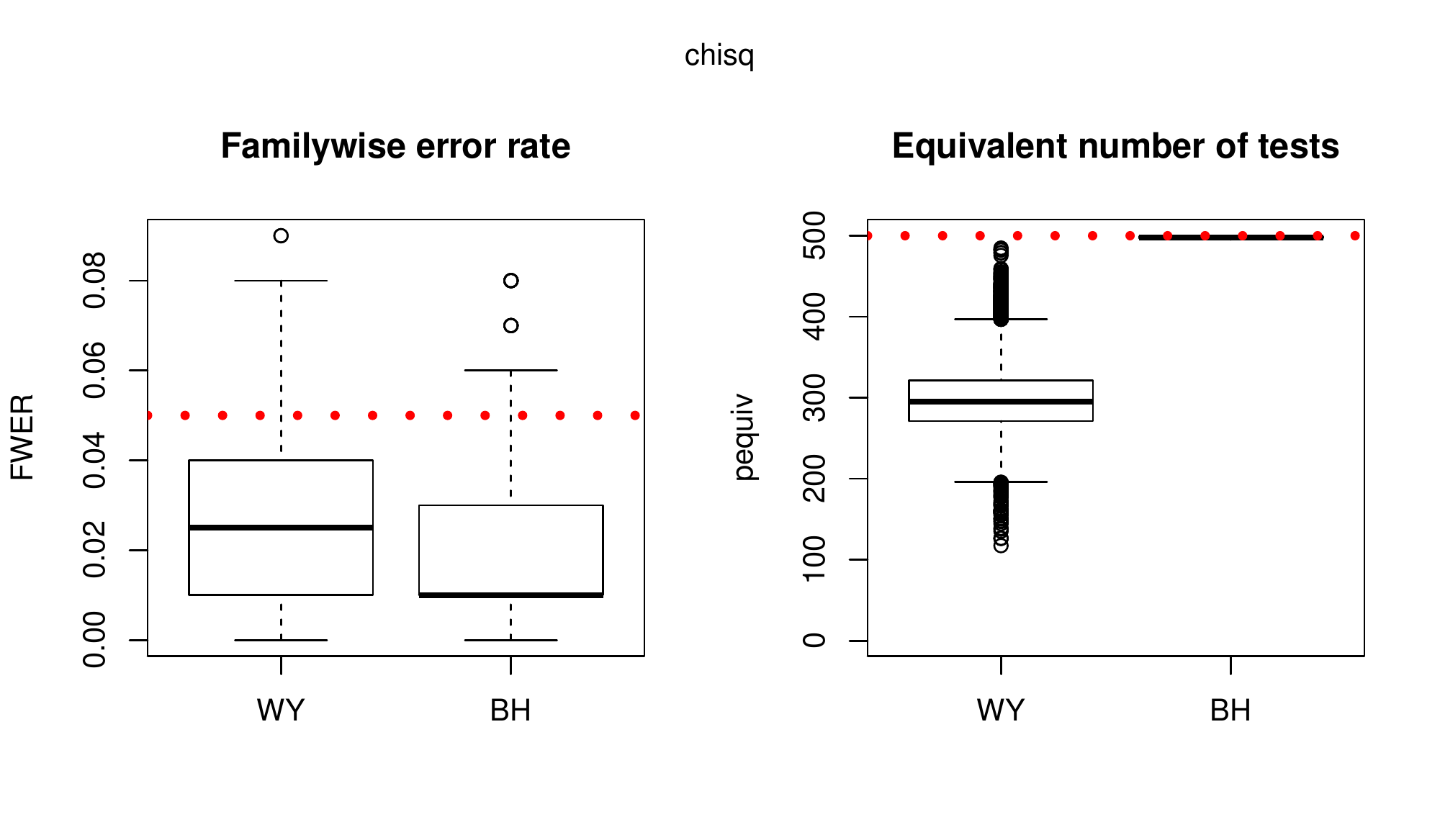}
  \caption{The same plots as Figure \ref{fig:powerfwer-toeplitzgauss}
    for the \textbf{homoscedastic Gaussian errors} (top) and Figure 
    \ref{fig:powerfwer-toeplitzchisq} for the \textbf{homoscedastic non-Gaussian
      errors} (bottom), but displaying the number of equivalent tests
    $p_{equiv}$ instead of the power. The actual number of hypotheses tested
    is highlighted by a red dotted horizontal line.}  
  \label{fig:powerfwer-toeplitz-pequiv}
\end{figure}

As another more signal-independent way to investigate multiple testing
performance, we compare the 
computed rejection regions. Unfortunately, the actual 
values of the rejection thresholds are often quite unintuitive to
compare. Instead, it can be more informative to invert the
Bonferroni-Holm adjustment rule to compute some \emph{equivalent
  number of 
  tests} which is essentially equivalent to the number of tests
under independence. The Westfall-Young procedure computes a 
rejection threshold $t_{rej}$ for the absolute value of the test statistic
and we can then compute the equivalent number of tests (with the Bonferroni
adjustment) $p_{equiv}$ as
\begin{equation}
  \label{eq:pequiv}
  p_{equiv} = \frac{\alpha}{2(1-\Phi(t_{rej}))},
\end{equation}
for controlling the familywise error rate at level $\alpha$, and with
$\Phi(.)$ the cumulative distribution function for
$\mathcal{N}(0,1)$. Improvements in rejection threshold
are then reflected  
in $p_{equiv}$ being a lot smaller than the actual number of hypotheses
tested $p$, while still properly controlling the error rates.


Looking at the rejection thresholds presented
in Figure \ref{fig:powerfwer-toeplitz-pequiv}, we can see that the
bootstrap does improve substantially over the original estimator with a
Bonferroni correction. Multiple testing
with the bootstrap is often equivalent to testing about 300 (independent)
tests with Bonferroni correction in comparison to the original 500.







\subsubsection{Real measurements design}
\label{subsubsec:realXpequiv}

We take design matrices from real data and simulate a linear model with
known signal and homoscedastic Gaussian errors. We look at all 6 signal
options described in Section \ref{subsec:varyingepsdist}.
\begin{figure}[!htb]
 \centering
\includegraphics[scale=0.6]{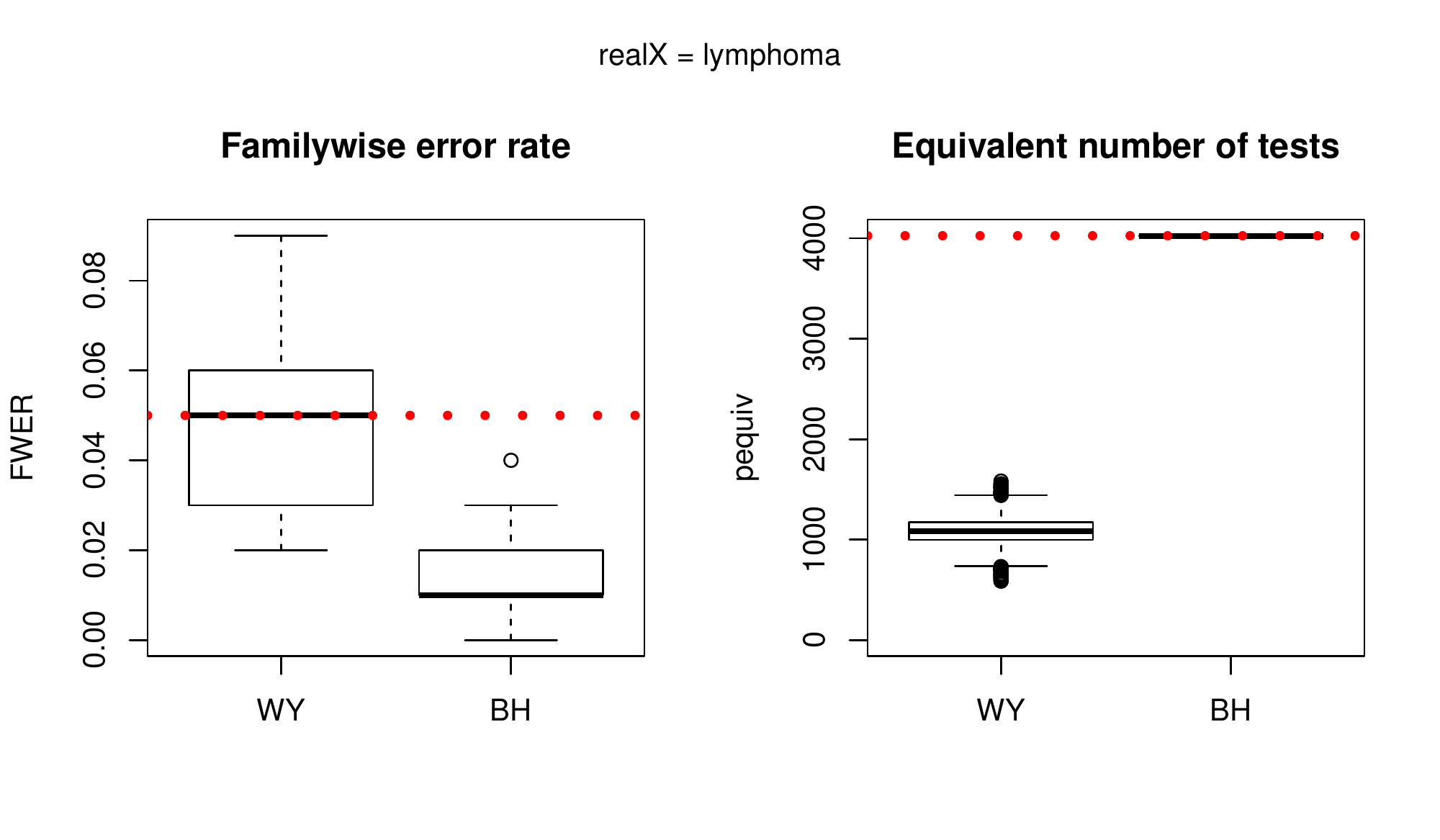}
 \caption{The same plot as in Figure \ref{fig:powerfwer-toeplitz-pequiv}
   but with design matrix coming from real measurements (lymphoma in this
   case) with simulated signal and \textbf{homoscedastic Gaussian errors.}}  
 \label{fig:powerfwer-toeplitz-pequiv-realX-lymphoma}
\end{figure}

For every signal type, we only look at 5 different seeds for generating the
coefficients and for the permutations of the coefficient vector (in
contrast to the typical 50 as used in Section
\ref{subsubsec:homoscedastic-gaussian}).  As usual, the familywise error
rates are computed based on 100 realizations of each model.

Boxplots of the familywise error rate and $p_{equiv}$ for the
\texttt{lymphoma} dataset can be found in Figure
\ref{fig:powerfwer-toeplitz-pequiv-realX-lymphoma}. The
median values of the equivalent number of tests and the FWER for all
the different designs are as follows:

\medskip\noindent 
{\centering
\begin{tabular}{lrrrrrrrr}
  \hline
 & dsmN71 & brain & breast & lymphoma & leukemia & colon & prostate & nci \\ 
  \hline
  Median $p_{equiv}$ WY & 1264 & 886 & 1162 & 1083 & 1230 & 655 & 2466 & 1289 \\ 
  Median $p_{equiv}$ BH & 4088 & 5596 & 7129 & 4025 & 3570 & 2000 & 6032 & 5243 \\ 
  Dimension p & 4088 & 5597 & 7129 & 4026 & 3571 & 2000 & 6033 & 5244 \\ 
   \hline
  \hline
  Median FWER WY & 0.02 & 0.06 & 0.05 & 0.05 & 0.05 & 0.03 & 0.06 & 0.03 \\ 
  Median FWER BH & 0.00 & 0.02 & 0.01 & 0.01 & 0.03 & 0.00 & 0.04 & 0.01 \\ 
   \hline
\end{tabular}}

\medskip\noindent
The bootstrap achieves substantial reductions in the median equivalent
number of tests for all datasets investigated here. 

We note that when studentizing the test-statistics with the robust
  standard error, the power gain with the
  bootstrap (Westfall-Young method) is often rather marginal. This is
  illustrated in Appendix \ref{app:multiplier-quantileest0p95}.

\subsubsection{Real data example}


We revisit a dataset about riboflavin production by bacillus subtilis
\citep{bumeka13}, already studied in \citet{pb13}, \citet{vdgetal13} and
\citet{dezeureetal14}. The dataset has dimensions $n=71$ $p=4088$ and
the original de-sparsified Lasso doesn't manage to reject any null
hypothesis $H_{0,j}$ at the 5\% significance level after multiple testing
correction with Bonferroni-Holm. 
\begin{figure}[!htb]
 \centering
\includegraphics[scale=0.5]{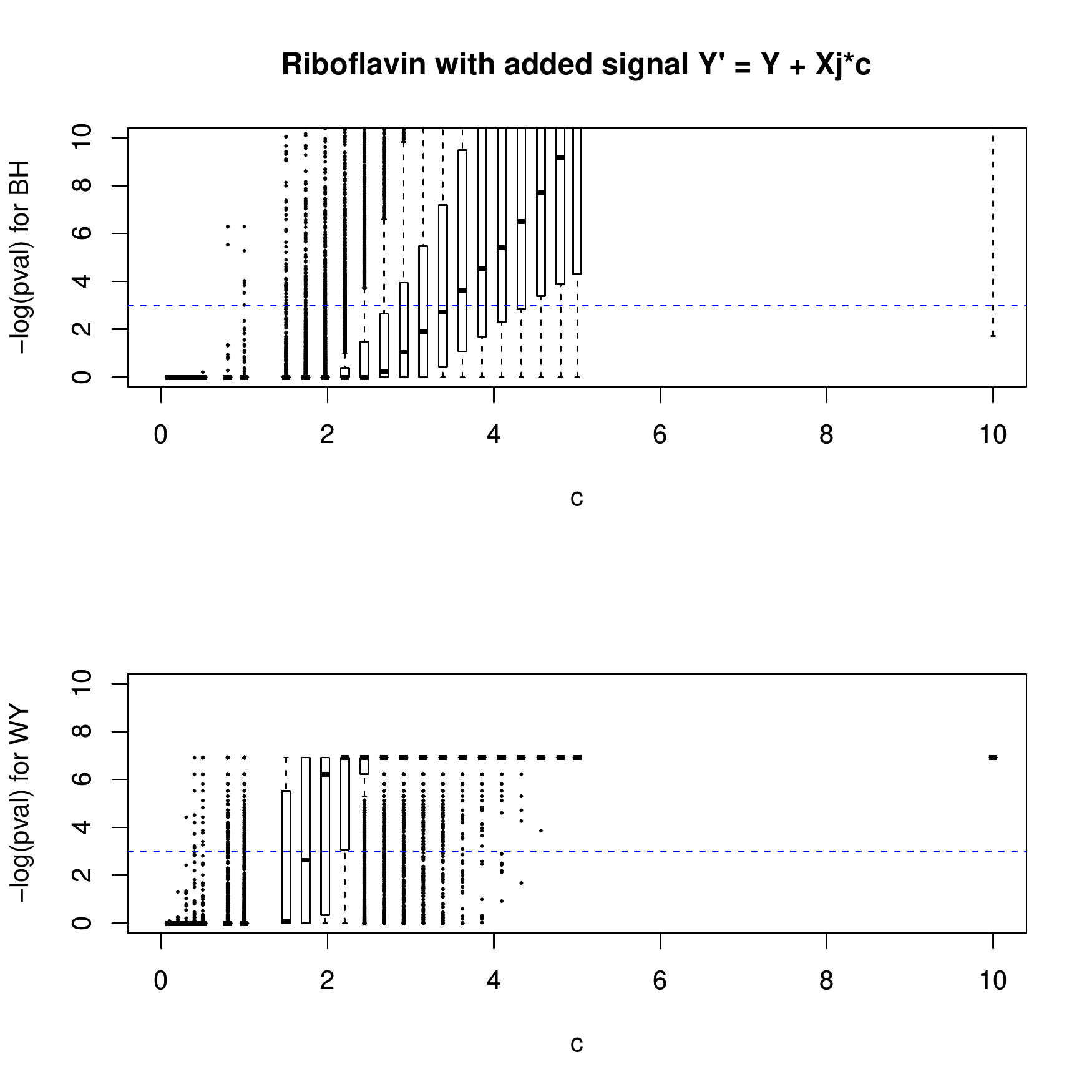}
\caption{Multiple testing corrected p-values for increasing artificial
   signal added to the real dataset Riboflavin (dsmN71). Signal is added
   only one variable at a 
   time $Y^{'} = Y + X_j c$, and only the p-value for that coefficient
   $p_j$ is stored for each such experiment and value of $c$. The values
   $-\log(p_j)$ for all different experiments ($j=1,\dots,p=4088$) are plotted
   in boxplots grouped by value of 
   $c$. A horizontal dashed blue line indicates the rejection threshold
   $0.05$. The bootstrap approach to multiple testing clearly has higher
   power as it picks up on the signal quicker. The bootstrap has a lower
   bound on the minimal achievable p-value due to the finite number of
   bootstrap samples $B=1000$, namely $-\log(1/1000) \approx 6.9$.}
  \label{fig:c-experiment}
\end{figure}

Despite the power gain that is possible with this design matrix (see
\texttt{dsmN71} in the Table in Section \ref{subsubsec:realXpequiv})
, the bootstrapped
estimator doesn't reject any hypotheses either with the Westfall-Young
procedure. 


We investigate what signal strength one would be able to detect in this real
dataset by adding artificial signal to the original responses. This is done
by adding a linear component $X_j c$ of increasing signal strength \texttt{c}
for a single variable $j$,
\begin{equation*}
  Y^{'} = Y + X_j c.
\end{equation*}
One can keep track of the p-value for this particular coefficient and
repeat the experiment for all possible columns of the design
$j=1,\dots,p$. Boxplots of this experiment can be found in Figure
\ref{fig:c-experiment}. 

The bootstrap results in smaller p-values for the same signal values. It
rejects the relevant null hypothesis almost all the time for signal values
above $c>2.5$. Note that we do not have access to replicates and
  therefore, we cannot determine the actual error rate.

\subsubsection{Discussion}

The bootstrap with the Westfall-Young (WY) multiple testing adjustment
leads to reliable familywise error control while providing a rejection
threshold which is far more powerful than using the Bonferroni
adjustment (in the case of homoscedastic errors), especially in presence
of dependence among the components for 
testing (while for heteroscedastic errors and when using the
robust standard error for studentization, the efficiency gain of WY
often seems less substantial).
Since the efficient WY adjustment is not adding additional
computational costs to the one from bootstrapping, such simultaneous
inference and WY multiple testing adjustment is highly recommended.

\section{Further considerations}

We discuss here additional points before providing some conclusions. 

\subsection{Model misspecification}\label{subsec.modelmisspec}

So far, the entire discussion has been for a linear model as in
\eqref{lin.mod} with a sparse regression vector $\beta^0$. The linearity is
not really a restriction: suppose that the true model would involve a
nonlinear regression function
\begin{eqnarray*}
Y_i = f^0(X_i) + \eps_i\ (i=1,\ldots n).
\end{eqnarray*}
The $n \times 1$ vector of the true function at the observed data values
$(f^0(X_1),\ldots ,f^0(X_n))^T$ can be represented as
\begin{eqnarray}\label{beta-solution}
(f^0(X_1),\ldots ,f^0(X_n))^T = \bx \beta^0
\end{eqnarray} 
for many possible solutions
$\beta^0$: this is always true as long as $\mathrm{rank}(\bx) = n$ which
typically holds in $p \ge n$ settings. The issue is whether there are
solutions $\beta^0$ in \eqref{beta-solution} which are sparse: a general
result to address this point is not available, see
\citet{misspecified-pbvdg2015}. 

It is argued in \citet{misspecified-pbvdg2015} that weak sparsity in terms
of $\|\beta^0\|_r$ for $0<r<1$ suffices to guarantee that the de-sparsified
Lasso has an asymptotic Gaussian distribution centered at $\beta^0$, as
described in Theorem \ref{th1b} or \ref{th2}. Thus, assuming that there is
a weakly sparse solution $\beta^0$ in \eqref{beta-solution} is relaxing the
requirement for $\ell_0$ sparsity. The presented theory for the bootstrap
could be adapted to cover the case for weakly sparse $\beta^0$.  

The interpretation of a confidence interval for $\beta^0$, based on the
Gaussian limiting distribution of the de-sparsified Lasso or using its
bootstrapped version as described in this paper, is that it covers all
$\ell_r\ (0<r<1)$ weakly sparse solutions $\beta^0$ which are solutions of
\eqref{beta-solution}. Thereby, we implicitly assume that there is at least
one such $\ell_r$ weakly sparse solution. 

\subsection{Random design} 

The distinction between fixed and random design becomes crucial for
misspecified models. If the true model with random design is linear, then
by conditioning on the covariables, the corresponding fixed design model is
again linear. And if the inference is correct conditional on $\bx$, it is
also correct unconditional for random design. If the true random design
model is nonlinear, one can look at the best projected random design linear
model: but then, when conditioning, the obtained projected fixed design
linear model has a bias (or non-zero conditional mean for the
error). In other words, conditioning on the covariables is not appropriate
when the model is wrong, and one should rather do unconditional inference in
a random design (best approximating) linear model; see
\citet{misspecified-pbvdg2015}.  

Thus, there are certainly situations where one would like to do
unconditional inference in a random design linear model, see also
\citet{freedman1981} who proposes the ``paired bootstrap'' in a
low-dimensional setting. The bootstrap which we discussed in this paper is for
fixed design only: for random design, one should resample the covariables
as well. Because of the latter the computational task becomes
much more demanding: for the de-sparsified Lasso, and when $p$ is large,
most of the computation is spent on computing all the residual vectors
$Z_1,\ldots ,Z_p$ which requires running the Lasso $p$ times. For
bootstrapping with fixed design, this computation has to be done only once
(since $Z_1,\ldots ,Z_p$ are deterministic values of the fixed design $\bx$);
with random design, it seems unavoidable to do it $B \approx 100-1'000$
times which would result in a major additional computational cost. 


\subsection{Conclusions}

We propose a residual, wild and paired bootstrap methodology for individual and
simultaneous inference in high-dimensional linear models with possibly
non-Gaussian and heteroscedastic errors. The bootstrap is used to
approximate the distribution of the de-sparsified Lasso, a regular
non-sparse estimator which is not exposed to the unpleasant super-efficiency
phenomenon. 

We establish asymptotic consistency for possibly simultaneous
inference for parameters in a group $G \subseteq \{1,\ldots ,p\}$ of
variables, where $p \gg n$ but $s_0 = o(n^{1/2}/\{\log(p) \log(|G|)\})$ and
$\log(|G|) = o(n^{1/7})$ with $s_0$ denoting the sparsity.  
The presented general theory is complemented by many empirical
results, demonstrating the advantages of our approach over other
proposals. Especially for 
simultaneous inference and multiple testing adjustment, the bootstrap is
very powerful. 

For homoscedastic errors, the residual and wild bootstrap perform
similarly. For heteroscedastic errors, the wild bootstrap is more natural
and can be used for simultaneous inference (whereas the residual bootstrap
fails to be consistent). Thus, for protecting against heteroscedastic
errors, the wild bootstrap seems to be the preferred method.
Our proposed procedures are implemented in the
\texttt{R}-package \texttt{hdi} \citep{hdipackage}. 

\paragraph{Acknowledgments.} We gratefully acknowledge visits at the
American Institute of Mathematics (AIM), San Jose, US, and at the Mathematisches Forschungsinstitut (MFO), Oberwolfach, Germany.

\bibliographystyle{apalike}
\bibliography{reference}

\begin{thebibliography}{}

\bibitem[Bickel et~al., 1998]{bicketal98}
Bickel, P., Klaassen, C., Ritov, Y., and Wellner, J. (1998).
\newblock {\em Efficient and Adaptive Estimation for Semiparametric Models}.
\newblock Springer.

\bibitem[Breiman, 1996]{brei96b}
Breiman, L. (1996).
\newblock Heuristics of instability and stabilization in model selection.
\newblock {\em Annals of Statistics}, 24:2350--2383.

\bibitem[B\"uhlmann, 2013]{pb13}
B\"uhlmann, P. (2013).
\newblock Statistical significance in high-dimensional linear models.
\newblock {\em Bernoulli}, 19:1212--1242.

\bibitem[B\"uhlmann et~al., 2014]{bumeka13}
B\"uhlmann, P., Kalisch, M., and Meier, L. (2014).
\newblock High-dimensional statistics with a view towards applications in
  biology.
\newblock {\em Annual Review of Statistics and its Applications}, 1:255--278.

\bibitem[B\"uhlmann and van~de Geer, 2011]{pbvdg11}
B\"uhlmann, P. and van~de Geer, S. (2011).
\newblock {\em Statistics for High-Dimensional Data: Methods, Theory and
  Applications}.
\newblock Springer.

\bibitem[B{\"u}hlmann and van~de Geer, 2015]{misspecified-pbvdg2015}
B{\"u}hlmann, P. and van~de Geer, S. (2015).
\newblock High-dimensional inference in misspecified linear models.
\newblock {\em Electronic Journal of Statistics}, 9:1449--1473.

\bibitem[Chatterjee and Lahiri, 2011]{chatter11}
Chatterjee, A. and Lahiri, S. (2011).
\newblock Bootstrapping {L}asso estimators.
\newblock {\em Journal of the American Statistical Association}, 106:608--625.

\bibitem[Chatterjee and Lahiri, 2013]{chatter13}
Chatterjee, A. and Lahiri, S. (2013).
\newblock Rates of convergence of the adaptive {LASSO} estimators to the oracle
  distribution and higher order refinements by the bootstrap.
\newblock {\em Annals of Statistics}, 41:1232--1259.

\bibitem[Chernozhukov et~al., 2013]{chernozhukov2013}
Chernozhukov, V., Chetverikov, D., and Kato, K. (2013).
\newblock Gaussian approximations and multiplier bootstrap for maxima of sums
  of high-dimensional random vectors.
\newblock {\em Annals of Statistics}, 41:2786--2819.

\bibitem[Dezeure et~al., 2015]{dezeureetal14}
Dezeure, R., B\"uhlmann, P., Meier, L., and Meinshausen, N. (2015).
\newblock High-dimensional inference: Confidence intervals, $p$-values and
  {R}-software hdi.
\newblock {\em Statistical Science}, 30:533--558.

\bibitem[Efron, 1979]{efron1979}
Efron, B. (1979).
\newblock Bootstrap methods: Another look at the jackknife.
\newblock {\em Annals of Statistics}, 7:1--26.

\bibitem[Eicker, 1967]{eicker1967limit}
Eicker, F. (1967).
\newblock Limit theorems for regressions with unequal and dependent errors.
\newblock In {\em Proceedings of the fifth Berkeley symposium on mathematical
  statistics and probability}, volume~1, pages 59--82.

\bibitem[Foygel~Barber and Cand\`{e}s, 2015]{foygcand14}
Foygel~Barber, R. and Cand\`{e}s, E.~J. (2015).
\newblock Controlling the false discovery rate via knockoffs.
\newblock {\em Annals of Statistics}, 43:2055--2085.

\bibitem[Freedman, 1981]{freedman1981}
Freedman, D.~A. (1981).
\newblock Bootstrapping regression models.
\newblock {\em Annals of Statistics}, 9:1218--1228.

\bibitem[Gine and Zinn, 1989]{gine1989}
Gine, E. and Zinn, J. (1989).
\newblock Necessary conditions for the bootstrap of the mean.
\newblock {\em Annals of Statistics}, 17:684--691.

\bibitem[Gine and Zinn, 1990]{gine1990}
Gine, E. and Zinn, J. (1990).
\newblock Bootstrapping general empirical measures.
\newblock {\em Annals of Probability}, 18:851--869.

\bibitem[Hall and Wilson, 1991]{hallwilson1991}
Hall, P. and Wilson, S.~R. (1991).
\newblock Two guidelines for bootstrap hypothesis testing.
\newblock {\em Biometrics}, 47:pp. 757--762.

\bibitem[Huber, 1967]{huber1967behavior}
Huber, P.~J. (1967).
\newblock The behavior of maximum likelihood estimates under nonstandard
  conditions.
\newblock In {\em Proceedings of the fifth Berkeley symposium on mathematical
  statistics and probability}, volume~1, pages 221--233.

\bibitem[Javanmard and Montanari, 2014]{jamo13b}
Javanmard, A. and Montanari, A. (2014).
\newblock Confidence intervals and hypothesis testing for high-dimensional
  regression.
\newblock {\em Journal of Machine Learning Research}, 15:2869--2909.

\bibitem[Liu and Singh, 1992]{liu1992efficiency}
Liu, R.~Y. and Singh, K. (1992).
\newblock Efficiency and robustness in resampling.
\newblock {\em Annals of Statistics}, 20:370--384.

\bibitem[Mammen, 1993]{mammenwild1993}
Mammen, E. (1993).
\newblock Bootstrap and wild bootstrap for high dimensional linear models.
\newblock {\em Annals of Statistics}, 21:255--285.

\bibitem[McKeague and Qian, 2015]{mckeague15}
McKeague, I.~W. and Qian, M. (2015).
\newblock {An Adaptive Resampling Test for Detecting the Presence of
  Significant Predictors}.
\newblock {\em Journal of the American Statistical Association},
  110:1422--1433.

\bibitem[Meier et~al., 2016]{hdipackage}
Meier, L., Dezeure, R., Meinshausen, N., M\"achler, M., and B\"uhlmann, P.
  (2016).
\newblock {\em hdi: High-Dimensional Inference}.
\newblock R package version 0.1-6.

\bibitem[Meinshausen, 2015]{meins13}
Meinshausen, N. (2015).
\newblock Group bound: confidence intervals for groups of variables in sparse
  high dimensional regression without assumptions on the design.
\newblock {\em Journal of the Royal Statistical Society, Series B},
  77:923--945.

\bibitem[Meinshausen and B{\"u}hlmann, 2006]{mebu06}
Meinshausen, N. and B{\"u}hlmann, P. (2006).
\newblock High-dimensional graphs and variable selection with the {L}asso.
\newblock {\em Annals of Statistics}, 34:1436--1462.

\bibitem[Meinshausen and B\"uhlmann, 2010]{mebu10}
Meinshausen, N. and B\"uhlmann, P. (2010).
\newblock {S}tability {S}election (with discussion).
\newblock {\em Journal of the Royal Statistical Society, Series B},
  72:417--473.

\bibitem[Meinshausen et~al., 2011]{memabu11}
Meinshausen, N., Maathuis, M.~H., and B\"uhlmann, P. (2011).
\newblock Asymptotic optimality of the {Westfall-Young} permutation procedure
  for multiple testing under dependence.
\newblock {\em Annals of Statistics}, 39:3369--3391.

\bibitem[Meinshausen et~al., 2009]{memepb09}
Meinshausen, N., Meier, L., and B{\"u}hlmann, P. (2009).
\newblock P-values for high-dimensional regression.
\newblock {\em Journal of the American Statistical Association},
  104:1671--1681.

\bibitem[Reid et~al., 2016]{reidtibsh13}
Reid, S., Tibshirani, R., and Friedman, J. (2016).
\newblock A study of error variance estimation in lasso regression.
\newblock {\em Statistica Sinica}, 26:35--67.

\bibitem[Rudelson and Zhou, 2013]{RudelsonZ13}
Rudelson, M. and Zhou, S. (2013).
\newblock Reconstruction from anisotropic random measurements.
\newblock {\em Information Theory, IEEE Transactions on}, 59:3434--3447.

\bibitem[Shah and B\"uhlmann, 2015]{shahpb15}
Shah, R. and B\"uhlmann, P. (2015).
\newblock Goodness of fit tests for high-dimensional models.
\newblock Preprint arXiv:1511.03334.

\bibitem[Shah and Samworth, 2013]{shah13}
Shah, R. and Samworth, R. (2013).
\newblock Variable selection with error control: another look at {S}tability
  {S}election.
\newblock {\em Journal of the Royal Statistical Society Series B}, 75:55--80.

\bibitem[van~de Geer et~al., 2014]{vdgetal13}
van~de Geer, S., B\"uhlmann, P., Ritov, Y., and Dezeure, R. (2014).
\newblock On asymptotically optimal confidence regions and tests for
  high-dimensional models.
\newblock {\em Annals of Statistics}, 42:1166--1202.

\bibitem[{van de Geer} et~al., 2011]{geer11}
{van de Geer}, S., B{\"u}hlmann, P., and Zhou, S. (2011).
\newblock The adaptive and the thresholded {L}asso for potentially misspecified
  models (and a lower bound for the {L}asso).
\newblock {\em Electronic Journal of Statistics}, 5:688--749.

\bibitem[Wasserman and Roeder, 2009]{WR08}
Wasserman, L. and Roeder, K. (2009).
\newblock {High dimensional variable selection}.
\newblock {\em Annals of Statistics}, 37:2178--2201.

\bibitem[Westfall and Young, 1993]{westyoung93}
Westfall, P. and Young, S. (1993).
\newblock {\em Resampling-based Multiple Testing: Examples and Methods for
  P-value Adjustment}.
\newblock John Wiley \& Sons.

\bibitem[White, 1980]{white1980heteroskedasticity}
White, H. (1980).
\newblock A heteroskedasticity-consistent covariance matrix estimator and a
  direct test for heteroskedasticity.
\newblock {\em Econometrica}, 48:817--838.

\bibitem[Wu, 1986]{wu1986jackknife}
Wu, C.-F.~J. (1986).
\newblock Jackknife, bootstrap and other resampling methods in regression
  analysis.
\newblock {\em Annals of Statistics}, 14:1261--1295.

\bibitem[Ye and Zhang, 2010]{YeZ10}
Ye, F. and Zhang, C.-H. (2010).
\newblock Rate minimaxity of the {L}asso and {D}antzig selector for the
  $\ell_q$ loss in $\ell_r$ balls.
\newblock {\em Journal of Machine Learning Research}, 11:3481--3502.

\bibitem[Zhang and Huang, 2008]{ZhangH08}
Zhang, C.-H. and Huang, J. (2008).
\newblock The sparsity and bias of the {L}asso selection in high-dimensional
  linear regression.
\newblock {\em Annals of Statistics}, 36:1567--1594.

\bibitem[Zhang and Zhang, 2014]{zhangzhang11}
Zhang, C.-H. and Zhang, S.~S. (2014).
\newblock Confidence intervals for low dimensional parameters in high
  dimensional linear models.
\newblock {\em Journal of the Royal Statistical Society, Series B},
  76:217--242.

\bibitem[Zhang and Cheng, 2016]{ZhangCheng2016}
Zhang, X. and Cheng, G. (2016).
\newblock Simultaneous inference for high-dimensional linear models.
\newblock {\em Journal of the American Statistical Association. Published
  online, DOI:10.1080/01621459.2016.1166114}.

\bibitem[Zhou, 2014]{zhou14}
Zhou, Q. (2014).
\newblock {Monte Carlo Simulation for Lasso-Type Problems by Estimator
  Augmentation}.
\newblock {\em Journal of the American Statistical Association},
  109:1495--1516.

\end{thebibliography}


\appendix
\section{Appendix}
We present here all the proofs and additional
empirical results. 

The proof is composed of four propositions, stating the 
consistency of variance estimates and Gaussian approximation of studentized statistics 
for the original estimator, paired bootstrap, wild bootstrap and xyz-paired bootstrap. 
The theorems then follow directly from the corresponding propositions. 

The following notation will be used. 
For any vectors $u = (u_1,\ldots,u_n)^T$ and $v = (v_1,\ldots,v_n)^T$, denote 
the mean of $u$ by ${\overline u} = n^{-1}\sum_{i=1}^n u_i$, 
the centered $u$ by $u_{\rm cent} = (u_1-{\overline u},\ldots,u_n-{\overline u})^T$, 
and the Hadamard product by $u\circ v = (u_1v_1,\ldots,u_nv_n)^T$.

\subsection{Proof of Theorem \ref{th1b} for homoscedastic errors}\label{subsec.proofth1b}

We remark first that the variance estimator in \eqref{est.vareps} is
asymptotically equivalent to $\hat{\sigma}_{\eps}^2 = n^{-1} \sum_{i=1}^n
(Y_i - (\bx\hat{\beta})_i)^2$ if $\hat{s} =O(s_0) = o(n)$. The latter
holds under the assumption (B3) (which, together with other assumptions,
ensures (A4)). For simplicity of the proofs, we consider here this
modified variance estimator with the factor $n^{-1}$. 


We first collect in the following proposition results on the original estimator in the 
more general heteroscedastic case. 
This will allow us to apply the proposition to the plug-in bootstrap estimator by checking the 
assumptions of the proposition under probability measure $\P^*$ for each bootstrap method.  
The proposition allows $\bx$ and $(Z_j, j\in G)$ to be random and dependent on the noise $\eps$, 
so that it can be applied to the xyz-paired bootstrap. 
To this end, we replace assumptions (A2), (A3), (A5) and (A6) by the following: 
\begin{description}
\item[(A2dep)]$\displaystyle 
\max_{k\neq j}\frac{|Z_j^TX_k|^2}{n\E\|Z_j\|_2^2} \le \lambda_X^2, 
\E[Z_j\circ \eps] = 0, 
\frac{\E\|\eps\circ Z_j\|_2^2}{n\vee \E\|Z_j\|_2^2} \ge L, 
\frac{\E\|\eps\circ Z_j\|_{2+ \delta}^{2 + \delta}}{(\E\|\eps\circ Z_j\|_2^2)^{1+ \delta/2}}\ll 1,\ \forall j\in G$.  
\item[(A3dep)] $(X_{ik},k\le p, \eps_i,Z_{j,i},j\in G)$ independent, 
  $\EE[\eps] = 0$, $\EE\|\eps\|_2^2/n = \sigma_{\eps}^2$, 
  $\EE\|\eps\|_{2+\delta}^{2+ \delta}/n \le C$. 
\item[(A5dep)] $\max_{j\le p}\|X_j\|_\infty \le C_X$, $\E[X_j\circ \eps] = 0$ for all $j\le p$. 
\item[(A6dep)] $\max_{j\in G}\|Z_j\|_\infty \le K$, $\max_{1\le i\le n}\E\,\eps_i^2 =O(n^{3/7})$, 
$\delta=2$, $\log(2|G|)\ll n^{1/7}$. 
\end{description}
Again, $\sigma_\eps$, $L$, $C$, $C_X$ and $K$ are positive constants 
bounded away from $0$ and $\infty$, 
$\delta\in (0,2]$ is fixed and the same in (A2dep) and (A3dep), 
and $\lambda_X \asymp \sqrt{\log(p)/n}$. 
It is clear that when $\bx$ and $(Z_j, j\in G)$ are deterministic, 
(A2), (A3) and (A5) directly imply (A2dep), (A3dep) and (A5dep), 
and (A3) and (A6) directly imply the first two requirements in (A6dep). 
We generalize $\omega_j$ and define $\omega_{j,k}$ as 
\bes
\omega_j = \sqrt{\E\|\eps\circ Z_j\|_2^2/n},\quad 
\omega_{j,k} = \E (\eps \circ Z_j)^T(\eps \circ Z_k)/n. 
\ees
Let $(\zeta_j,j=1,\ldots,p)$ be a Gaussian vector with 
\bes
\E\,\zeta_j=0,\ \E\,\zeta_j\zeta_k = \frac{\omega_{j,k}}{\omega_j\omega_k} 
= \hbox{\rm corr}\Big(\eps \circ Z_j,\eps \circ Z_k\Big). 
\ees

\begin{prop}\label{prop1}
Assume (A1), (A2dep), (A3dep) and (A5dep). Then, 
\begin{eqnarray}\nonumber 
& |\widehat{s.e.}_j/{s.e.}_j -1| = |\hsigma_{\veps}/\sigma_{\veps} -1|
= O_P(n^{-{\delta}/(2+{\delta})}) + o_P(1)\big/\big\{\log(p)\log(2|G|)\big\},\ \forall j\le p,
\\ \nonumber 
& \left|\widehat{s.e.}_{\mathrm{robust},j}/s.e._{\mathrm{robust},j}-1\right|
= \left|\hat{\omega}_j/\omega_j -1\right| = o_P(1)\ \hbox{for each $j\in G$.}
\end{eqnarray}
Let $T_j = \sgn(Z_j^TX_j)(\hb_j-\beta_j^0)/\widehat{s.e.}_{\mathrm{robust},j}$. If $|G|=O(1)$, then
\bel{lyapunovclt} 
\sup_{(t_j, j\in G)} \Big|\P\left[T_j \le t_j, j\in G\right] 
- \P\left[\zeta_j \le t_j, j\in G\right] \Big|=o_P(1). 
\eel
If (A6dep) holds, then 
$\max_{j\in G}\left|\hat{\omega}_j/\omega_j -1\right| = o_P(1)/\log^2(2|G|)$ and 
\bel{clt-for-max}
\sup_{c \in \R}\big|\PP\big[\max_{j \in G} h_j(T_j) \le c\big] - \PP\big[\max_{j \in G}h_j(\zeta_j) \le
    c\big]\big| = o(1) 
\eel
for any combination of functions $h_j(t)=t$, $h_j(t)=-t$ or $h_j(t)=|t|$. 
Moreover, (\ref{lyapunovclt}) and (\ref{clt-for-max}) hold under respective assumptions for 
$T_j = \sgn(Z_j^TX_j)(\hb_j-\beta_j^0)/\widehat{s.e.}_j$ 
in the homoscedastic case where $\E[\eps_i|Z_j, j\in G]=0$ and $\E[\eps_i^2|Z_j,j\in G] = \sigma_\eps^2$ 
for all $i\le n$. 
\end{prop}

Although $\widehat{s.e.}_j$ is a consistent estimator of ${s.e.}_j$, 
$\widehat{s.e.}_j\neq s.e._{\mathrm{robust},j}$ in general. 
Thus, $\widehat{s.e.}_j$ may not properly normalize $\hb_j$ without the homoscedasticity assumption. 
Meanwhile, $\widehat{s.e.}_j^*$ always properly 
normalize the residual bootstrapped $\hb^*_j$ as stated later in Proposition \ref{th1}.

\medskip
Proof: 
It follows from (A3dep) and the Marcinkiewicz-Zygmund inequality that 
\bel{mz-moment}
\E\Big|\,\|\eps\|_2^2 - n\sigma^2_{\eps}\Big|^{(1+\delta/2)}
= O(1)\E\Big|\sum_{i=1}^n(\veps_i^2-\E\veps_i^2)^2\Big|^{1/2+\delta/4}
= O(1)\E\|\eps\|_{2+\delta}^{2+\delta} = O(n). 
\eel
Because Nemirovski's inequality still applies as in Ex.14.3 of \cite{pbvdg11} under the weaker assumptions 
(A3dep) and (A5dep), 
\bel{A5prime}
\E\|\bx^T\eps\|_\infty \le O(C_X\sqrt{\log p})\sqrt{\E\|\eps\|_2^2} = O(\sqrt{n\log p}). 
\eel
As $\heps-\eps = \bx(\hbeta-\beta^0)$, 
$|\eps^T(\heps-\eps)| \le \|\bx^T\eps\|_\infty\|\hbeta-\beta^0\|_1 =o_P(n^{1/2})$ by (A1). 
By (A1) and (A5dep),  
\bel{heps-infty-bd}
\|\hepscent - \eps\|_\infty\vee\|\heps - \eps\|_\infty 
\le \big|{\overline{\eps}}\big|+2\|\bx\|_\infty\|\hbeta-\beta^0\|_1
\le O_P(n^{-1/2})+o_P(1)/\sqrt{\log(p)\log(2|G|)}. 
\eel
This and (\ref{mz-moment}) yield the first statement as 
$\widehat{s.e.}_j/{s.e.}_j = \hsigma_{\veps}/\sigma_{\veps}$ and $\hsigma_\eps^2=\|\heps\|_2^2/n$ by definition. 

The second statement follows in the same manner by (\ref{heps-infty-bd}), (A2dep) and 
\bel{A2prime}
\E\Big|\,\|\eps\circ Z_j\|_2^2 - n\omega_j^2\Big|^{(1+\delta/2)}
= O(1)\E\|\eps\circ Z_j\|_{2+\delta}^{2+\delta} = o\big((n\omega_j^2)^{1+\delta/2}\big). 
\eel

For the normal approximation (\ref{lyapunovclt}) and (\ref{clt-for-max}), define 
\bes
\xi_{i,j} = \frac{Z_{j;i}^T\veps_i}{\omega_j},\quad 
\xi_j = \sum_{i=1}^n \frac{\xi_{i,j}}{n^{1/2}} = \frac{Z_j^T\eps}{n^{1/2}\omega_j}.
\ees
By (A3dep) and (A2dep), $\{Z_{j;i}\eps_i, i\le n\}$ are independent variables satisfying the Lyapunov condition, 
so that $\xi_j\to N(0,1)$. 
Furthermore, for $|G|=O(1)$, 
any linear combination $\sum_{j\in G}a_j\xi_j$ converges in distribution to the corresponding Gaussian 
linear combination $\sum_{j\in G}a_j\zeta_j$ 
by the Lyapunov CLT when $\E(\sum_{j\in G}a_j\zeta_j)^2>0$ and converges in probability 
to $\sum_{j\in G}a_j\zeta_j = 0$ otherwise. Because $\Var(\zeta_j)=1$, this is equivalent to 
\begin{eqnarray}\label{lyapunovclt-1}
\sup_{(t_j, j\in G)} \Big|\P\left[\xi_j \le t_j, j\in G\right] 
- \P\left[\zeta_j \le t_j, j\in G\right] \Big|=o_P(1). 
\end{eqnarray}

Next we bound $T_j-\xi_j$. 
Let $\Delta_j^{(1)} = \sum_{k\neq j} Z_j^TX_k(\beta^0_k-\hbeta_k)/(\sqrt{n}{\hat\omega}_j)$. 
By (\ref{desparsLasso-constr}),  
\bes
\hb_j - \beta_j^0 = \frac{Z_j^TY}{Z_j^TX_j} - \sum_{k\neq j} \frac{Z_j^TX_k}{Z_j^TX_j}\hbeta_k - \beta_j^0
= \frac{Z_j^T\veps + \Delta_j^{(1)}\sqrt{n}{\hat\omega}_j}{Z_j^TX_j}. 
\ees 
Let $\Delta_j^{(2)} = (\omega_j/{\hat\omega}_j-1)\xi_j$.  
By the definition of $\widehat{s.e.}_{\mathrm{robust},j}$ and simple algebra, 
\bes
T_j = \frac{Z_j^T\veps + \Delta_j^{(1)}\sqrt{n}{\hat\omega}_j}{\sqrt{n}{\hat\omega}_j}
= \xi_j + \Delta_j^{(1)}+\Delta_j^{(2)}. 
\ees
It follows from (A1), the first requirement in (A2dep), the consistency of ${\hat\omega}_j$ 
and (\ref{lyapunovclt-1}) that 
\bes
\max_{j\in G}\left| T_j - \xi_j\right|=\max_{j\in G}\Big|\Delta_j^{(1)} + \Delta_j^{(2)}\Big| 
= O_P\big(\sqrt{n}\lambda_X\|\hbeta-\beta^0\|_1\big) + o_P(1) = o_P(1). 
\ees
This and (\ref{lyapunovclt-1}) yield the CLT (\ref{lyapunovclt}) for $\{T_j, j\in G\}$. 

Now we impose the additional assumption (A6dep).  We note that 
\bel{omega_j-bd}
L \le \omega_j^2 = \E\|\eps\circ Z_j\|_2^2/n\le K^2C^{1/(1+\delta/2)}
\eel
by (A2dep), (A3dep) and (A6dep). Again, as $\E\|\eps\|_4^4/n\le C$, Nemirovski's inequality gives
\bel{A2-Z_j-eps}
& \E\max_{j,k\in G}\Big|(\eps\circ Z_j)^T(\eps\circ Z_k)^T - n\omega_{j,k}\Big|
\le K^2\sqrt{8Cn\log(2|G|)}, 
\\ \nonumber & \max_{j\in G}\Big|(\eps\circ Z_j)^T((\heps-\eps)\circ Z_j))^T\Big|
\le \max_{j,k}\big|(Z_j\circ Z_j\circ X_k)^T\eps\big|\,\|\hbeta-\beta^0\|_1 = o_P(n^{1/2}). 
\eel
As $\|(\heps-\eps)\circ Z_j\|_2^2/n \le O_P(\|\heps-\eps\|_\infty^2) = o_P(1)/\{\log p\log(2|G|)\}$, we have 
\bes
|{\hat\omega}_j^2/\omega_j^2-1| = O_P\big(\sqrt{\log(2|G|)/n}\big)+\frac{o_P(1)}{\log p\log(2|G|)}
= o_P\big(\log^{-2}(2|G|)\big). 
\ees
Moreover, as $\E\max_{j\in G}|Z_j^T\eps/\sqrt{n}| = O(\sqrt{\log(2|G|)})$, 
\bel{bound1} 
\max_{j\in G}\left| T_j - \xi_j\right| 
\le O_P\big(\sqrt{n}\lambda_X\|\hbeta-\beta^0\|_1\big)
+ \frac{o_P(1)}{\log^{3/2}(2|G|)} = \frac{o_P(1)}{\sqrt{\log(2|G|)}}. 
\eel

To prove (\ref{clt-for-max}), we note that the covariance structure of $\xi_j$ is the same as that of $\zeta_j$, 
\bes
\E\big(\xi_j\xi_k\big)
= \frac{1}{n}\sum_{i=1}^n \E \big(\xi_{i,j}\xi_{i,k}\big)
= \frac{\E(\eps\circ Z_j)^T(\eps\circ Z_k)}{n\omega_j\omega_k} 
= \E\,\zeta_j\zeta_k. 
\ees
The anti-concentration inequality in Lemma 2.1 of \citet{chernozhukov2013} asserts that 
\bes
\Delta = \frac{o(1)}{\log^{1/2}(2|G|)} \ \Rightarrow\ 
\sup_{c \in \R}\PP\big[ c\le \max_{j \in G}h_j(\zeta_j) \le c+ \Delta \big] = o(1). 
\ees
Thus, as the differences $T_j-\xi_j$ are negligible by (\ref{bound1}), 
(\ref{clt-for-max}) is a consequence of 
\bel{clt-for-max-xi}
\sup_{c \in \R}\big|\PP\big[\max_{j \in G} h_j(\xi_j) \le c\big] - \PP\big[\max_{j \in G}h_j(\zeta_j) \le
    c\big]\big| = o(1). 
\eel
Moreover, (\ref{clt-for-max-xi})
can be established by Theorem 2 in \citet{chernozhukov2013}
provided proper fourth moments and $\ell_\infty$ bounds for $\xi_{i,j}$ exist. 
As $|\xi_j|=\max(\xi_j,-\xi_j)$, $h_j(t)=|t|$ is allowed. 

Because $\delta=2$ under (A6dep), (A6dep) and (A3dep) provide the fourth moment bound 
\bes
\max_{j\in G}\E\sum_{i=1}^n\frac{|\xi_{i,j}|^4}{n} =\max_{j\in G} \frac{1}{n}\E\sum_{i=1}^n
\left|\frac{Z_{j,i}\eps_i}{\omega_j}\right|^4
\le \frac{K^4 \E\|\eps\|_4^4}{n\omega_j^4} = O(1). 
\ees  
Thus, for any $\ell_\infty$ bound $u(\gamma)=u(\gamma,X,Y)$ satisfying 
\bes
\max\left\{\P\Big[ \max_{i\le n, j\in G}|\xi_{i,j}| > u(\gamma)\Big] 
, \sum_{i\le n, j\in G}\P\left[ N\left(0,\frac{\E(Z_{j,i}\eps_i)^2}{\omega_j^2}\right) > u(\gamma)\right]\right\} \le \gamma, 
\ees
Theorem 2 of \citet{chernozhukov2013} asserts that the left-hand side of (\ref{clt-for-max-xi}) 
is no grater than 
\bes
O(1)\left\{n^{-1/8}(\log(|G|n/\gamma))^{7/8}+n^{-1/2}(\log(|G|n/\gamma))^{3/2}u(\gamma)+\gamma\right\}. 
\ees
Similar to the bound for the fourth moment, we have 
\bes
\E\max_{i\le n, j\in G}|\xi_{i,j}| 
= \E\left(\max_{i\le n, j\in G}\frac{|Z_{j,i}\eps_i|}{\omega_j}\right)
= O(\E\|\eps\|_4) = O(n^{1/4}). 
\ees
By (A6dep), $\max_{j\in G} \E(Z_{j,i}\eps_i)^2/\omega_j^2 = O(1)\max_{i\le n}\E\eps_i^2 = O(n^{3/7})$. 
These $\ell_\infty$ bounds provide  
\bes
u(\gamma) = O(1)\big\{n^{1/4} +n^{3/14} \log^{1/2}(n|G|)\big\}
\ees 
for certain $\gamma=o(1)$. Thus, \eqref{clt-for-max-xi} holds via the conditions $\log(|G|)\ll n^{1/7}$.\hfill$\Box$

\medskip
As a next step, we present the counter part of Proposition \ref{prop1} for 
the residual bootstrap. 

\begin{prop}\label{th1}
Assume (A1)-(A5) with $\E\eps_i^2=\sigma^2_\eps$ for all $i\le n$. 
Let $\P^*$ represent the residual bootstrap of $\hat{\eps}_{\mathrm{cent}}$. 
Then, ${s.e.}_j = s.e._{\mathrm{robust},j}$, 
\bes
& \displaystyle \widehat{s.e.}_j^*/{s.e.}_j = \hsigma_{\veps}^*/\sigma_{\veps} 
= 1 + \frac{O_{P^*}}{n^{{\delta}/(2+{\delta})}}+ \frac{o_{P^*}(1)}{\log(p)\log(2|G|)}\ \forall j\le p \hbox{ in probability,}
\cr & \left|\widehat{s.e.}_{\mathrm{robust},j}^*/s.e._{\mathrm{robust},j}-1\right|
= \left|\hat{\omega}_j^*/\omega_j -1\right| = o_{P^*}(1)\ \hbox{in probability for each $j\in G$.}
\ees
Let $T_j^* = \sgn(Z_j^TX_j)(\hb^*_j-\hbeta_j)/\widehat{s.e.}^*_{j}$ and 
$(\zeta_j, j\in G)$ be as in Proposition \ref{prop1}. 
If $|G|=O(1)$, then  
\bel{clt-boot}
\sup_{(t_j, j\in G)} \Big|\P^*\left[T_j^* \le t_j, j\in G\right] 
- \P\left[\zeta_j \le t_j, j\in G\right] \Big|=o_P(1). 
\eel
If (A6) holds, then 
\bel{clt-max-boot}
\sup_{c \in \R}\big|\PP^*\big[\max_{j \in G} h_j(T_j^*) \le c\big] - \PP\big[\max_{j \in G}h_j(\zeta_j) \le
    c\big]\big| = o_{P}(1)
\eel
for any combination of functions $h_j(t)=t$, $h_j(t)=-t$ or $h_j(t)=|t|$. 
\end{prop}

Proof: It follows from (\ref{mz-moment}) and (\ref{heps-infty-bd}) 
that the bootstrap analogue of (A3dep) holds:
\bel{A3-1*}
\left|\EE^*\|\eps^*\|_2^2/n - \sigma^2_\eps\right| 
= \left|\|\hat{\eps}_{\mathrm{cent}}\|_2^2/n - \sigma^2_\eps\right| 
= o_P(1),\ \max_{i\le n}\EE^*|\eps_i^*|^{2+ {\delta}} 
= \|\hat{\eps}_{\mathrm{cent}}\|_{2+{\delta}}^{2+{\delta}}/n
= O_P(1). 
\eel
Therefore, because $\sigma_i^2=\sigma^2_\eps$ is the same for all $i$ 
and $Z_j$ and $\bx$ are unchanged from the original in the residual bootstrap, 
we have the $\P^*$ analogue of all the conditions in Proposition \ref{prop1}. 
Moreover, 
$\sqrt{n}\omega_j = \|Z_j\|_2\sigma_\eps$, $n\omega_{j,k}=Z_j^TZ_k\sigma^2_\eps$, 
and the correlation structure of 
\bes
Z_j^T\eps^*/\sqrt{\Var^*(Z_j^T\eps^*)}
= Z_j^T\eps^*/\{\|Z_j\|_2(\|\hat{\eps}_{\mathrm{cent}}\|_2/\sqrt{n})\}
\ees
is the same as that of  $\zeta_j$. Proposition \ref{th1} follows. \hfill$\Box$

\medskip
\paragraph{Proof of Theorem \ref{th1b}.}
Because the Gaussian vector $\zeta$ in (\ref{lyapunovclt}) and (\ref{clt-for-max}) 
is identical to the one in (\ref{clt-boot}) and (\ref{clt-max-boot}), 
the conclusions follow from Propositions \ref{prop1} and \ref{th1}. 
\hfill$\Box$

\medskip
Besides the proof we note that the estimated standard errors are all
consistent and asymptotically equivalent:
\begin{eqnarray*}
\widehat{s.e}_j \sim \sqrt{\hbox{\rm Asym.Var}(\hat{b}_j)}
\sim \sqrt{\hbox{\rm Asym.Var}^*(\hat{b}^*_j)} \sim  \widehat{s.e.}^*_j,
\end{eqnarray*}
where ``$\sim$'' denotes asymptotic equivalence (the ratio converging to
one), and we omit here details regarding the measure $\P$ or $\P^*$ and
that statements hold ``in probability'' only. 

\subsection{Proof of Theorem \ref{th2} for heteroscedastic
  errors}\label{subsec.proofth2}

\medskip
\paragraph{Proof of Theorem \ref{th2}.} 
The statements follow directly from (\ref{lyapunovclt}) and (\ref{clt-boot}), as $\Var(\zeta_j)=1$. 
\hfill$\Box$

\medskip
We note that 
\begin{eqnarray*}
\widehat{s.e}_{\mathrm{robust},j} \sim \sqrt{\hbox{\rm Asym.Var}(\hat{b}_j)} 
\not\sim \sqrt{\hbox{\rm Asym.Var}^*(\hat{b}^*_j)} \sim  \widehat{s.e.}^*_{\mathrm{robust},j},
\end{eqnarray*}
This happens because the bootstrap is \emph{not} mimicking the
heteroscedastic errors (but constructs i.id. errors instead). But since we
approximate only the studentized pivotal quantity $(\hat{b}_j -
\beta^0_j)/\widehat{s.e}_{\mathrm{robust},j}$ with the bootstrap analogue,
we still obtain asymptotic consistency for the bootstrap to approximate the
studentized pivot. Note that the wild bootstrap or paired xyz-bootstrap
(see Sections \ref{subsec.multiplbtrp} and \ref{subsec.hetresbootstrap})
would indeed provide asymptotic equivalence of the
original and bootstrapped estimated standard errors, see Theorem
\ref{th.max}.

\subsection{Proof of Theorem \ref{th.max} for the wild 
and xyz-paired bootstrap}\label{subsec.proofthmax}

We first provide an analogue of Proposition \ref{prop1} for the wild bootstrap.  

\begin{prop}\label{prop3}
Assume (A1)-(A5). Let $\P^*$ represent the wild bootstrap. Then, 
\bes
\left|\widehat{s.e.}_{\mathrm{robust},j}^*/s.e._{\mathrm{robust},j}-1\right|
= \left|\hat{\omega}_j^*/\omega_j -1\right| = o_{P^*}(1)\ \hbox{ in probability for each $j\in G$.}
\ees
Let $T_j^* = \sgn(Z_j^TX_j)(\hb_j^* - \hbeta_j)/\widehat{s.e.}_{\mathrm{robust},j}^*$. 
If $|G|=O(1)$, then
\bes
\sup_{(t_j, j\in G)} \Big|\P^*\left[T_j^* \le t_j, j\in G\right] 
- \P\left[\zeta_j \le t_j, j\in G\right] \Big|=o_P(1). 
\ees
where $(\zeta_j, j\in G)$ is as in Proposition \ref{prop1}. 
If (A6) holds, then 
\bes
\sup_{c \in \R}\big|\PP^*\big[\max_{j \in G} h_j(T_j^*) \le c\big] - \PP\big[\max_{j \in G}h_j(\zeta_j) \le
    c\big]\big| = o_P(1) 
\ees
for any combination of functions $h_j(t)=t$, $h_j(t)=-t$ or $h_j(t)=|t|$. 
\end{prop}

Proof: It follows from (\ref{heps-infty-bd}) and (\ref{A2prime}) that for $|G|=O(1)$
\bel{wild-cov-bd}
\max_{j\in G,k\in G}\left| \frac{(Z_j\circ\hepscent)^T(Z_k\circ\hepscent)}{n\omega_j\omega_k} 
- \frac{\omega_{j,k}}{\omega_j\omega_k} \right| \le \frac{o_P(1)}{\log(2|G|)}. 
\eel
For unbounded $|G|$, (\ref{wild-cov-bd}) follows from 
(\ref{omega_j-bd}) and (\ref{A2-Z_j-eps}) under (A6). 

As $\eps^{*W}_i$ are i.i.d. variables under $\P^*$, 
(\ref{A3-1*}) gives the $\P^*$ analogue of (A3dep): $\E\big[\eps^{*W}_i\big]=0$, 
\bes
\E\big(\eps^{*W}_i\big)^2 = \|\hepscent\|_2^2/n = \sigma^2_\eps+o_P(1),\ 
\E\big|\eps^{*W}_i\big|^{2+\delta} = \big(\E|W_1|^{2+\delta}\big)\|\hepscent\|_{2+\delta}^{2+\delta}/n=O_P(1). 
\ees 
Therefore, as $X_k$ and $Z_j$ are unchanged from the original ones in wild bootstrap 
and the $\P^*$ analogue of (A1) is (A4), we have the $\P^*$ analogue of all conditions of 
Proposition \ref{prop1}. It follows that 
\bes
& \displaystyle \left|\hat{\omega}_j^*/(\|Z_j\circ\hepscent\|_2/\sqrt{n}) -1\right| 
= o_{P^*}(1)\ \hbox{ in probability for each $j\in G$,}
\cr & \sup_{(t_j, j\in G)} \Big|\P^*\left[T_j^* \le t_j, j\in G\right] 
- \P^*\left[\zeta^*_j \le t_j, j\in G\right] \Big|=o_P(1), 
\ees
for $|G|=O(1)$ and a centered Gaussian vector $(\zeta^*_j, j\in G)$ with covariance structure 
\bes
\E^*\zeta_j^*\zeta_k^* 
= \frac{\E^*(Z_j\circ\eps^*)^T(Z_k\circ\eps^*)}{\sqrt{\E^*\|Z_j\circ\eps^*\|_2^2\E^*\|Z_k\circ\eps^*\|_2^2}}
= \frac{(Z_j\circ\hepscent)^T(Z_k\circ\hepscent)}{\|Z_j\circ\hepscent\|_2\|Z_k\circ\hepscent\|_2}. 
\ees
These and (\ref{wild-cov-bd}) yield the first two statements as 
$\E\zeta_j\zeta_k = \omega_{j,k}/(\omega_j\omega_k)$. 
Moreover, 
\bes
\sup_{c \in \R}\Big|\PP^*\big[\max_{j \in G} h_j(T_j^*) \le c\big] - \PP^*\big[\max_{j \in G}h_j(\zeta^*_j) \le
    c\big]\Big| = o_P(1), 
\ees
under the additional condition (A6), 
so that the conclusion for $\max_{j\in G}h_j(T_j)$ follows directly 
from a comparison between the distributions of $\max_{j\in G}h_j(\zeta^*_j)$ under $\P^*$ 
and $\max_{j\in G}h_j(\zeta_j)$ under $\P$
through (\ref{wild-cov-bd}) and Lemma 3.1 of \cite{chernozhukov2013}. 
\hfill$\Box$

\medskip
Next, we provide an analogue of Proposition \ref{prop1} for the xyz-paired bootstrap.  

\begin{prop}\label{prop4}
Assume (A1)-(A5) with $\delta=2$. Suppose $\log p = O(n^{1/2})$. 
Let $\P^*$ represent the xyz paired-bootstrap. Then, 
\bes
\left|\widehat{s.e.}_{\mathrm{robust},j}^*/s.e._{\mathrm{robust},j}-1\right|
= \left|\hat{\omega}_j^*/\omega_j -1\right| = o_{P^*}(1)\ \hbox{ in probability for each $j\in G$.}
\ees
Let $T_j^* = \sgn((Z_j^*)^TX_j^*)(\hb_j^* - \hbeta_j)/\widehat{s.e.}_{\mathrm{robust},j}^*$. 
If $|G|=O(1)$, then
\bes
\sup_{(t_j, j\in G)} \Big|\P^*\left[T_j^* \le t_j, j\in G\right] 
- \P\left[\zeta_j \le t_j, j\in G\right] \Big|=o_P(1). 
\ees
where $(\zeta_j, j\in G)$ is as in Proposition \ref{prop1}. 
If (A6) holds, then 
\bes
\sup_{c \in \R}\big|\PP^*\big[\max_{j \in G} h_j(T_j^*) \le c\big] - \PP\big[\max_{j \in G}h_j(\zeta_j) \le
    c\big]\big| = o_P(1) 
\ees
for any combination of functions $h_j(t)=t$, $h_j(t)=-t$ or $h_j(t)=|t|$. Moreover, 
\bes
\P^*\big[ \sgn((Z_j^*)^TX_j^*) \neq \sgn(Z_j^TX_j), \forall\, j \in G\big] =o_P(1),
\ees
provided that $\sqrt{\log(2|G|)} = o_P(1)\min_{j\in G}(|Z_j^TX_j|/\|Z_j\|_2)$. 
\end{prop}

\medskip
Proof: We shall think about bootstrap sampling of the entire rows 
$({\hat\bx},\hY,{\hat\bz},\bx,Y,Z_j,j\le p)$ and denote by $(U)^*$ the bootstrapped $U$. 
Although we shall be careful as $Z_j^*=(\hZ_j)^*\neq (Z_j)^*$, this should lead to no confusion 
as we always name the original variables inside the parentheses. 

The main task of the proof is to establish the $\P^*$ analogue of (A2dep), (A5dep) and (A6dep). 
Note that the $\P^*$ analogue of (A1) is (A4). 
Because the elements of $\eps^*$ are i.i.d. random elements of $\hepscent$ 
as in residual bootstrap, we have already verified the $\P^*$ analogue of (A3dep) in (\ref{A3-1*}). 
 
We shall study properties of $\hX_k$ and $\hZ_j$. Recall that 
\bes
\hX_k = X_k - a_k{\hat\eps}_{\rm cent},\ 
\hZ_j = Z_j - b_j{\hat\eps}_{\rm cent}, 
\ees
with $a_k = (X_k^T\hepscent)/\|\hepscent\|_2^2$ and $b_j = (Z_j^T\hepscent)/\|\hepscent\|_2^2$. 
We need to use the fact that $\hbeta$ is the Lasso estimator. 
By (\ref{A5prime}) and the basic inequality \citep[Lemma 6.1]{pbvdg11}, 
\bes
\|\hepscent-\eps\|_2^2/n\le \|\bx(\hbeta-\beta^0)\|_2^2/n + {\overline\eps}^2 
\le (\|\bx^T\eps/n\|_\infty+\lambda)\|\hbeta-\beta^0\|_1 +O_P(1/n) 
= o_P(n^{-1/2}). 
\ees
Thus, by (\ref{A3-1*}), (\ref{A5prime}) and the condition $\log p\ll \sqrt{n}$,  
\bes
\hbox{$\max_{k\le p}$}|a_k|
= O_P\big(\|\bx^T\eps/n\|_\infty+\|\hepscent - \eps\|_2/n^{1/2}\big)
= o_P(n^{-1/4}). 
\ees
This gives the $\P^*$ analogue of (A5dep) as $\E^*(X_k^*\circ\eps^*)_i = \hX_k^T\hepscent/n=0$ for all $i\le n$ and 
\bes
\hbox{$\max_{k\le p}$}\|X^*_k\|_\infty\le C_X+o_P(n^{-1/4})\|\hepscent\|_4 = C_X+o_P(1). 
\ees
Similarly, the $\P^*$ analogue of (A6dep) holds along with $\max_{j\in G}|b_j|=o_P(n^{-1/4})$ 
under the additional condition (A6), as $\E^*(\eps^*_i)^2 = \|\hepscent\|_2^2/n = \sigma^2_\eps+o_P(1)$. 

Now, under (A6), we prove the $\P^*$ analogue of (A2dep) and 
\bel{xyz-cov-bd}
\max_{j\in G,k\in G}\left| \frac{(\hZ_j\circ\hepscent)^T(\hZ_k\circ\hepscent)}{n\omega_j\omega_k} 
- \frac{\omega_{j,k}}{\omega_j\omega_k} \right| \le \frac{o_P(1)}{\log(2|G|)}. 
\eel
Since $\hZ_j\circ\hepscent = Z_j\circ\hepscent - b_j\hepscent\circ\hepscent$ 
and $\max_{j\in G}|b_j|=o_P(n^{-1/4})$, 
\bes
\max_{j\in G}\|\hZ_j\circ\hepscent - Z_j\circ\hepscent\|_2
\le o_P(n^{-1/4})\|\hepscent\|_4^2 = o_P(n^{1/4}), 
\ees
so that (\ref{xyz-cov-bd}) follows from (\ref{wild-cov-bd}) and the condition $\log(|G|)\ll n^{1/7}$.  
As $L\le \|Z_j\|_2^2/n\le K$ and $\max_j\|Z^*_j\|_\infty\le K+o_P(1)$, 
the second and fourth moment requirements in the $\P^*$ analogue of (A2dep) follow
respectively from (\ref{xyz-cov-bd}) and the $\P^*$ analogue of (A3dep). 
By Nemirovski's inequality, 
\bel{z_jx_k}
&& \E^*\max_{j\in G,k\le p}\Big|\{(Z_j^*)^TX_k^* - Z_j^TX_j\}/(n^{1/2}\|Z_j\|_2)\Big|
\cr &\le&\sqrt{8\log(2p|G|)/n}\max_k\|\hX_k\|_\infty 
+ \max_{j\in G,k\le p}\Big|(\hZ_j^T\hX_k - Z_j^TX_j)/(n^{1/2}\|Z_j\|_2)\Big|
\cr &\le&O_P(1)\sqrt{\log(2p)/n} + O_P(1)\max_{j\in G,k\le p}\Big|a_jb_k/(n^{1/2}\|Z_j\|_2)\Big|
\cr &\le&O_P(1)\sqrt{\log(2p)/n},
\eel
where $\sqrt{\log(2p)/n}$ on the right-hand side can be replaced by $\sqrt{\log(2|G|)/n}$ when 
the maximum on the left-hand size is taken over $j\in G$ and $k=j$. 
Thus, as $\E^*[(Z_j^*\circ \eps^*)_i] = \hZ_j^T\hepscent/n=0$ 
and $\max_{j\in G,k\neq j}|Z_j^TX_j/n|\le\lambda_X$ by (A2), we have the $\P^*$ analogue of (A2dep). 

We still want to prove (\ref{xyz-cov-bd}) and the $\P^*$ analogue of (A2dep) for $|G|=O(1)$ 
without assuming (A6). As $b_j = (Z_j^T\hepscent)/\|\hepscent\|_2^2$, 
(A2) and (\ref{A3-1*}) imply 
\bes
|b_j| \le O_P(1)\big(\|Z_j\|_2/n + \|Z_j\|_2\|\hepscent-\eps\|_2/n \big) 
= o_P\big(n^{-1/4}\big)\|Z_j\|_2/\sqrt{n} = o_P\big(\omega_j/n^{1/4}\big). 
\ees
Thus, by (\ref{heps-infty-bd}), the proof still works for (\ref{xyz-cov-bd}) and the $\P^*$ analogue of (A2dep). 

As we have proved the last statement about the sign agreement between $(Z_j^*)^TX_j^*$ 
and $Z_j^TX_j$ via (\ref{z_jx_k}) and the $\P^*$ analogue of the conditions of 
Proposition \ref{prop1}, 
the other statements of the proposition follow from Proposition \ref{prop1} in the same manner 
as in the proof of Proposition \ref{prop3}. 
\hfill$\Box$

\paragraph{Proof of Theorem \ref{th.max}.}
The theorem is a direct consequence of Propositions \ref{prop3} and \ref{prop4}. \hfill$\Box$

\subsection{Additional simulation results}
\label{app:additionalsimulations}

\subsubsection{Quantile estimation with the Gaussian multiplier bootstrap
  and the residual bootstrap}\label{app:multiplier-quantileest0p95}
We compare different bootstrap methods for estimating the 95\% quantile of
the distribution of  
\begin{equation*}
  \max_{j \in \{1,\dots,p\}} |T_j|  \: \: \: \mbox{under} \: \: \:
  H_{0,\mathrm{complete}},\ \ T_j = \hat{b}_j/\widehat{s.e.}_j.
\end{equation*}

The data is generated with a single Toeplitz type design matrix and a single
choice of $U(0,2)$ signal vector. The ground truth is computed by fitting
the de-sparsified Lasso $10^5$ times on newly generated pure noise $Y =
\eps$ and taking the 95\% empirical quantile. The bootstrap estimates
are computed for 500 realizations of the linear model by computing $B=1000$
bootstrap samples each time. We therefore have 500 estimates for the
quantile from each method that can be plotted in a boxplot.

\begin{figure}[!htb]
  \centering
\includegraphics[scale=0.6]{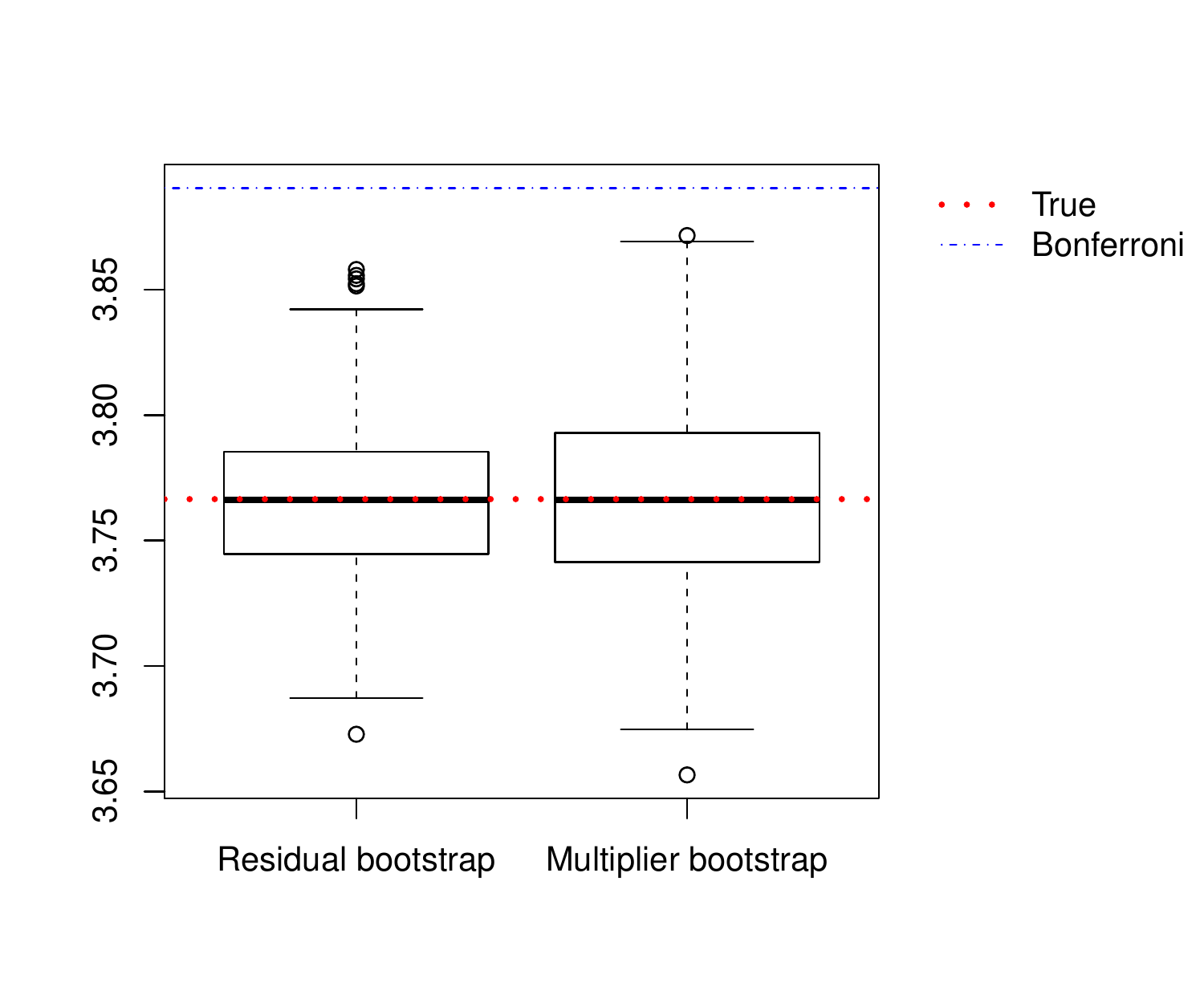}
\caption{Boxplots comparing the estimates of the multiplier bootstrap and
  the residual bootstrap for the 95\% quantile of the distribution $\max_{j
    \in \{1,\dots,p\}} |T_j|  \: \: \: \mbox{under} \: \: \:
  H_{0,\mathrm{complete}}$. The boxplots are based on 500 estimates,
  each computed for a different realization of the model. \textbf{The dimensions are
    $n=100,\ p=500$ for a matrix of Toeplitz design type. The errors are homoscedastic
    Gaussian $\eps \sim \mathcal{N}_n(0,I_n)$.} The horizontal red line
  denotes the true value of the quantile, the blue line denotes the
  corresponding Bonferroni rejection threshold.}
\label{fig:q95-experiment}
\end{figure}

\begin{figure}[!htb]
  \centering
\includegraphics[scale=0.6]{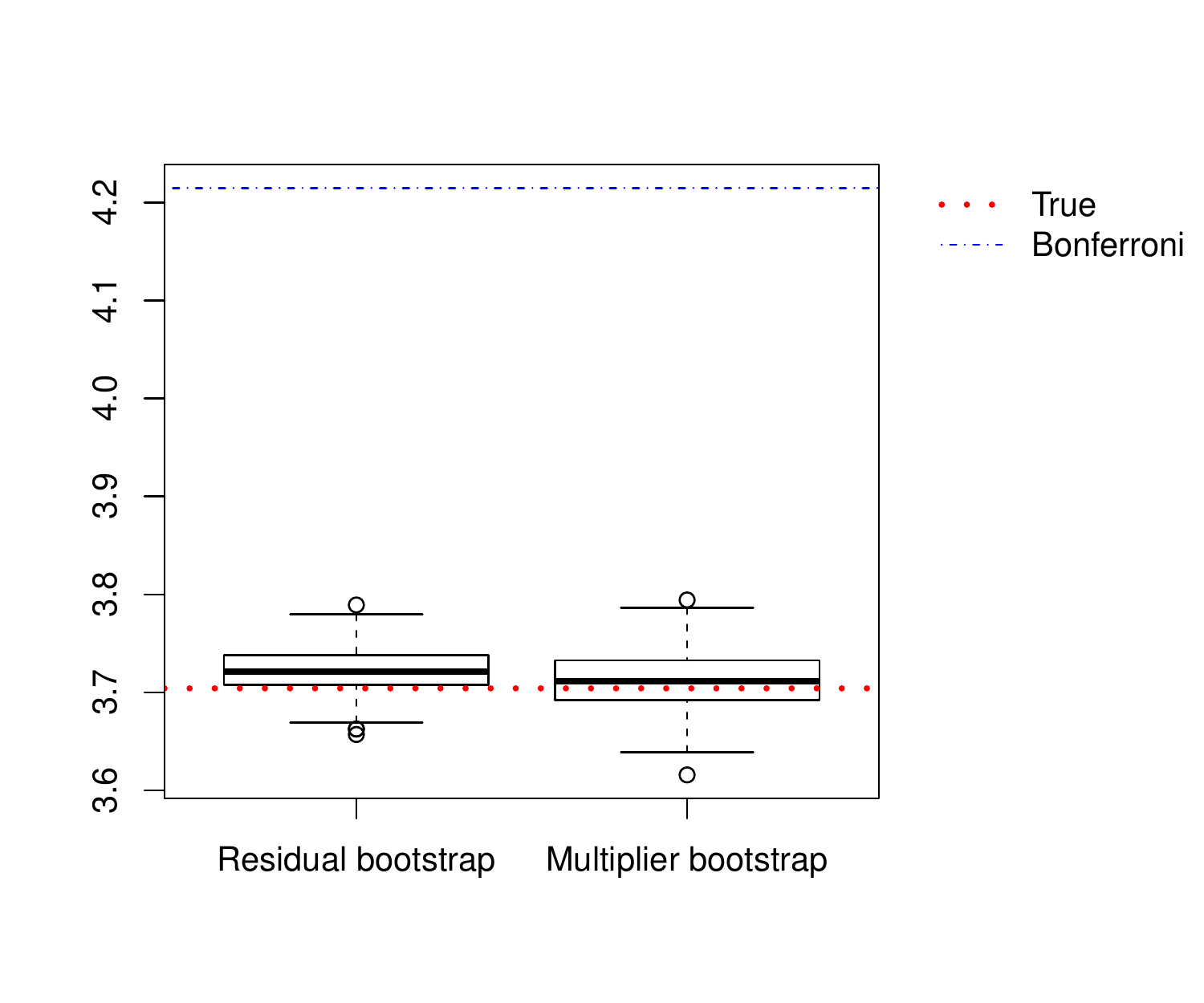}
\caption{Boxplots comparing the estimates of the multiplier bootstrap and
  the residual bootstrap for the 95\% quantile of the distribution $\max_{j
    \in \{1,\dots,p\}} |T_j|  \: \: \: \mbox{under} \: \: \:
  H_{0,\mathrm{complete}}$. The boxplots are based on 500 estimates,
  each computed for a different realization of the model. \textbf{The dimensions are
    $n=30,p=2000$ for a matrix of Toeplitz design type. The errors are homoscedastic
    chi-squared as described in Section
    \ref{subsubsect:homoscedasticnongaus}.} The horizontal red line
  denotes the true value of the quantile, the blue line denotes the
  corresponding Bonferroni rejection threshold.} 
  \label{fig:q95-experiment-chisq}
\end{figure}

We first compare the Gaussian multiplier bootstrap to the residual bootstrap.
The results for dimensions $n=100,p=500$ and Gaussian noise $\eps \sim
\mathcal{N}_n(0,I_n)$ can be found in Figure \ref{fig:q95-experiment}. Both
bootstrapping methods seem to be equally good. The results for dimensions
$n=30,p=2000$ and centered $\chi_1^2$ errors as described in  
Section \ref{subsubsect:homoscedasticnongaus} can be found in Figure
\ref{fig:q95-experiment-chisq}. Again, there seems to be hardly any
difference between the bootstrapping methods.

\begin{figure}[!htb]
  \centering
\includegraphics[scale=0.6]{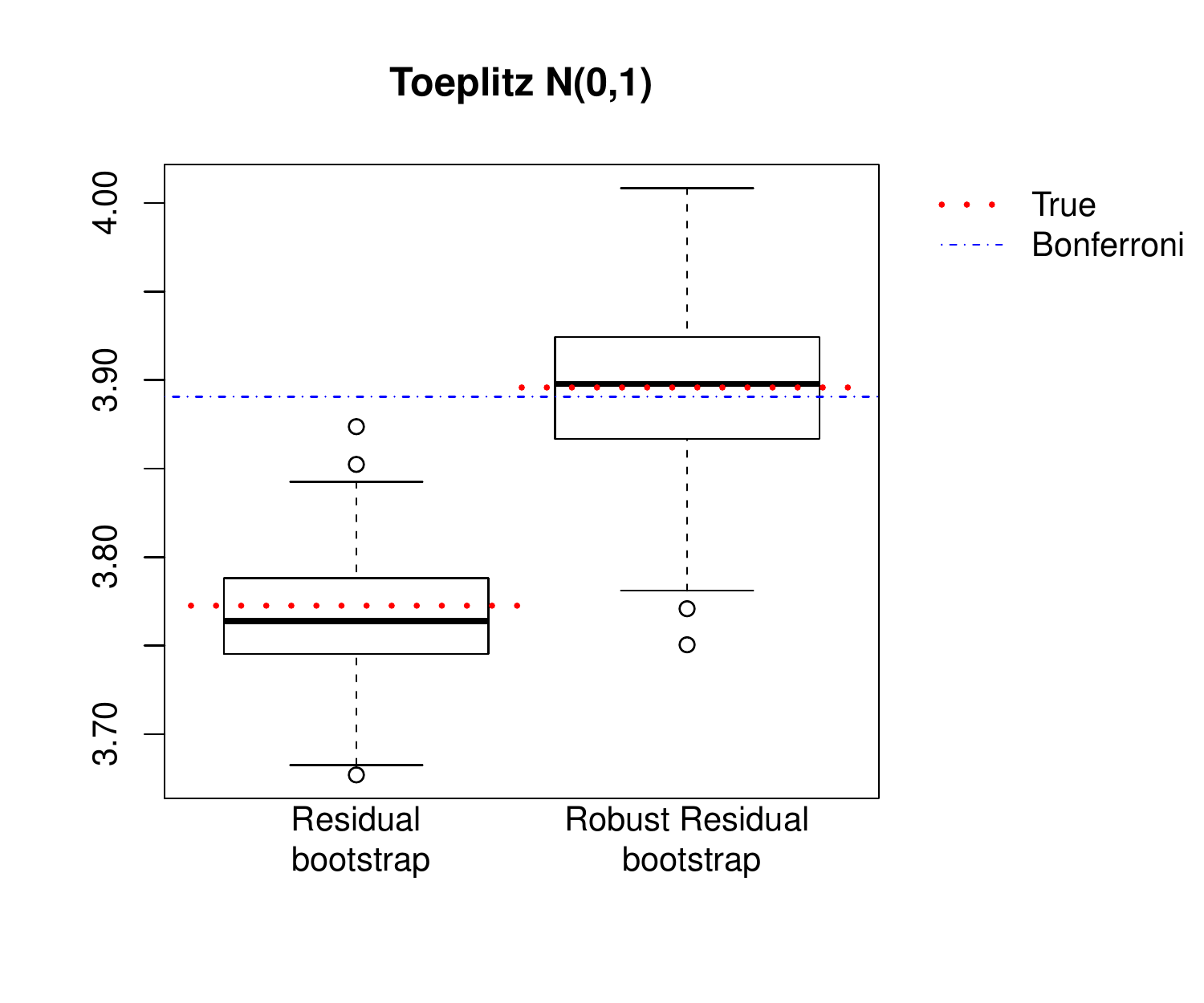}
\caption{Same as Figure \ref{fig:q95-experiment} but comparing the
  estimates of the non-robust studentized estimator to the robust version. Due
  to the difference in the test statistics, the underlying true quantiles
  are different. The
  residual bootstrap is used.} 
\label{fig:q95-experiment-homoscedastic-robustvsnot}
\end{figure}

\begin{figure}[!htb]
  \centering
\includegraphics[scale=0.6]{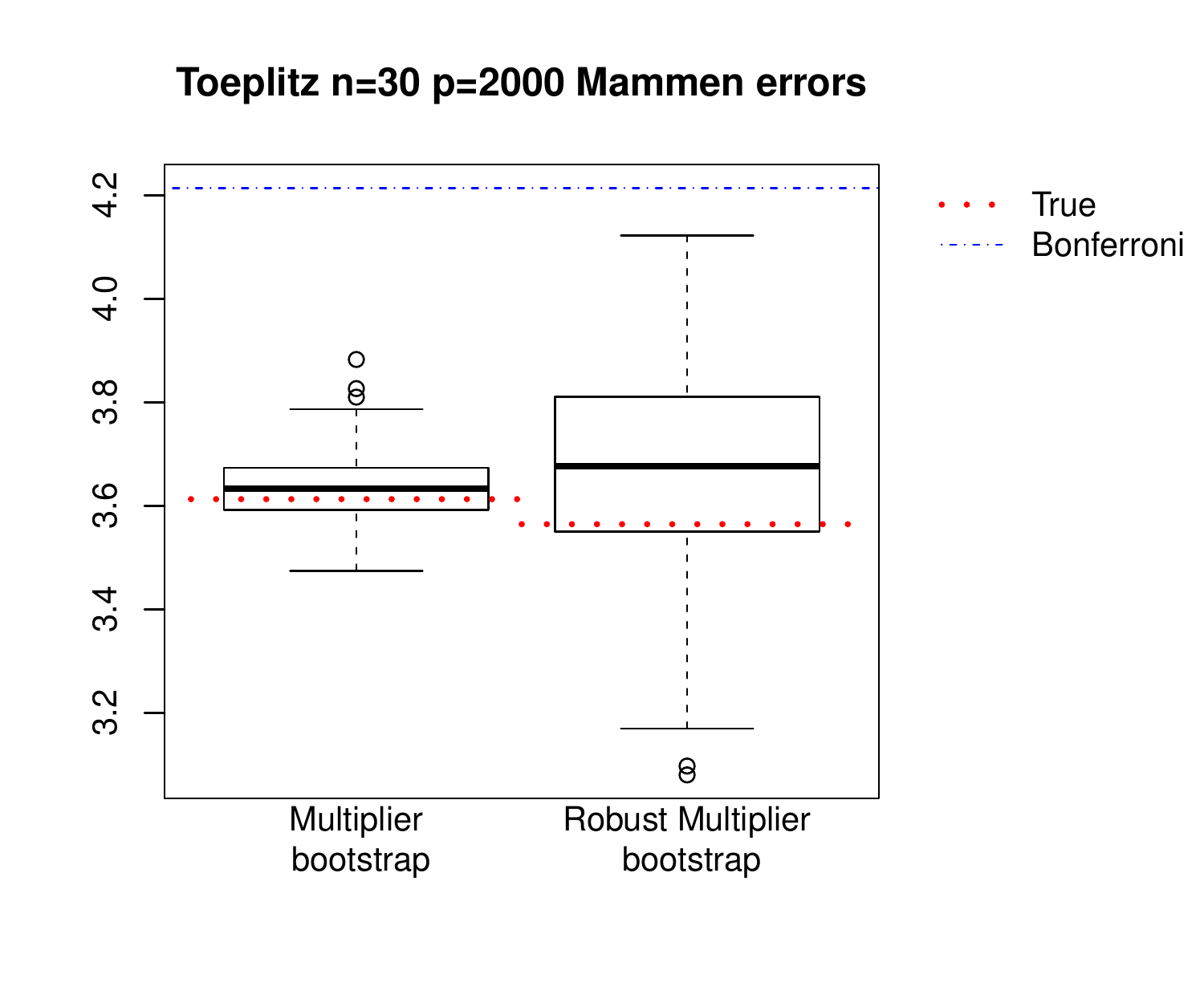}
  \caption{Same as Figure
    \ref{fig:q95-experiment-homoscedastic-robustvsnot}, but for a Toeplitz 
    design matrix of different dimensions ($n=30,p=2000$) and with
    heteroscedastic errors as described in Section
    \ref{subsubsect:heteroscedasticnongaus}. The multiplier bootstrap is
    used for the estimator with either the robust or non-robust
    studentization. The Gaussian multiplier wild bootstrap is used (the
    non-robust studentization does not lead to $N(0,1)$ scaling of the
    components in the maximum).}
\label{fig:q95-experiment-heteroscedastic-robustvsnot}
\end{figure}

Next, we compare the robust version of the test statistic $T_j =
\hat{b}_j/\widehat{s.e.}_{\mathrm{robust},j}$ to the one with non-robust
studentization, for both the
  bootstrap approaches. In Figure
  \ref{fig:q95-experiment-homoscedastic-robustvsnot}  
  the comparison of the residual bootstrap methods can be found for the
  homoscedastic linear model used in Figure 
  \ref{fig:q95-experiment}. Figure \ref{fig:q95-experiment-heteroscedastic-robustvsnot}
  compares both versions of the multiplier bootstrap for heteroscedastic
  errors (taken from Section \ref{subsubsect:heteroscedasticnongaus}), with
  a design matrix 
  of type Toeplitz of dimensions $n=30,p=2000$. The robust version of the
  bootstrap performs well in the heteroscedastic example. For the
  homoscedastic case, the Bootstrap for the robust test-statistic doesn't
  seem to gain over Bonferroni. 

\subsubsection{All results for the multiplier bootstrap}
\label{app:additionalmultiplier}

We compare the residual bootstrap approach to the wild bootstrap for all
the results from Section \ref{subsec:varyingepsdist}. 

\begin{figure}[!htb]
  \centering
\includegraphics[scale=0.6]{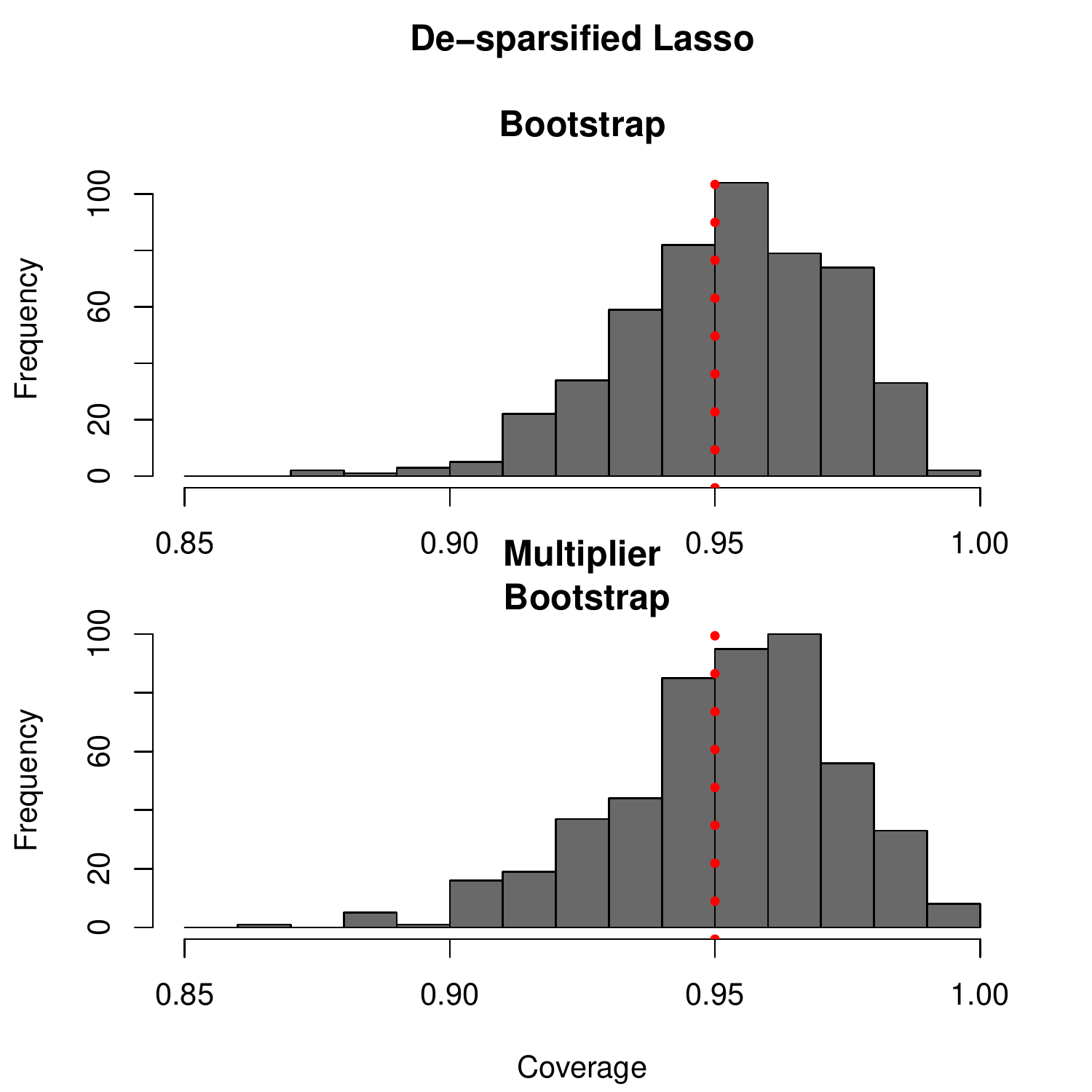}
  \caption{Same plot as the homoscedastic Gaussian results in Figure
    \ref{fig:cihist-toeplitzgauss}, but showing also the 
    multiplier bootstrap.}
  \label{fig:cihist-toeplitzgaussmultiplier}
\end{figure}

\begin{figure}[!htb]
  \centering
\includegraphics[scale=0.75]{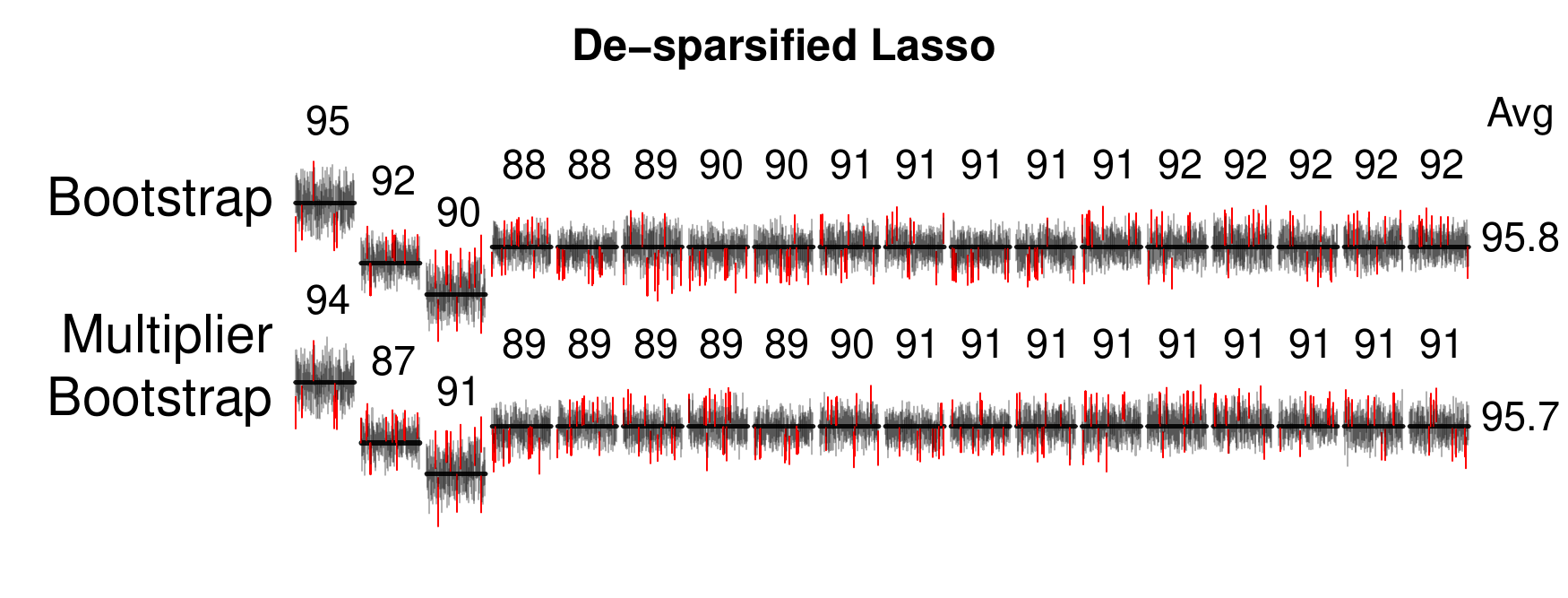}
  \caption{Same plot as the homoscedastic Gaussian results in Figure
    \ref{fig:ci-toeplitzgauss}, but showing also the 
    multiplier bootstrap.} 
  \label{fig:ci-toeplitzgaussmultiplier}
\end{figure}

\begin{figure}[!htb]
  \centering
\includegraphics[scale=0.6]{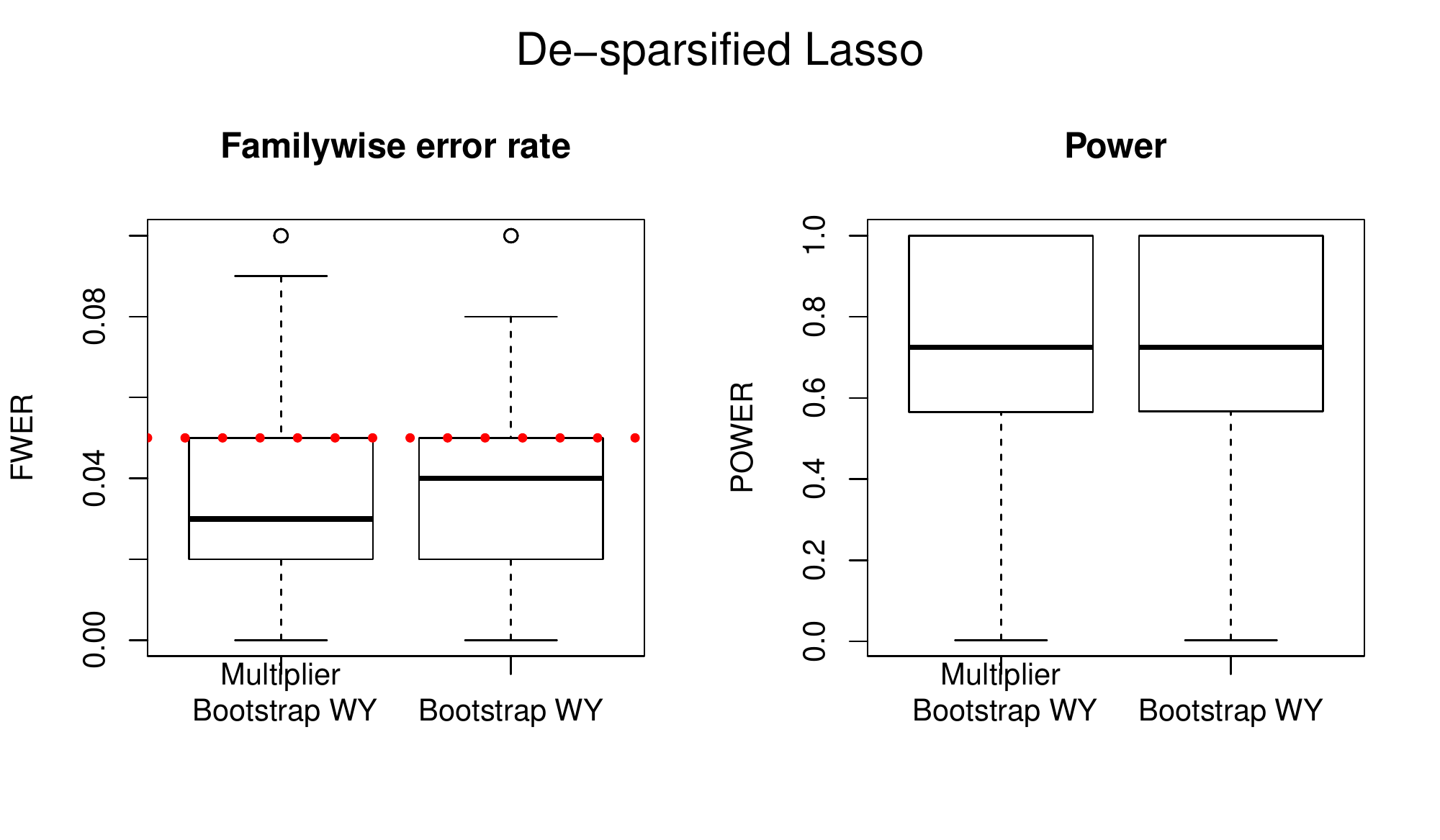}
  \caption{Same plot as the homoscedastic Gaussian results in Figure
    \ref{fig:powerfwer-toeplitzgauss}, but showing also the 
    multiplier bootstrap.}  

  \label{fig:powerfwer-toeplitzgaussmultiplier}
\end{figure}

\begin{figure}[!htb]
  \centering
\includegraphics[scale=0.6]{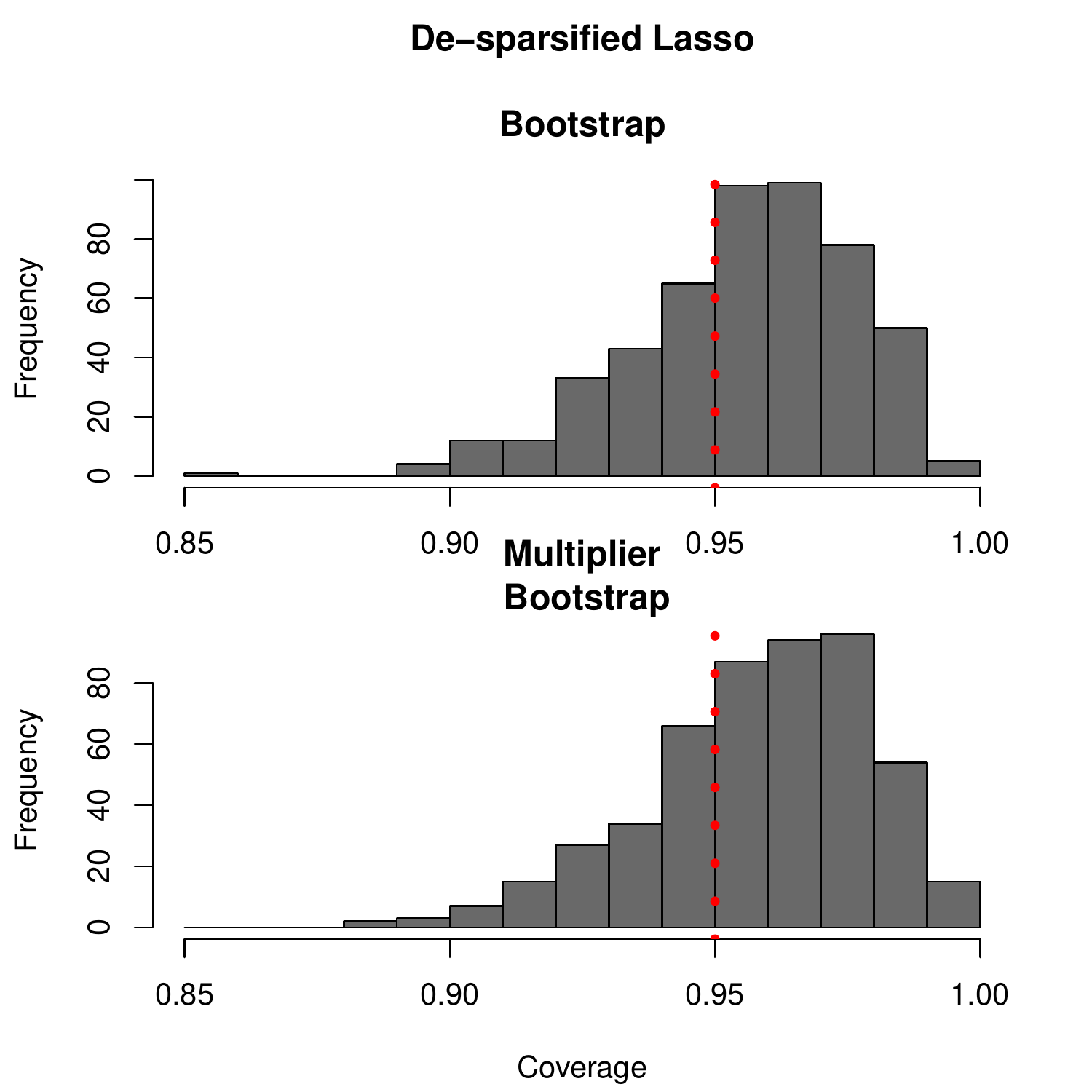}
  \caption{The same plot as Figure \ref{fig:cihist-toeplitzgauss} but for
    \textbf{homoscedastic chi-squared errors} and showing also the 
    multiplier bootstrap.} 
  \label{fig:cihist-toeplitzchisqmultiplier}
\end{figure}

\begin{figure}[!htb]
  \centering
  \includegraphics[scale=0.8]{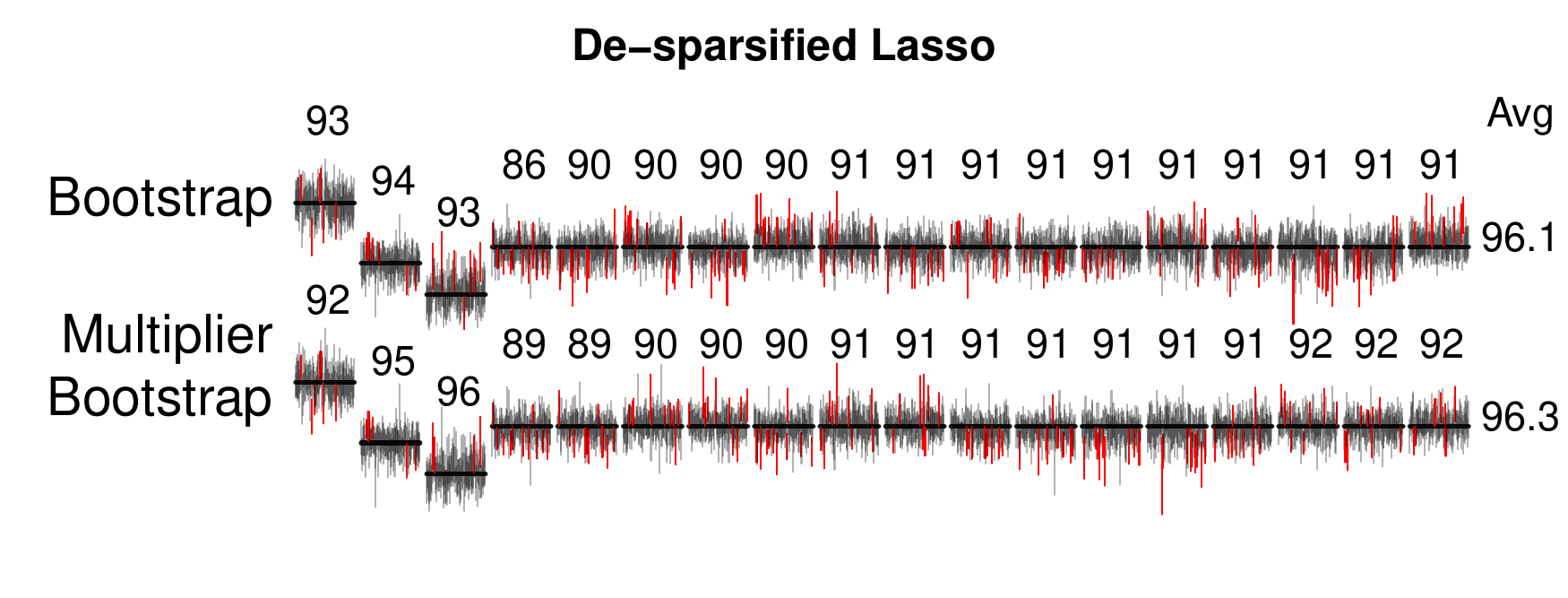}
  \caption{The same plot as Figure \ref{fig:ci-toeplitzgauss} but for
    \textbf{homoscedastic chi-squared errors} and showing also the 
    multiplier bootstrap.} 
  \label{fig:ci-toeplitzchisqmultiplier}
\end{figure}

\begin{figure}[!htb]
  \centering
\includegraphics[scale=0.6]{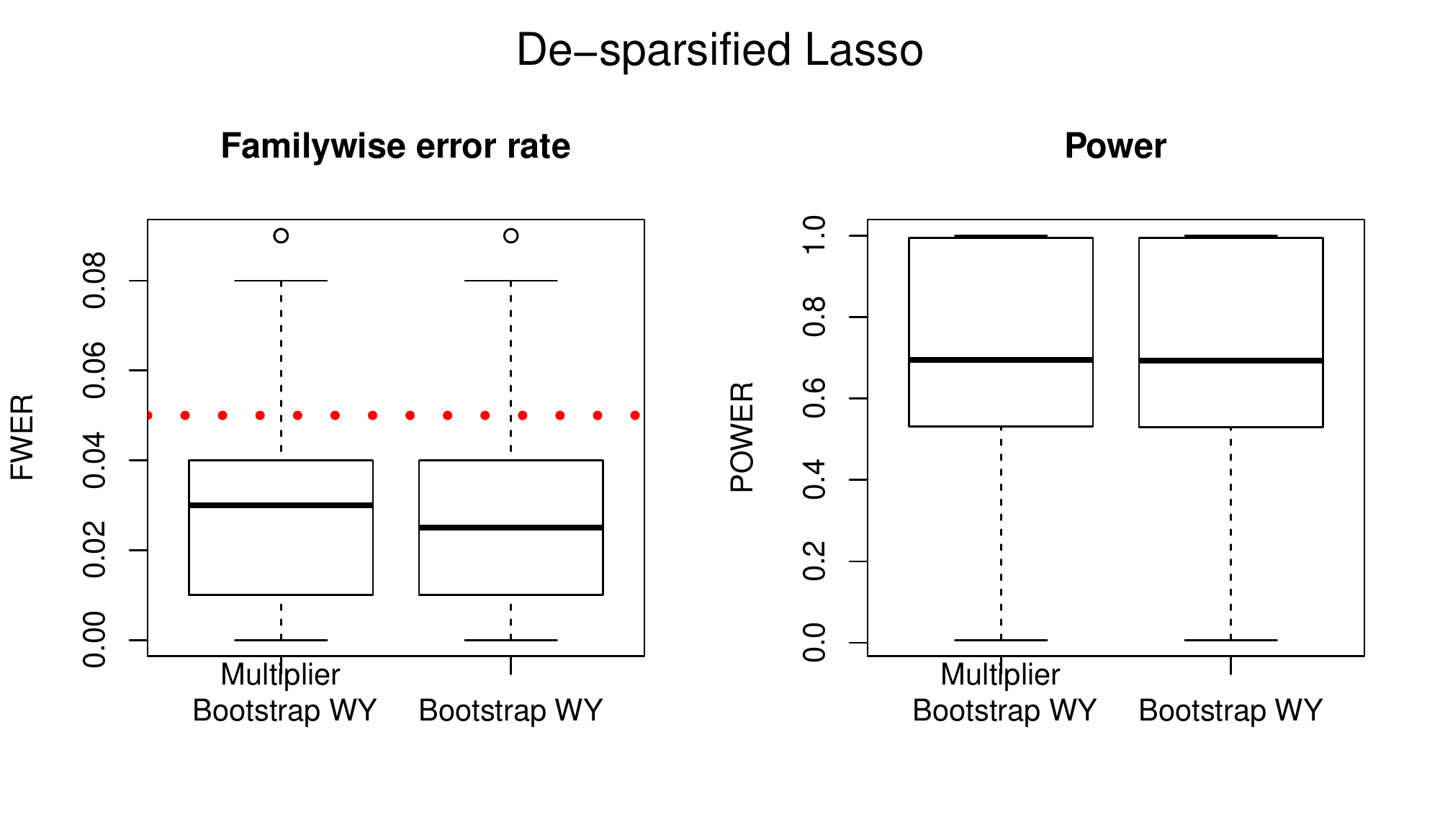}
\caption{The same plot as Figure \ref{fig:powerfwer-toeplitzgauss} but for
    \textbf{homoscedastic chi-squared errors} and showing also the 
    multiplier bootstrap.}
  \label{fig:powerfwer-toeplitzchisqmultiplier}
\end{figure}

\begin{figure}[!htb]
  \centering
\includegraphics[scale=0.6]{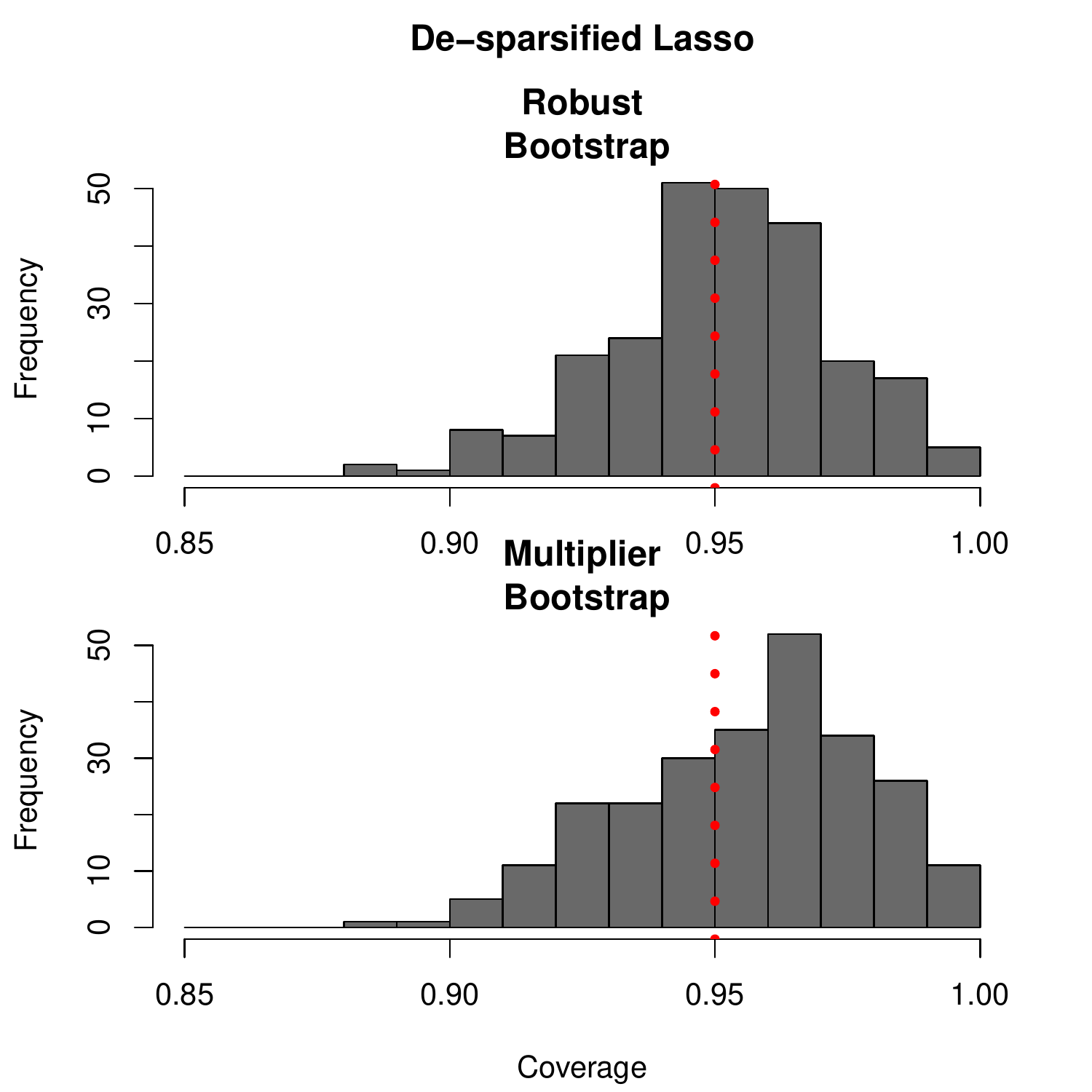}
  \caption{The same plot as Figure \ref{fig:cihist-toeplitzgauss} but for
    \textbf{heteroscedastic non-Gaussian errors} and showing also the 
    multiplier bootstrap (the latter with non-robust studentization).} 
  \label{fig:cihist-heteroscedmultiplier}
\end{figure}

\begin{figure}[!htb]
  \centering
\includegraphics[scale=0.8]{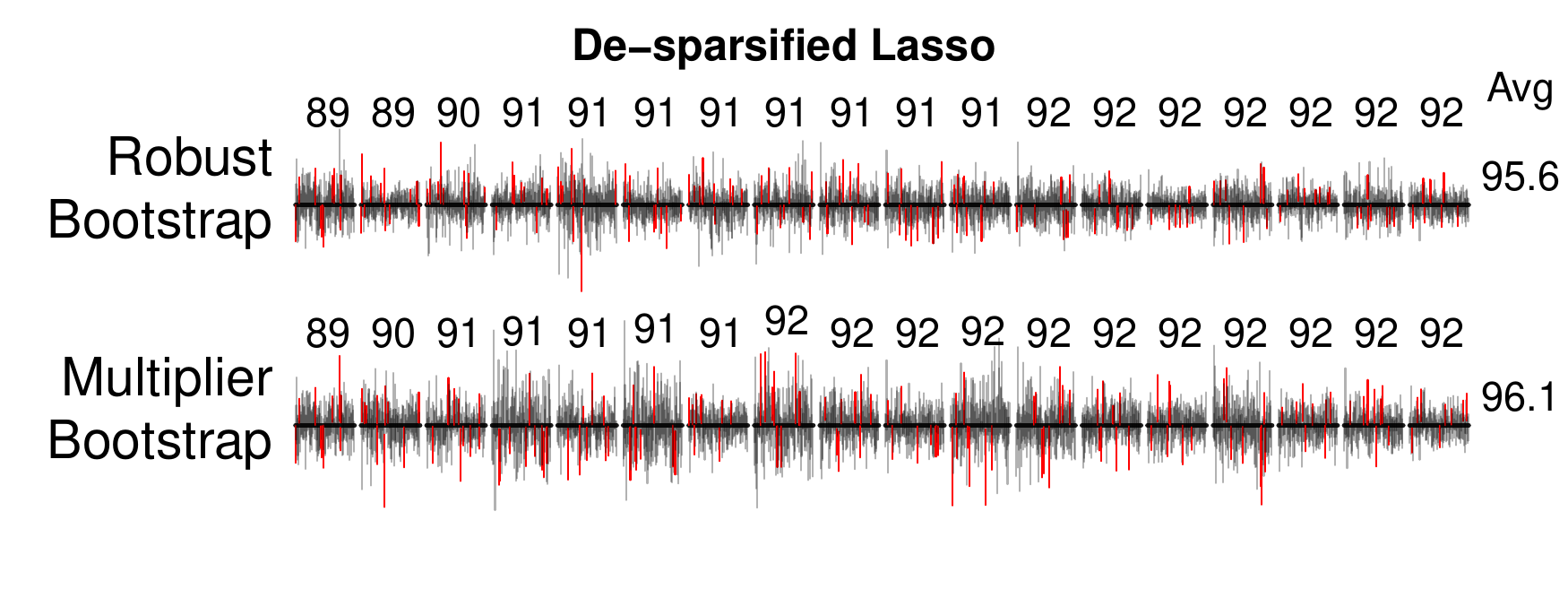}
  \caption{The same plot as Figure \ref{fig:ci-toeplitzgauss} but for
    \textbf{heteroscedastic non-Gaussian errors} and showing also the 
    multiplier bootstrap (the latter with non-robust studentization).} 
  \label{fig:ci-heteroscedmultiplier}
\end{figure}

\begin{figure}[!htb]
  \centering
\includegraphics[scale=0.8]{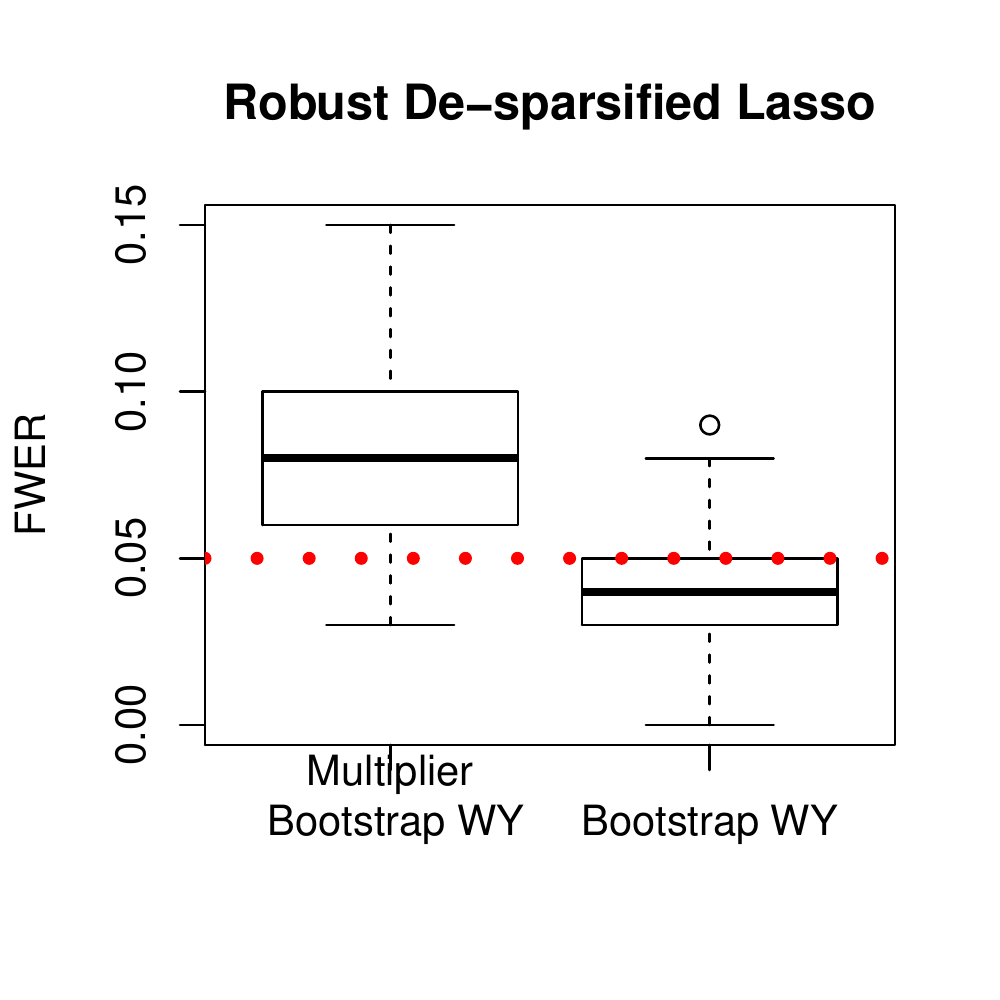}
  \caption{The same plot as Figure \ref{fig:powerfwer-toeplitzgauss} but for
    \textbf{heteroscedastic non-Gaussian errors} with robust
      studentization and showing also the multiplier bootstrap.}
  \label{fig:powerfwer-heteroscedmultiplier}
\end{figure}

For the homoscedastic Gaussian errors, the confidence intervals comparison
can be found in Figures \ref{fig:cihist-toeplitzgaussmultiplier} and
\ref{fig:ci-toeplitzgaussmultiplier}, the multiple testing results can be
found in Figure \ref{fig:powerfwer-toeplitzgaussmultiplier}. For the
homoscedastic non-Gaussian errors, the confidence intervals 
comparison can be found in Figures \ref{fig:cihist-toeplitzchisqmultiplier}
and \ref{fig:ci-toeplitzchisqmultiplier}, the multiple testing 
results can be found in Figure
\ref{fig:powerfwer-toeplitzchisqmultiplier}. For the heteroscedastic
non-Gaussian errors, the confidence intervals 
comparison can be found in Figures \ref{fig:cihist-heteroscedmultiplier}
and \ref{fig:ci-heteroscedmultiplier}, the multiple testing 
results can be found in Figure \ref{fig:powerfwer-heteroscedmultiplier}. 

The differences in performance seem very minimal. The wild bootstrap has no
problem dealing with heteroscedastic errors, it performs similar to the
robust bootstrap approach in our example.

\subsubsection{Homoscedastic non-Gaussian errors - Robust estimators}
\label{app:additionalrobust}
\begin{figure}[!htb]
  \centering
\includegraphics[scale=0.6]{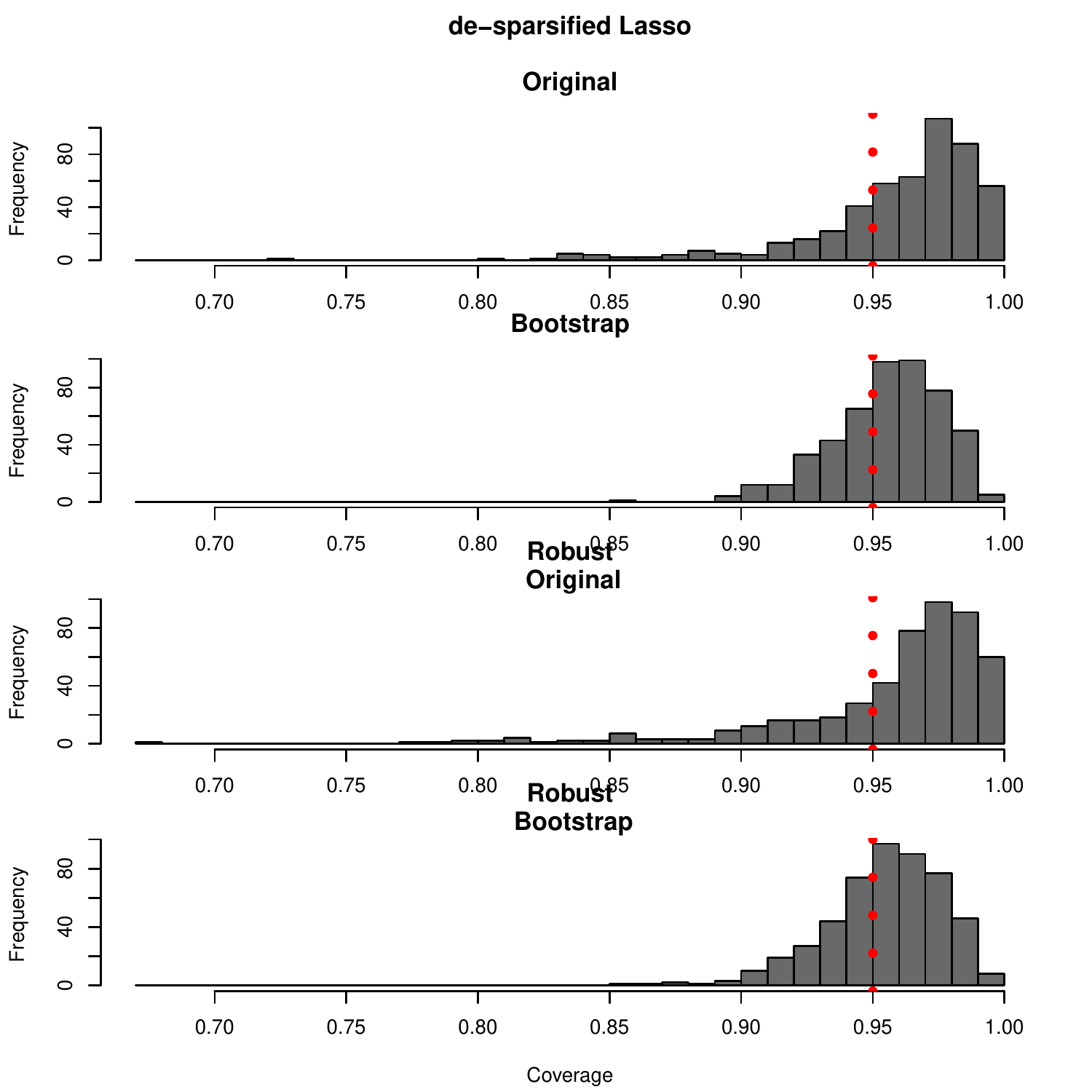}
  \caption{The same plot as Figure \ref{fig:cihist-toeplitzchisq} with
    \textbf{homoscedastic chi-squared errors} but also
    including the robust alternatives of the estimators.} 
  \label{fig:cihist-toeplitzchisq-alsorobust}
\end{figure}

\begin{figure}[!htb]
  \centering
\includegraphics[scale=0.7]{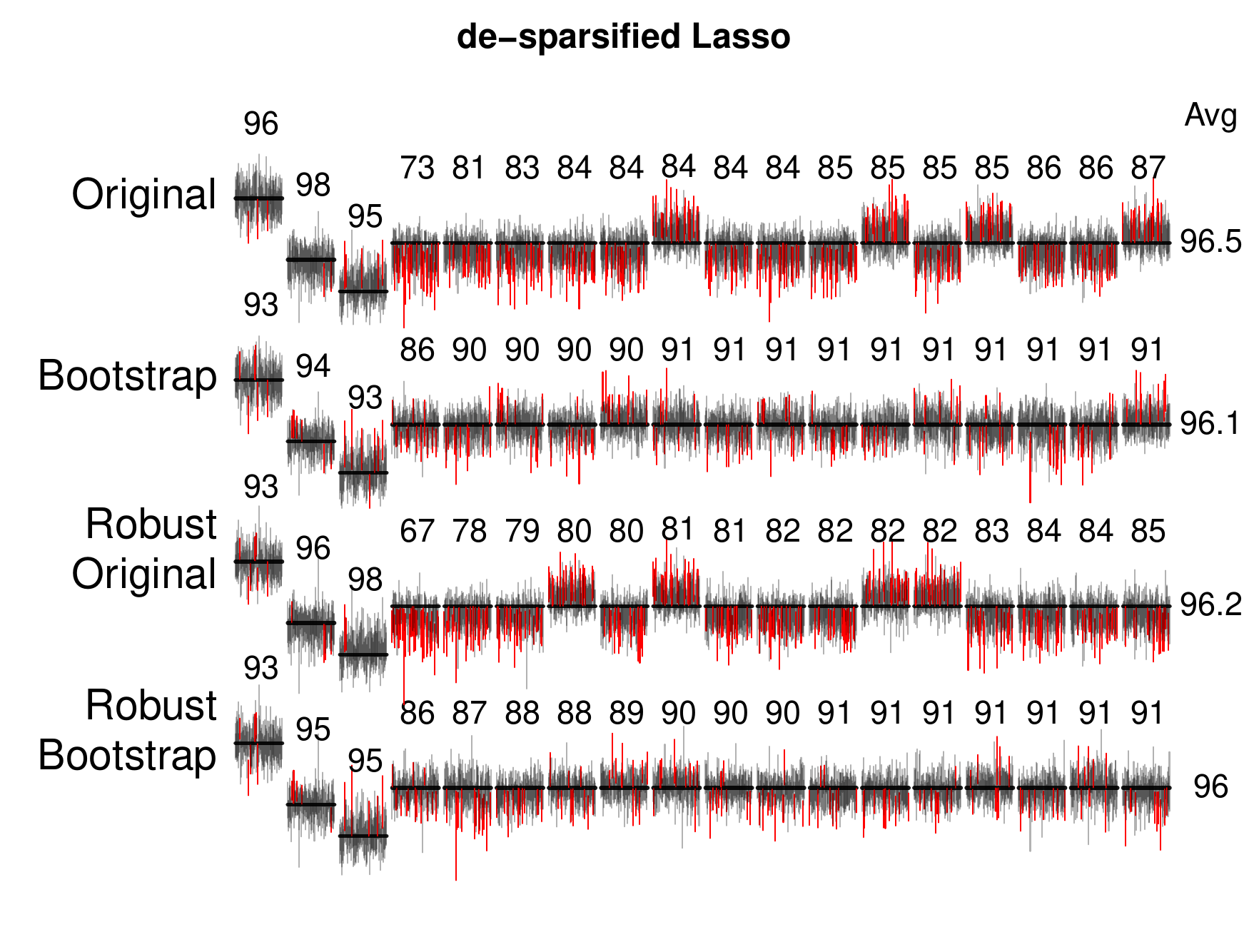}
  \caption{The same plot as Figure \ref{fig:ci-toeplitzchisq} with
    \textbf{homoscedastic chi-squared errors} but also
    including the robust alternatives of the estimators.  The original
    estimator (both robust 
    and non robust) has quite some bias for a few coefficients, which
    results in a lower than desired coverage for those coefficients.} 
  \label{fig:ci-toeplitzchisq-alsorobust}
\end{figure}

We present in Figures \ref{fig:cihist-toeplitzchisq-alsorobust} and
\ref{fig:ci-toeplitzchisq-alsorobust} some results about individual
inference when using the robust studentization in presence of homoscedastic
errors: an efficiency loss is not really visible.

\end{document}